UNIVERSITY OF CALIFORNIA

Los Angeles

**The Structures, Functions, and Evolution of Sm-like Archaeal Proteins (SmAPs)**

A dissertation submitted in partial satisfaction of the
requirements for the degree Doctor of Philosophy in
Biochemistry and Molecular Biology

by

Cameron Mura

2002

The dissertation of Cameron Mura is approved.

_______________________________
James U. Bowie

_______________________________
Allan Tobin

_______________________________
David Eisenberg, Committee Chair

University of California, Los Angeles

2002

*To anyone*

*with whom*

*I've discussed*

*science*

# TABLE OF CONTENTS

# LIST OF FIGURES AND TABLES

## Chapter 1: Introduction to RNA metabolism and Sm proteins

## Chapter 2: The crystal structure of a heptameric archaeal Sm protein: Implications for the eukaryotic snRNP core

**Chapter 3: The oligomerization and ligand-binding properties of Sm-like archaeal proteins (SmAPs)**

**Chapter 4: The structure and potential function of an archaeal homolog of survival protein E (SurEα)**

**Chapter 5: The crystal structure of a Nudix protein from *Pyrobaculum aerophilum* reveals a dimer with two intersubunit β-sheets**

## ACKNOWLEDGEMENTS

All of the research in this dissertation was conducted in the laboratory of my thesis advisor – David Eisenberg. I am greatly indebted to him for continuous support of my scientific endeavors, and allowing me the room to freely investigate proteins that I find interesting. His encouragement was obvious from my first contact with him (as a rotation student), and has been constant over the past several years. I would also like to thank the other two members of my dissertation committee – Jim Bowie and Allan Tobin. Their enthusiasm and interest in biochemistry has always been apparent and inspirational.

Also, I thank Jim Bowie, William Gelbart, and Todd Yeates for fielding various structural biology, physical chemistry, or math questions I have had over the years. Doug Black and Guillaume Chanfreau have been immensely helpful with my forays into the RNA world of Sm proteins. I appreciate the receptiveness of these professors. The dedicated crystallization efforts of Anna Kozhukhovsky cannot go unmentioned. The crystallographic expertise of Dan Anderson, Duilio Cascio, and Mike Sawaya has been indispensable – without their help I would have been cutoff at the knees. Other members of the UCLA structural biology community who have provided encouragement or scientific advice include: Mari Gingery, Celia C.W. Goulding, Rob Grothe, Magdalena Ivanova, Gary Kleiger, Sangho Lee, Yanshun Liu, Jeanne Perry, and Ioannis Xenarios. I especially thank Celia, Duilio, Mari, and Marcín Apostol for providing an indescribable work environment over the past couple years.

For their pivotal roles in my education prior to graduate school, I would like to thank my high school chemistry teacher (Dorothy Flanagan) and the chemistry advisor at Georgia Tech who introduced me to crystallography (Loren D. Williams). Also, all of my family has been supportive of my academic interests. I especially thank my parents (Aubin and Pari Mura) for promotion of my interests in nature since roughly age seven, and my brother (Cambiz Mura) for sharing my interests in and enthusiasm for math and science. My wife (Linda Columbus) is a talented biochemist who also provided much needed encouragement during my polyglutamine era.

I had the pleasure of collecting most of the crystallographic data for this dissertation at beamlines 5.0.2 of the Advanced Light Source (Berkeley National Lab) and X8C of the National Synchrotron Light Source (Brookhaven National Lab). The funding agencies that supported various portions of this research include a National Science Foundation Graduate Research Fellowship (NSF 01-146) and a UCLA Dissertation Year Fellowship. Their financial support made this work possible.

Chapter 2 of this dissertation is used with permission of the National Academy of Sciences, and is a reprint of the following article: *The crystal structure of a heptameric archaeal Sm protein: Implications for the eukaryotic snRNP core* by Mura, C., Cascio, D., Sawaya, M.R., and Eisenberg, D. *Proc. Natl. Acad. Sci. USA* (2001), *98*, 5532-5537.

Chapter 3 of this dissertation is an adapted pre-print of the following manuscript: *The oligomerization and ligand-binding properties of Sm-like archaeal proteins (SmAPs)* by Mura, C., Kozhukhovsky, A., Gingery, M., Phillips, M., and Eisenberg, D. (prepared for submission to the journal *Protein Science*). I would like to thank Drs. Duilio Cascio and Michael Sawaya (UCLA) for diffraction data collection at the synchrotron.

Chapter 4 of this dissertation is an adapted pre-print of the following manuscript: *The structure and potential function of an archaeal homolog of survival protein E (SurEα)* by Mura, C., Katz, J., Clarke, S.G. and Eisenberg, D. (prepared for submission to the *Journal of Molecular Biology*). The following people were helpful in this work: Drs. Duilio Cascio and Michael Sawaya (UCLA) provided crystallographic advice; Dr. Thomas Schneider (University of Göttingen) assisted with the ESCET program; and Dr. Sorel Fitz-Gibbon (UCLA) provided the phosmid vector containing the *Pae surE* gene. Prof. Eisenberg directed and supervised the research that forms the basis of this chapter.

Chapter 5 is an adaptation of the following published article: *Structure of a Nudix protein from* Pyrobaculum aerophilum *reveals a dimer with two intersubunit β-sheets* by Wang, S., Mura, C., Sawaya, M.R., and Eisenberg, D. *Acta Crystallographica* (2002), *D58*, 571-578. I thank Dr. Hanjing Yang (UCLA) for helpful Nudix protein discussions.

| | |
|---|---|
| 1975 | Born in Shiraz, Iran |
| 1996 | B.S. in Chemistry with Highest Honors<br>Georgia Institute of Technology |
| 1996-1997 | Teaching Assistant<br>Department of Chemistry & Biochemistry<br>University of California, Los Angeles |
| 1996-1999 | Predoctoral Fellow<br>National Science Foundation Graduate Fellowship<br>Department of Chemistry & Biochemistry<br>University of California, Los Angeles |
| 2001-2002 | Predoctoral Fellow<br>UCLA Dissertation Year Fellowship<br>Molecular Biology Institute<br>University of California, Los Angeles |

## PUBLICATIONS

**Oxyanion-mediated inhibition of serine proteases**

S.R. Presnell, G.S. Patil, C. Mura, K.M. Jude, J.M. Conley, J.A. Bertrand, C.-M. Kam, J.C. Powers, & L.D. Williams. *Biochemistry* (1998), *37*, 17068-17081

**The crystal structure of a heptameric archaeal Sm protein: Implications for the eukaryotic snRNP core**

C. Mura, D. Cascio, M.R. Sawaya, & D. Eisenberg. *Proc. Natl. Acad. Sci. USA* (2001), *98*, 5532-5537

**Structure of a Nudix protein from *Pyrobaculum aerophilum* reveals a dimer with two intersubunit β-sheets**

S. Wang, C. Mura, M. Sawaya, & D. Eisenberg. *Acta Crystallographica Sect. D* (2002), *D58*, 571-578

**The structure and potential function of an archaeal homolog of survival protein E (SurEα)**

C. Mura, J. Katz, S. Clarke, & D. Eisenberg. (prepared for submission to *J. Mol. Biol.*)

**The oligomerization and ligand-binding properties of Sm-like archaeal proteins**

C. Mura, A. Kozhoukhovsky, M. Gingery, M. Phillips, & D. Eisenberg.
(prepared for submission to *Prot. Sci.*)

**Synthetic *versus* biological catalysis of an unnatural transformation: Comparisons of chemzyme, ribozyme, and abzyme catalysis of a biphenyl stereoisomerization**

X. Wang, C. Mura, A. Shvets, S.C. Zimmerman, & K.N. Houk.
(in preparation for *J. Org. Chem.*)

**ABSTRACT OF THE DISSERTATION**

**The Structures, Functions, and Evolution of Sm-like Archaeal Proteins**

by

Cameron Mura

Doctor of Philosophy in Biochemistry and Molecular Biology

University of California, Los Angeles, 2002

Professor David Eisenberg, Chair

Sm proteins were discovered nearly 20 years ago as a group of small antigenic proteins ($\approx$ 90-120 residues). Since then, an extensive amount of biochemical and genetic data have illuminated the crucial roles of these proteins in forming ribonucleoprotein (RNP) complexes that are used in RNA processing, *e.g.*, spliceosomal removal of introns from pre-mRNAs. Spliceosomes are large macromolecular machines that are comparable to ribosomes in size and complexity, and are composed of uridine-rich small nuclear RNPs (U snRNPs). Various sets of seven different Sm proteins form the cores of most snRNPs. Despite their importance, very little is known about the atomic-resolution structure of snRNPs or their Sm cores. As a first step towards a high-resolution image of snRNPs and their hierarchic assembly, we have determined the crystal structures of archaeal homologs of Sm proteins, which we term Sm-like archaeal proteins (SmAPs).

Beginning with a *Pyrobaculum aerophilum* (*Pae*) structural genomics pilot project, we determined the structure of *Pae* SmAP1. This structure provided the first direct evidence for a toroid-shaped Sm homoheptamer at the snRNP core, and provided many insights and implications for SmAP evolution and RNA binding in Sm cores. Then, in order to extend these results, we solved the structure of *Pae* SmAP1 and a heptameric methanobacterial SmAP (*Mth* SmAP1) bound to uridine-5'-monophosphate (UMP); the uracil bases line the heptamer pore in the *Mth* ligand-bound structure, and suggest a more specific model for RNA binding than we were able to propose earlier.

Further characterization of the oligomerization and ligand-binding properties of *Mth* and *Pae* SmAP1s has allowed us to conclude that: *(i)* SmAPs form several oligomers besides the archetypal heptamer, including 14-mers and fibrillar polymers; *(ii) Mth* SmAP1 and *Archaeoglobus fulgidus* (*Afu*) SmAP1 recognize uracil bases in a nearly identical manner, suggesting a conserved RNA-binding mode for SmAPs; and *(iii) Pae* and *Mth* SmAP1s gel-shift supercoiled DNA, perhaps by nonspecific binding to single-stranded DNA. Our sequence analyses shed light on the evolution of Sm proteins: the SmAP module is a phylogenetically well-conserved domain that probably gave rise to modern (eukaryotic) Sm *hetero*heptamers *via* gene duplication and neutral drift. Crystal structure determinations for *Pae* SmAP2 and SmAP3 proteins are currently in progress, and will deepen our understanding of Sm protein function and evolution.

As part of the same *Pae* structural genomics project, we solved two structures that are unrelated to the SmAP work: an archaeal homolog of survival protein E (*Pae* SurEα) and a putative *Pae* Nudix protein. The final two chapters of this dissertation describe these structures and their significance.

**A synopsis of the dissertation**

The research reported in this dissertation is primarily concerned with Sm-like archaeal proteins (SmAP), which are archaeal homologs of eukaryotic Sm proteins. The Sm family is a broad, phylogenetically well-conserved set of proteins that function in several aspects of RNA processing, as described in an introductory chapter on the background and significance of Sm proteins in RNA metabolism (Chapter 1). Eukaryotic Sm and Sm-like (Lsm) proteins bind to small nuclear RNAs to form the core of several ribonucleoprotein (RNP) particles, most notably the uridine-rich small nuclear RNPs (U snRNP) that constitute the spliceosome (which in turn splices introns out of pre-mRNA). Since the biological function of Sm proteins has been most characterized in the context of U snRNPs and their role in spliceosome assembly and intron excision, the emphasis in Chapter 1 is on what is known about the canonical human and yeast Sm proteins and their involvement in intron splicing. However, the vast diversity of RNA metabolism – together with the central role of Sm proteins in forming the cores of U snRNPs – makes it likely that Sm, Lsm, and SmAP proteins are involved in a wide range of RNP complexes in addition to U snRNPs, and may justify the generalization of our SmAP results to include non-splicing functions (*e.g.*, Sm-like proteins as RNA chaperones).

Over a decade of biochemical and genetic data suggests that a heteroheptamer of Sm proteins is the biologically relevant species at the core of eukaryotic snRNPs. In order to obtain a high-resolution image of the structure and function of Sm proteins, we determined the crystal structure of SmAP1 from *Pyrobaculum aerophilum* (*Pae*) at a resolution of 1.75 Å (Chapter 2 is a reprint of the article which describes this structure). *Pae* SmAP1 was found to form a homoheptameric ring perforated by a cationic pore, thus providing the first direct evidence for such an assembly in eukaryotic snRNPs. Additionally, the structure: *(i)* showed that *Pae* SmAP homodimers are structurally similar to two human Sm heterodimers; *(ii)* supported a gene duplication model of Sm protein evolution; and *(iii)* suggested features that may be important in RNA binding (such as the cationic pore). In order to extend or generalize these results, we then studied another archaeal Sm protein – *Methanobacterium thermautotrophicum* (*Mth*) SmAP1.

Chapter 3 describes the oligomerization and ligand-binding properties of *Mth* and *Pae* SmAP1s (in preparation for publication). The *Mth* SmAP1 structure was determined in three crystal forms, each with different heptamer packings. In one of the forms an *Mth* SmAP1 14-mer co-crystallized with uridine-5'-monophosphate (UMP), and showed that our earlier, speculative model for RNA binding is probably incorrect (intercalation of uracil bases between conserved pore side chains suggests that RNA may wrap *around* the pore, not thread *through* it). The five *Pae* and *Mth* crystal structures contain various small molecules bound in what appear to be conserved ligand-binding sites. We fortuitously discovered that *Pae* and *Mth* SmAP1 gel-shift negatively supercoiled DNA. In addition to presenting these ligand binding properties, Chapter 3 describes the following features of the oligomerization of SmAPs: *(i) Pae* SmAP1 forms disulfide-bonded 14-mers, whereas *Mth* SmAP1 is almost exclusively heptameric *in vitro*; *(ii) Pae* SmAP1 forms sub-heptameric states when its inter-subunit disulfide bonds are reduced; *(iii)* both *Pae* and *Mth* SmAP1 polymerize into polar fibers by the head-to-tail stacking of heptamers. Our crystal structures of *Pae* SmAP1 in two crystal forms and *Mth* SmAP1 in three crystal forms corroborate these novel oligomerization and polymerization properties of SmAPs.

The final two chapters describe two other crystal structures that we determined – an archaeal homolog of survival protein E (*Pae* SurEα, Chapter 4) and an archaeal homolog of Nudix proteins (Chapter 5). These proteins are unrelated to the SmAP work, and were solved in the course of a *P. aerophilum* structural genomics pilot project (although a recent report of the Sm-like properties of *E. coli* Hfq protein suggests a weak link between Sm proteins and SurE – see the introduction in Chapter 3). One of the interesting findings of this work was that crystalline *Pae* SurEα is an inhomogeneous mixture of domain swapped and non-domain swapped dimers. The account of the SurEα structure in Chapter 4 is an adaptation of a manuscript submitted for publication, and the description of the *Pae* Nudix structure in Chapter 5 is adapted from a published article (see chapter title pages for citations). The work reported in Chapter 5 was done in collaboration with Dr. Shuishu Wang of UCLA. After creating a *Pae* Nudix M16L point mutant by site-directed mutagenesis, I over-expressed, purified, and crystallized this protein in a $P2_1$ form ("Native-2" in Chapter 5). I then used the structure of Dr. Wang's $P2_12_12_1$ "Native-1" dimer as a molecular replacement search model for the Nudix tetramer found in the asymmetric unit of the $P2_1$ form.

The Appendix provides some of the more useful scripts that were used in the research of this dissertation. All of these utilities were written in either the UNIX C shell or the Perl scripting language, and a brief description of each script is provided at the beginning of the Appendix. Two notes regarding these programs are: *(i)* the **scripted_glrf.sh**, **alter.pl**, and **process_bigrun.pl** trio uses a published program (GLRF) to calculate cross-rotation functions that are systematically varied over the integration radius and resolution limits of diffraction data, and processes the output in a user-friendly format; and *(ii)* the rare codon calculator accepts query DNA sequences at the following URL: `http://www.doe-mbi.ucla.edu/cgi/cam/racc.html`.

As a final note on the SmAP work, sequence analysis suggests that *P. aerophilum* has two other Sm-like proteins, *Pae* SmAP2 and SmAP3. We have cloned and purified these proteins, and their structure determinations are in progress for the following reasons: *(i)* SmAP2 and SmAP3 co-crystallized with UMP, so their likely RNA-binding sites may be revealed and compared to known uracil binding sites; *(ii)* due to its Loop-4 insertion, the SmAP3 sequence more closely resembles certain eukaryotic Sm proteins than does any other SmAP; *(iii)* the SmAP2 and SmAP3 paralogs may provide insight into gene duplication as a mechanism for the evolution of eukaryotic Sm *hetero*heptamers; *(iv)* the collection of *Pae* SmAP1, SmAP2, and SmAP3 crystal structures will provide a high-resolution picture of the entire Sm protein complement of *Pae*, and serve as a starting point for further biochemical and biophysical experiments that will address the function of SmAPs in archaea (which presumably do not contain snRNPs). Progress towards the crystal structures of these two new SmAPs is not reported in this dissertation, although we are far along: diffraction data to 2.0 Å have been processed from native SmAP2 crystals, and phases have been calculated to 2.0 Å for derivatized SmAP3 crystals (see pg. 32). Preliminary results suggest that SmAP2 and SmAP3 are also heptamers, and these structures are anticipated within the next few months.

**Chapter 1:**

**Introduction to RNA metabolism and Sm proteins**

*Overview of RNA metabolism*

RNA metabolism is an extremely complex and multifaceted realm of cellular life. It includes the synthesis, degradation, post-transcriptional modification, and utilization of a variety of RNA species for biochemical tasks that range from the very general (protein synthesis *via* "messenger RNAs") to the highly specific (RNA interference *via* "small interfering RNAs"). The known classes of RNA species include the familiar messenger RNA (mRNA), transfer RNA (tRNA), and ribosomal RNA (rRNA) forms, as well as a multitude of novel, unconventional types of RNA that are only now being identified and characterized. The growing number of categories of RNA includes: *(i)* small nuclear RNAs (snRNAs) that are involved in pre-mRNA processing; *(ii)* small nucleolar RNAs (snoRNAs) that are involved in pre-rRNA processing; *(iii)* RNAs that are responsible for pre-tRNA 5' end processing, such as RNase P RNA; and *(iv)* a recently discovered class of small (≈ 20-30-nucleotide), non-coding microRNAs (miRNA) that attenuate post-transcriptional processing of mRNA *via* small interfering RNAs (siRNAs) or small temporal RNAs (stRNAs)[1,2,3] (see review in ref. 4). Precise regulation of these RNA metabolic processes by other RNAs, proteins, or small molecule ligands is another characteristic feature of RNA metabolism.

In fact, the complexity and diversity of the reactions in which RNA engages has led to the "RNA World" hypothesis, in which RNA can act as substrate, product, catalyst (ribozyme), or structural cofactor (see ref. 5 for a treatise on the RNA world). The basic idea behind this view of the origin of life is that pre-biotic, self-replicating RNAs existed, and that some of these RNAs – because they could copy themselves imperfectly – formed a platform for gradual evolution into the current RNA-DNA-protein world *via* early living systems based only on RNA and protein. Presumably, this intermediate RNA/protein-only world was the birthplace of many of the fundamental ribonucleoprotein (RNP) assemblies that are utilized in modern organisms (*e.g.*, the ribosome). The existence of RNP complexes in archaea, eubacteria, and eukaryotes underscores their importance (and early appearance) in the evolution and biochemistry of each of these three kingdoms.

A modular approach to studying RNPs and their roles in RNA metabolism is shown in Fig. 1.1. Some of the known forms of RNA processing are given in the top level of this hierarchy: *(i)* rRNA processing by nucleolar snoRNPs (reviewed in ref. 6); *(ii)* RNase P-based splicing and maturation of tRNA (reviewed in ref. 7); *(iii)* hairpin formation and processing of the 3' end of histone mRNA by U7 snRNP (ref. 8 and references therein); *(iv)* mRNA decapping and decay;[9] and *(v)* chromosome maintenance by the RNA component of telomerase.[10] Each of these processes employs RNPs that contain Sm or Sm-like (Lsm) proteins. Indeed, a central theme of Fig. 1.1 – and the motivation behind this dissertation – is that many RNA processing events (on a cellular scale) can be traced back to the Sm proteins (on a molecular scale). The roles of Sm proteins in the spliceosomal removal of introns from precursor mRNAs have been investigated more than in any of the other types of RNA processing shown in Fig. 1.1. Therefore, the remainder of this chapter will focus on mRNA processing by the spliceosome – the large macromolecular machine that catalyzes intron excision.

*Spliceosome-mediated excision of introns and mRNA processing*

Maturation of primary pre-mRNA transcripts into a final form that is ready for protein translation is a multi-step process that involves 5' cap methylation of the mRNA, 3' cleavage and polyadenylation, and intron excision. Introns are remarkably pervasive in eukaryotes; since being discovered in adenovirus by Sharp and coworkers over twenty years ago,[11] these non-coding, intervening sequences have been found in all three major kingdoms (eukaryotes, eubacteria, and archaea) and in all three major types of RNA (mRNA, tRNA, and rRNA). Such vast variation in the phylogenetic distribution and genomic location of introns is matched by the variation in known mechanisms for intron processing. For example, bacterial rRNA and tRNA introns are self-splicing (groups I, II, see refs. 12, 13 and references therein), while eukaryotic rRNA is processed by the snoRNPs of the nucleolus.[14] For pre-mRNA maturation, the last processing step (intron excision) occurs on the spliceosome.

The spliceosome is a large, transiently stable assembly of uridine-rich small nuclear RNPs (U snRNPs). Roughly the same size as the ribosome, this large RNP particle has a mass of $\approx 5 \times 10^6$ Da and sedimentation coefficient of 60S in one of its several catalytic states (Fig. 1.2).[15] It ligates two exons with concomitant release of the intron as a lariat structure, and it achieves this by catalyzing two successive trans-esterification reactions. These reactions occur on the pre-mRNA at the intron/exon junctions (*i.e.*, the 5' and 3' splice sites (SS)) and at a strictly conserved branch-point adenosine within the intron (and just upstream of an optional polypyrimidine region). In the first step, a catalytically active form of the spliceosome results from the dissociation of the U1 and U4 snRNPs; at this point, the spliceosome is composed of the U2, U5, and U6 snRNPs (the "A1" state in Fig. 1.2*(a)*). Bound by the A1 spliceosome, the 2' hydroxyl group of the branch-point adenosine attacks the phosphate center at the 5' SS and results in (or is driven by) formation of the "A2-2" state of the spliceosome. In the second step, the free 3' hydroxyl group at the 5' SS attacks the terminal phosphate of the 3' SS, resulting in ligation of the exons and release of the intron lariat. Completion of these steps with single nucleotide precision results in an mRNA with a correctly registered protein-encoding sequence.

The accepted view of spliceosome assembly (Fig. 1.2*(a)*) stipulates that several of the steps outlined above involve large-scale conformational changes and remodeling of the spliceosome, and that the active sites that catalyze these reactions exist only transiently (*i.e.*, they are not pre-formed as in most protein enzymes). However, new results from Abelson and colleagues refute this notion: their recent isolation of stable yeast "penta-snRNPs" (U1-U2-U4/U5•U6)[16] devoid of pre-mRNA suggests that a pre-formed spliceosome may exist irrespective of pre-mRNA binding. Indeed, several fundamental features of this catalytic process are unclear. One of the most intriguing questions is if the spliceosome is a ribozyme – *i.e.*, do U2, U5, or U6 snRNA mediate catalysis, or is one of the >80 spliceosomal proteins catalytically active? Similarities between the pre-mRNA splicing reaction and the self-splicing group II introns, as well as the $Mg^{2+}$-dependence of spliceosome activity, suggest that spliceosomal snRNA may be catalytic (reviewed in ref. 17). Other major questions are concerned with the spliceosome assembly pathway,[18] the structural roles of its snRNP constituents,[19] and regulation of the splicing cycle by non-spliceosomal factors. Since there is no evidence of a direct catalytic role for the Sm proteins in splicing, next we will consider the structural roles of Sm proteins in assembling the U snRNPs that form the spliceosome.

***Archetypal RNPs of the spliceosome: The assembly and structure of U snRNPs***

As shown in the two middle layers of Fig. 1.1, there are several types of spliceosomes, and an even more bewildering array of snRNPs (reviewed in refs. 20, 21). For instance, there is a minor class of spliceosomes that process the less abundant U12-type introns, and which are formed from U11, U12, U4atac, and U6atac snRNPs (instead of the corresponding U1, U2, U4, and U6 snRNPs of the major spliceosome).[21] Since the major spliceosome is the most thoroughly studied, the five core spliceosomal snRNPs (U1, U2, and the U4/U5•U6 tri-snRNP) are the best understood. A growing body of data suggests that many – if not most – structural characteristics are conserved between the U2/U12, U4/U4atac, *etc*. pairs of homologous snRNPs (ref. 19 and references therein).

U snRNP biogenesis in higher eukaryotes is illustrated in Fig. 1.3, which demonstrates the central role of Sm proteins in this pathway. Except for U6, all of the U snRNAs are transcribed by RNA polymerase II, modified with an $N^7$-monomethylguanosine ($m^7G$) cap, and then exported from the nucleus as part of a multi-protein export complex (U6 snRNA is transcribed by pol. III, acquires a 5' γ-monomethyl cap, and assembles with Lsm proteins entirely within the nucleus). Binding of cytoplasmic Sm heteromers to the exported snRNA occurs next (Fig. 1.3), and may be modulated by recently discovered interactions between some Sm proteins and the survival of motor neurons (SMN) protein complex.[22] A recent deluge of results has shown the symmetric dimethylation of arginine residues in some of the RG dipeptide repeats of Sm[23,24,25] and Lsm[26] proteins by a putative "methylosome",[27] and such post-translational methylation may be of great importance in targeting the Sm heteromers to the SMN complex. Formation of the U snRNP Sm core complex apparently triggers two more modifications to the snRNA (3' trimming and cap hypermethylation to 2,2,7-trimethyl-guanosine ($m_3$G)) that cause the snRNP Sm core to bind to a complex of snurportin-1 and importin-β. Along with an uncharacterized Sm core nuclear localization signal (NLS) receptor, this large import complex transits to the nucleus. Once there, the import complex dissociates, final nucleotide modifications are made to the snRNA (such as pseudouridylation), and U snRNP-specific proteins bind to give a mature U snRNP.

To simplify our understanding of this complicated biogenesis pathway, each U snRNP can be thought of as an RNA-protein complex composed of two parts: the respective U snRNAs (U1, U2, *etc*.), and several (up to dozens) proteins. Dissection of the snRNPs is illustrated in the two bottom layers of Fig. 1.1. The protein components fall into two classes: snRNP-specific proteins, *e.g.*, U2A' and U2B'' proteins of the U2 snRNP, and core proteins that are common to each snRNP (reviewed in ref. 28, 19). The snRNP-specific proteins most likely mediate highly specific RNA-RNA, protein-RNA, and protein-protein interactions, and function in ways that are unique to a given snRNP. Examples of such putative functions, chiefly inferred from sequence homology to known proteins, include putative DEAD- or DxxH-box RNA helicases, unwindases, GTPases, peptidyl *cis/trans* isomerases, and many RNA recognition motif (RRM)-containing proteins. In contrast, the function(s) of the shared core snRNP proteins – the Sm or Lsm proteins – are presumably much less specific.

### The biological functions of Sm proteins: formation of U snRNP cores

The primary structural or catalytic function of Sm proteins is still uncertain. Since being discovered as a group of eight small ($\approx$ 90-110 amino acids) antigenic proteins in autoimmune diseases such as systemic lupus erythematosus,[29,30] an extensive amount of biochemical and genetic data has shown that cytoplasmic Sm proteins assemble with snRNAs to form core snRNP complexes as described above and in Fig. 1.3. Sm proteins bind to short single-stranded regions of snRNA that are usually flanked by two stem-loop structures (Fig. 1.4). The consensus Sm binding site is the short uridine-rich sequence $PuAU_{\approx4-6}GPu$ (Pu = purine), although this selectivity is not very stringent and there is redundancy in Sm•RNA binding.[31] Sm binding is highly sensitive to modification of the flanking stem-loop structures and the snRNA Sm site, and even varies from one snRNA to another (Fig. 1.4).[32] The Lsm complex binds at the single-stranded 3' terminus of U6 snRNA (Fig. 1.4(c)), thus illustrating the variation in local secondary structure for Sm and Lsm binding sites. Stepwise binding to snRNA occurs with the Sm D1•D2 and E•F•G heteromers associating concomitantly to yield a "subcore" snRNP complex.[33] The final component to join the Sm complex (B/B'•D3) triggers the hypermethylation of the cap that, along with the Sm complex, forms a bipartite nuclear localization signal.

In addition to these two poorly characterized methylation and localization functions, Sm proteins probably mediate critical RNA-RNA and RNA-protein interactions near the snRNP core. Recent work by Zhang *et al.* showed that the extended and highly charged C-terminal tails of human Sm B, D1, and D3 are involved in sequence-independent binding to pre-mRNA substrate, and may stabilize U1 snRNP•pre-mRNA interactions.[34,35] An example of the possible role of Sm snRNP cores in recruiting other proteins to the assembling spliceosome is that the 70K protein of the U1 snRNP can be chemically cross-linked to the Sm B and D2 proteins.[36] Lying at the core of snRNPs, Sm proteins undoubtedly engage in a vast network of protein-protein and protein-RNA interactions. Thorough reviews by Will and Lührmann[37,19] have summarized what few interactions are known for Sm proteins, and suggest putative ones.

*Phylogenetic conservation of Sm proteins and their broader significance*

The Sm protein motif – traditionally described as Sm1 and Sm2 signature sequences joined by a variable linker – is strongly conserved in many species.[38] Stimulated by the current flood of genomic sequences, database searches show that Sm proteins are not exclusive to metazoans or other higher eukaryotes; indeed, several new Sm protein homologs have been found in eukaryotic species as divergent from humans as yeast[39] and trypanosomes.[40] Sm homologs also have been found in several archaeal species,[41] suggesting an ancient lineage of Sm proteins. An exciting recent discovery is that the *E. coli* Hfq protein (a host factor required for bacteriophage replication) is an Sm-like protein that preferentially binds to uridine-rich RNA[42,43] (along these lines, the *Herpesvirus saimiri* virus produces RNA transcripts that recruit host Sm proteins[44]). Thus, the phylogenetic diversity of Sm proteins is broader than was initially thought, including eubacteria as well as eukaryotes and archaea. These new Sm proteins expand the realm of possible Sm protein functions beyond splicing alone.

Biochemical and structural studies imply an ancient origin for Sm and Lsm proteins. Several sets of Lsm proteins have been discovered and characterized in organisms already known to have Sm proteins.[45] Electron microscopic and biochemical characterization of some of these proteins has verified their similarity to canonical Sm proteins,[46,38] and this dissertation treats them as roughly equivalent. The deep evolutionary origin of Sm proteins is further substantiated by the similarity of all known crystal structures of Sm proteins and their homologs (see Chapters 2 and 3).

***What is known about RNA processing in the archaea?***

Little is known about RNA processing in the archaea – particularly mRNA processing – mainly due to the lack of a convenient, genetically-manipulable model organism. However, many introns have been found in archaeal tRNA and rRNA genes (reviewed in ref. 47). Archaeal tRNA introns typically occur in the anticodon loop, while rRNA introns occur at diverse locations. Whereas bacterial introns are usually self-splicing (*e.g.*, group I introns), several forms of archaeal intron removal resemble their eukaryotic counterparts in terms of a protein requirement, *e.g.*, endonuclease-mediated splicing of archaeal tRNA introns (reviewed in ref. 7) or rRNA processing.[48] Recent discovery of archaeal homologs of U3 snoRNP proteins suggests that snoRNP-based rRNA processing may be another shared feature between archaea and eukaryotes.[49] Archaeal RNA processing other than intron removal is also beginning to be characterized, *e.g.*, tRNA 5'- and 3'- end processing.[50,51] Another RNA processing complex that may be conserved between archaea and eukaryotes is the exosome, a large complex of RNA exonucleases, RNA-binding proteins, and RNA helicases that mediates the 3'→5' degradation of many RNA species (including mRNA).[52]

It is generally assumed that archaea do not have standard, spliceosomal U snRNP-like particles, since archaeal mRNA processing is so poorly characterized and it is not known whether or not their pre-mRNAs contain introns. However, a recent report provides the first evidence for archaeal mRNA introns: the gene of an archaeal homolog of eukaryotic centromere-binding factor 5 (Cbf5p) was found to contain an intron that is spliced *in vivo* (as detected by reverse transcriptase-PCR).[53] The intron/exon boundaries in this gene are predicted to adopt bulge-helix-bulge motifs, which are the motifs recognized by the splicing endonucleases involved in processing of archaeal pre-tRNAs and rRNAs. Although the regulation and diversity of RNA metabolism in archaea may not be as sophisticated as in eukaryotes, these examples illustrate the complexities of archaeal RNA processing. The central role of the highly conserved Sm proteins in eukaryotic mRNA processing suggests that archaeal RNA processing may utilize Sm-like archaeal proteins in similar RNP assemblies (snRNP-like or otherwise).

### *Seeking an atomic-resolution understanding of snRNP cores and Sm proteins*

Until the crystal structure determinations of human Sm D1•D2 and D3•B heterodimers by Kambach *et al*. in 1999,[54] there was no atomic-level information for the structure of the Sm core complex. Several lines of biochemical and genetic data provided indirect evidence that the Sm core is a hetero-oligomer of seven Sm proteins. Ultrastructural investigations of U snRNP core particles by electron microscopy suggested that the Sm[55] and Lsm[46] cores are composed of a "doughnut-shaped heteromer" (Fig. 1.5*(a)*). The gradual realization that Sm and Lsm proteins are always found in groups of at least seven subtypes within the genome of a given organism lends credence to this structural model. The homoheptameric nature of an Sm-like protein from the archaeon *Archaeoglobus fulgidus* (*Afu*) was recently established by multivariate statistical analysis of electron micrographs (significantly, in biochemical assays this *Afu* Sm protein specifically bound oligo(U) RNAs).[56]

The fold of each human Sm monomer consists of a strongly bent five-stranded antiparallel β-sheet capped by a short N-terminal α helix (Fig. 1.5*(b)*). The human Sm D1•D2 and B•D3 structures show that the four monomers have nearly identical 3D structures to within 1 Å root mean-squared deviation (rmsd) for Cα atoms. Moreover, the D1•D2 and B•D3 heterodimers are nearly identical. The structure of the Sm-Sm protein interface in both heterodimers is conserved, and is created by the hydrogen-bonded juxtaposition of one β strand from each monomer. Kambach *et al*. speculated that the heterodimeric Sm crystal structures form the starting point for a seven-membered Sm ring arrangement, and created the first atomic model for the Sm core heteromer by extrapolation from the heterodimer structures.[54,57]

Efforts toward atomic-resolution structures of snRNP Sm cores have been highly successful over the past two years. In the low-resolution range, Stark *et al*. were able to construct a 10-Å resolution map of the entire U1 snRNP by cryo-electron microscopy.[58] Several known U1 components were identified (Fig. 1.5*c, d*), and these results underscored the importance of the Sm heptamer at the snRNP core. The solution structure of the SMN Tudor domain (which interacts with Sm proteins to form snRNP cores)[22] has provided an unexpected result: the SMN and Sm monomers have the same fold, and nearly identical structures. This raises the intriguing possibility that the SMN protein interacts with the Sm complex by forming mixed heteromers.

Finally, direct evidence for the structure of snRNP cores was provided by the crystal structures of three different heptameric Sm-like archaeal proteins, solved concurrently by Collins *et al*.,[59] Mura *et al*.,[41] and Törö *et al.*[60] Details of these structure determinations and motivation for current and future structural experiments are provided in Table 1.1. The determination and interpretation of these SmAP structures is the main subject of this dissertation. Elucidation of the crystal structure of an intact snRNP is the logical next step, and the U1 snRNP – which is one of the simplest and best-characterized snRNPs – may be a good candidate. Given the usual advantages of working with thermostable/archaeal proteins for structural studies, it also will be worthwhile to pursue the identification and characterization of possible archaeal snRNP-like particles.

## Figure and Table legends

**Figure 1.1: A modular approach to RNA metabolism.** Placing the Sm protein family in a biochemical context underscores the central importance of these proteins in RNA metabolism. This figure shows how several RNA processing events (on a cellular scale) can be traced back hierarchically to the Sm proteins (on a molecular scale). One of the most characterized and well-studied examples is the excision of introns from pre-mRNA, which can be dissected as: intron splicing $\Rightarrow$ major spliceosome $\Rightarrow$ U1, U2, U4/U6, and U5 snRNPs $\Rightarrow$ Sm core of snRNPs. This figure is by no means comprehensive (*i.e.*, not all of the known connections are shown), and new, non-mRNA splicing related examples are being discovered continuously.

**Figure 1.2: The spliceosomal cycle and the complexity of pre-mRNA processing *via* intron excision.** The spliceosome is a large, transiently stable macromolecular machine, and the complexity of its catalytic cycle is diagrammed in *(a)*. The precursor mRNA is shown at the top, with its branch point adenosine and splice sites noted. The upstream (5') and downstream (3') exons are denoted by E1 and E2, and are shown as ligated product mRNA in the bottom left corner (the intron lariat is shown above that). The five spliceosomal snRNPs are illustrated as gray-shaded shapes, and various states of this dynamic assembly are denoted by their yeast or mammalian labels (*e.g.*, "CC (E)" = commitment complex). An electron micrograph of a spliceosome bound to β-globin pre-mRNA is shown in *(b)*. Note the large dimensions of this particle, which is roughly the same size as the ribosome. Panel *(a)* was adapted from ref. 21 and *(b)* is from ref. 61.

**Figure 1.3: The central role of Sm proteins in snRNP assembly.** This diagram of the U1 snRNP biogenesis pathway exemplifies the key roles of Sm proteins in snRNP assembly (adapted from ref. 19). Some of the involved proteins are shown as shaded ovals, and the U1 snRNA is drawn as thin lines. A recently discovered interaction between the SMN protein and Sm D1 and D3 proteins is shown as a dashed arrow. The key assembly step with snRNA is shown in the dashed box to the right. Assembly of the snRNP Sm core complex triggers hypermethylation of the $m^7G$ guanosine cap to a trimethylated state ($m_3G$). Together with subsequent association of the Sm core complex and several other proteins (*e.g.*, snurportin-1), this results in nuclear import and the final stages of snRNP maturation.

**Figure 1.4: Secondary structures of some snRNAs and their Sm binding sites.** Predicted secondary structures are shown for a sample of spliceosomal snRNAs: U1 *(a, b)*, U2 *(a)*, and base-paired U4•U6 *(c)*. The shaded U1 snRNA stem-loops in *(a)* are colored in agreement with the cryo-EM reconstruction shown in Figures 5*(c, d)*. The uridine-rich Sm binding sites are indicated in each panel, as well as a few other features (*e.g.*, 5' guanosine caps or the branch point adenosine). The snRNA consensus sequence for Sm binding is $PuAU_{\approx4-6}GPu$ (Pu = purine); however, the stringency of this site is not very strict, as illustrated by the interrupting guanosine in the U1 site *(b)*. Note that the snRNA may base pair with both intronic and exonic elements *(a)*, and that the Sm core may mediate RNA-RNA as well as RNA-protein interactions. These secondary structures were taken from Burge *et al.*[21]

**Figure 1.5: Progress towards an atomic-resolution understanding of eukaryotic Sm proteins and snRNPs.** Panels *(a) – (d)* trace the progress in our knowledge of the structures of snRNPs and their Sm cores. This work has extended from the low resolution, ultrastructural studies that first suggested that Sm proteins form ring-shaped oligomers *(a)*, to a recent 10-Å resolution reconstruction of the U1 snRNP by cryo-electron microscopy *(c, d)*. The first Sm protein crystal structure that was determined – that of the human D1•D2 heterodimer – is shown in *(b)*. A model of the human Sm heptamer was built from this structure (the entire heptamer is unstable without snRNA and was not crystallized).[54] The experimental U1 snRNP envelope shown in *(c)* permits docking of various structural elements, such as a hypothetical model of the human Sm heptamer *(d)*. Although it is not based on direct experimental evidence, this model allows the Sm core of U1 snRNP to be identified unambiguously, and underscores the likely importance of the Sm heptamer in nucleating snRNP assembly. The EM is from Achsel *et al.*,[46] and *(c)* and *(d)* are from Stark *et al.*[58]

**Table 1.1: A summary of all known Sm and SmAP structures.** A comprehensive list of Sm and SmAP structures is provided. Details of a given structure determination are provided (resolution, PDB code, heptamer packings in various space groups), along with its significance. The first Sm protein structure was reported by Kambach *et al.*,[54] and the first heptameric structures of Sm-like proteins were revealed concurrently by Collins *et al.*,[59] Mura *et al.*,[41] and Törö *et al.*[60] As outlined in the *Synopsis* (pg. 5), structure determinations for *Pae* SmAP2 and SmAP3 are in progress.

**Figure 1.1: A modular approach to RNA metabolism.**

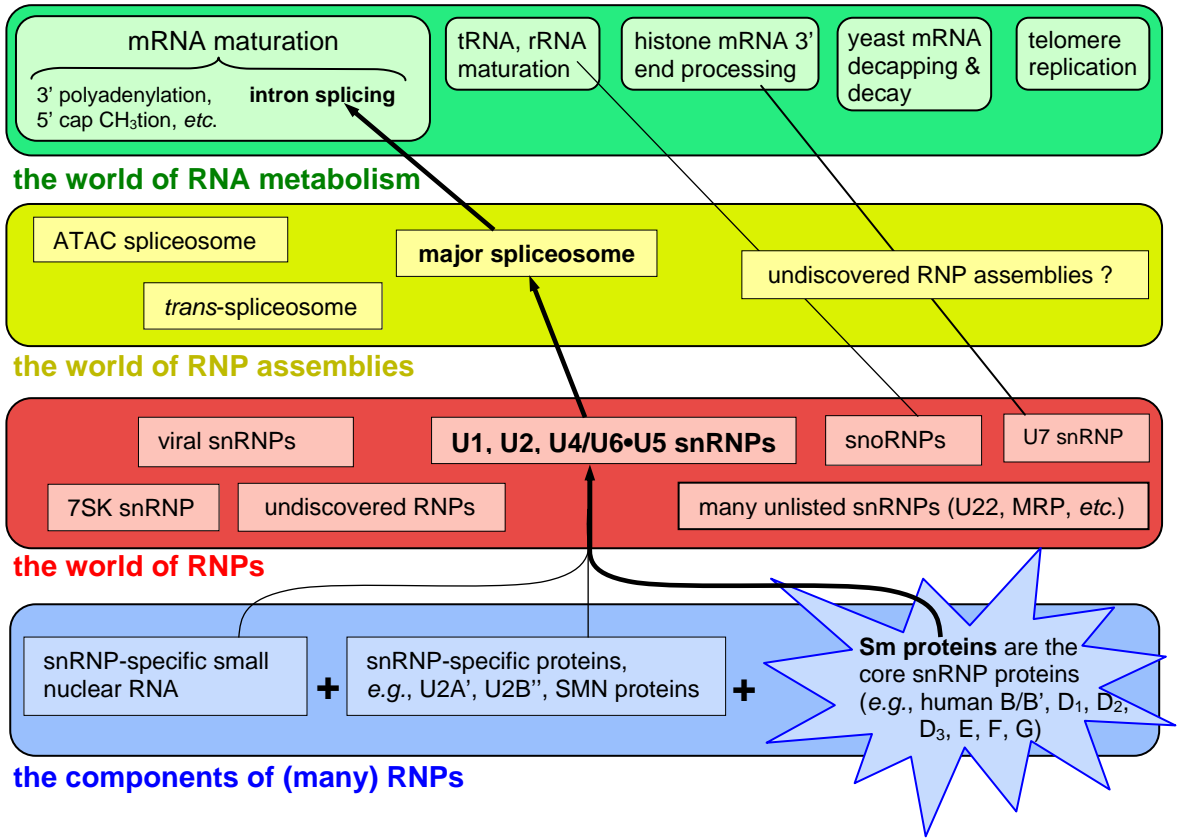

**Figure 1.2: The spliceosomal cycle and the complexity of pre-mRNA processing *via* intron excision.**

(a)

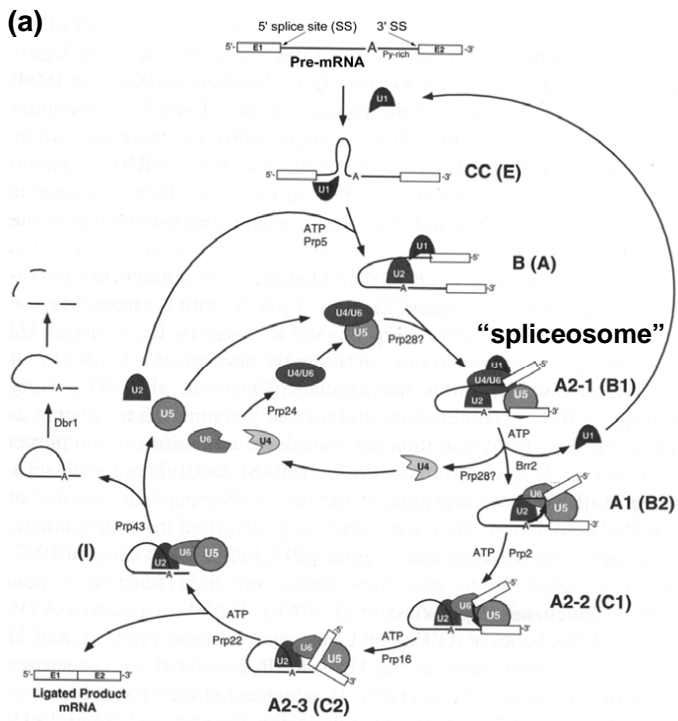

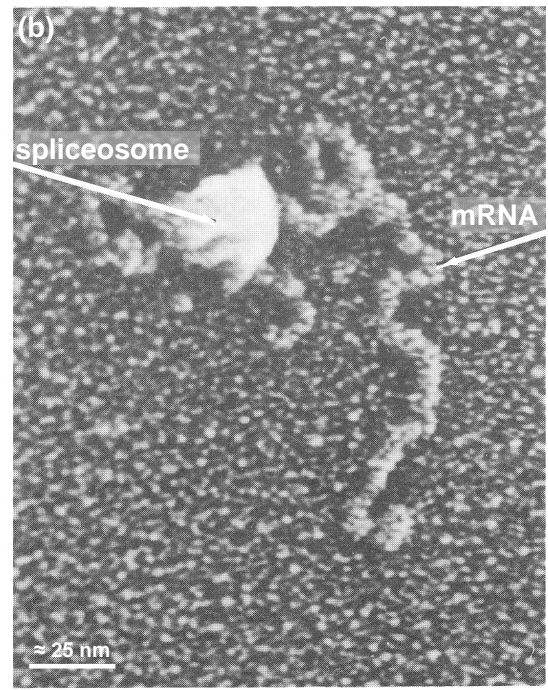

**Figure 1.3: The central role of Sm proteins in snRNP assembly.**

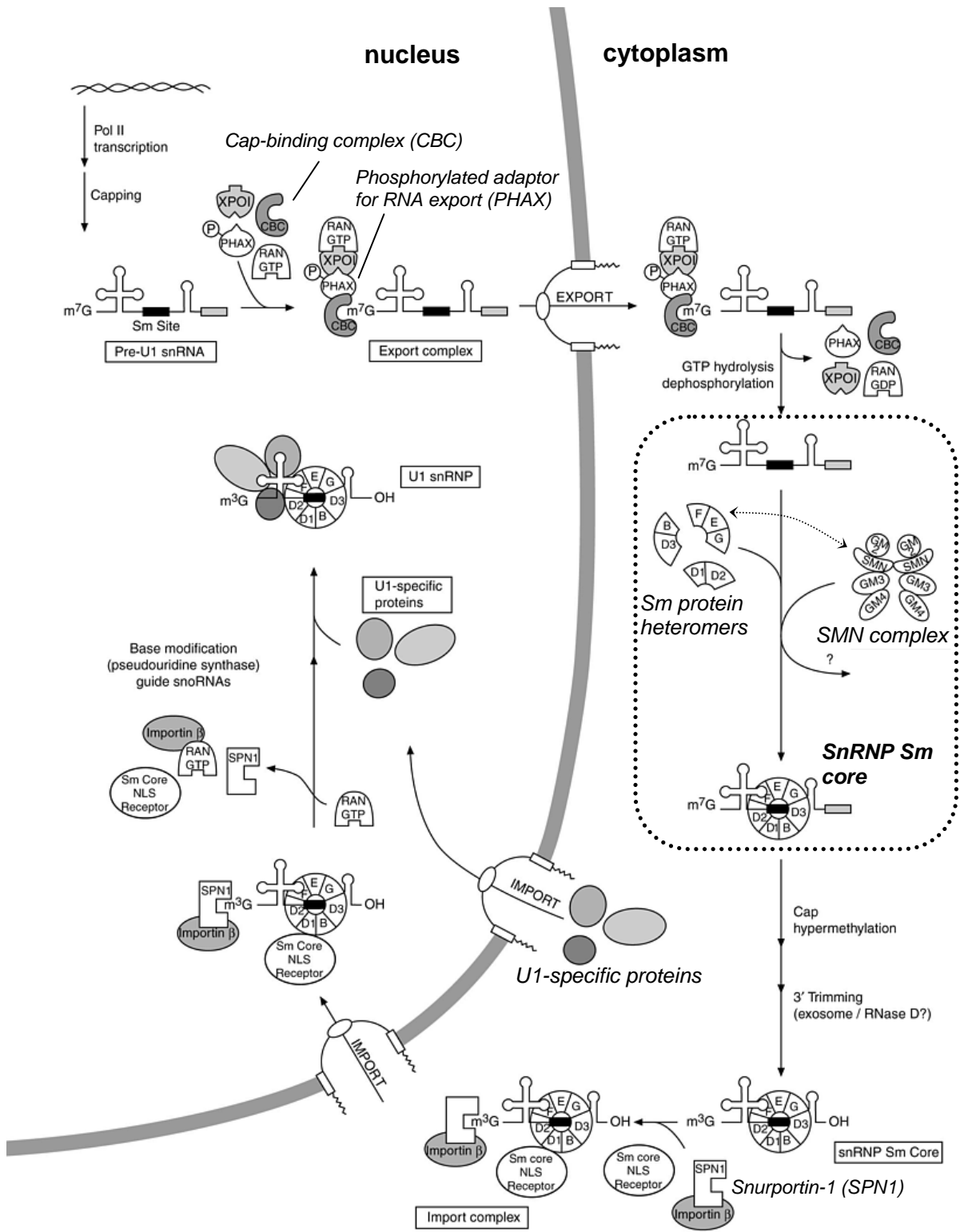

**Figure 1.4: Secondary structures of some snRNAs and their Sm binding sites.**

**(a)**

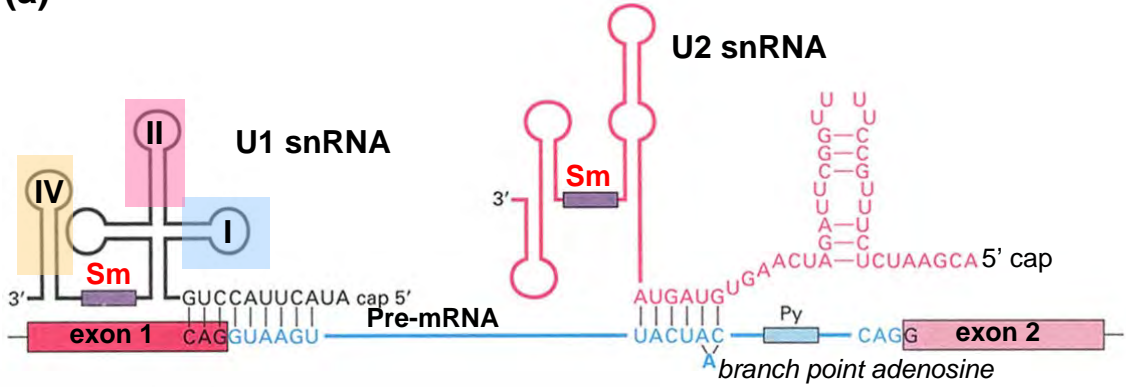

**(b)**

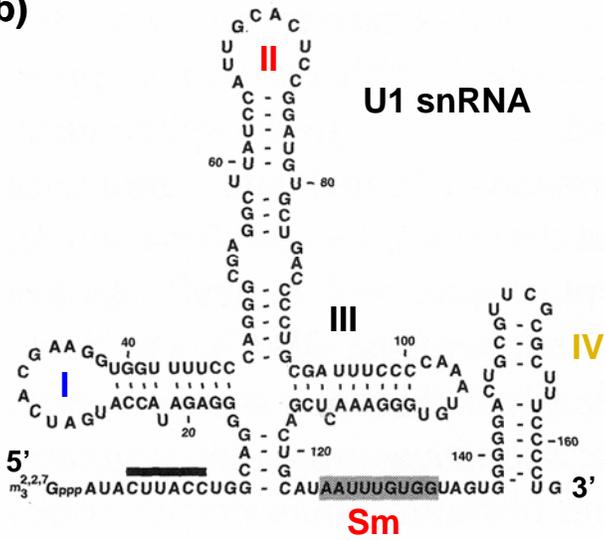

**(c)**

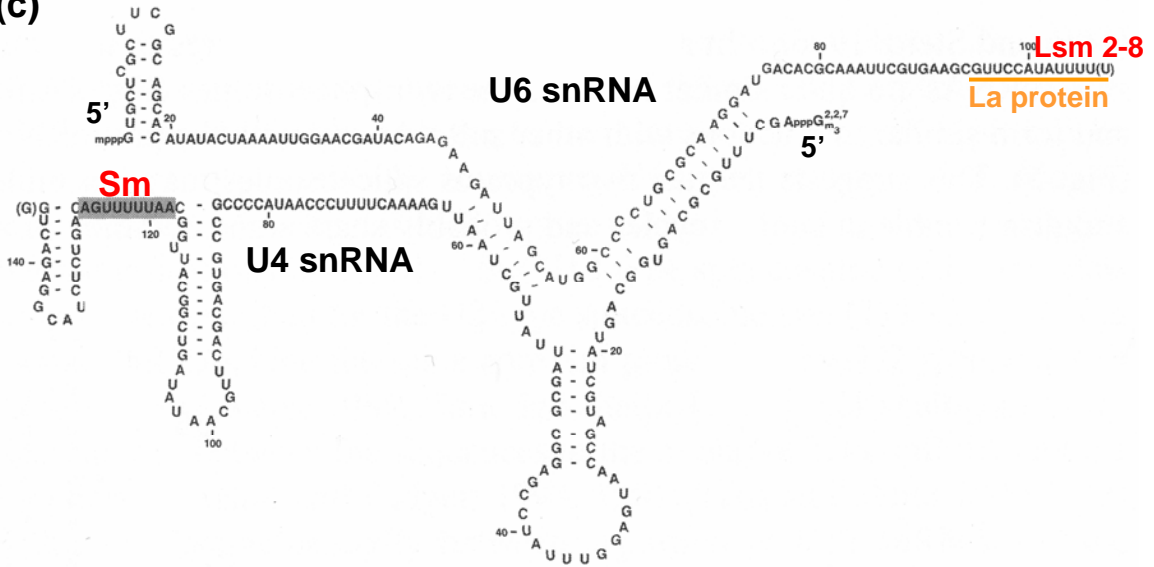

**Figure 1.5: Progress towards an atomic-resolution understanding of eukaryotic Sm proteins and snRNPs.**

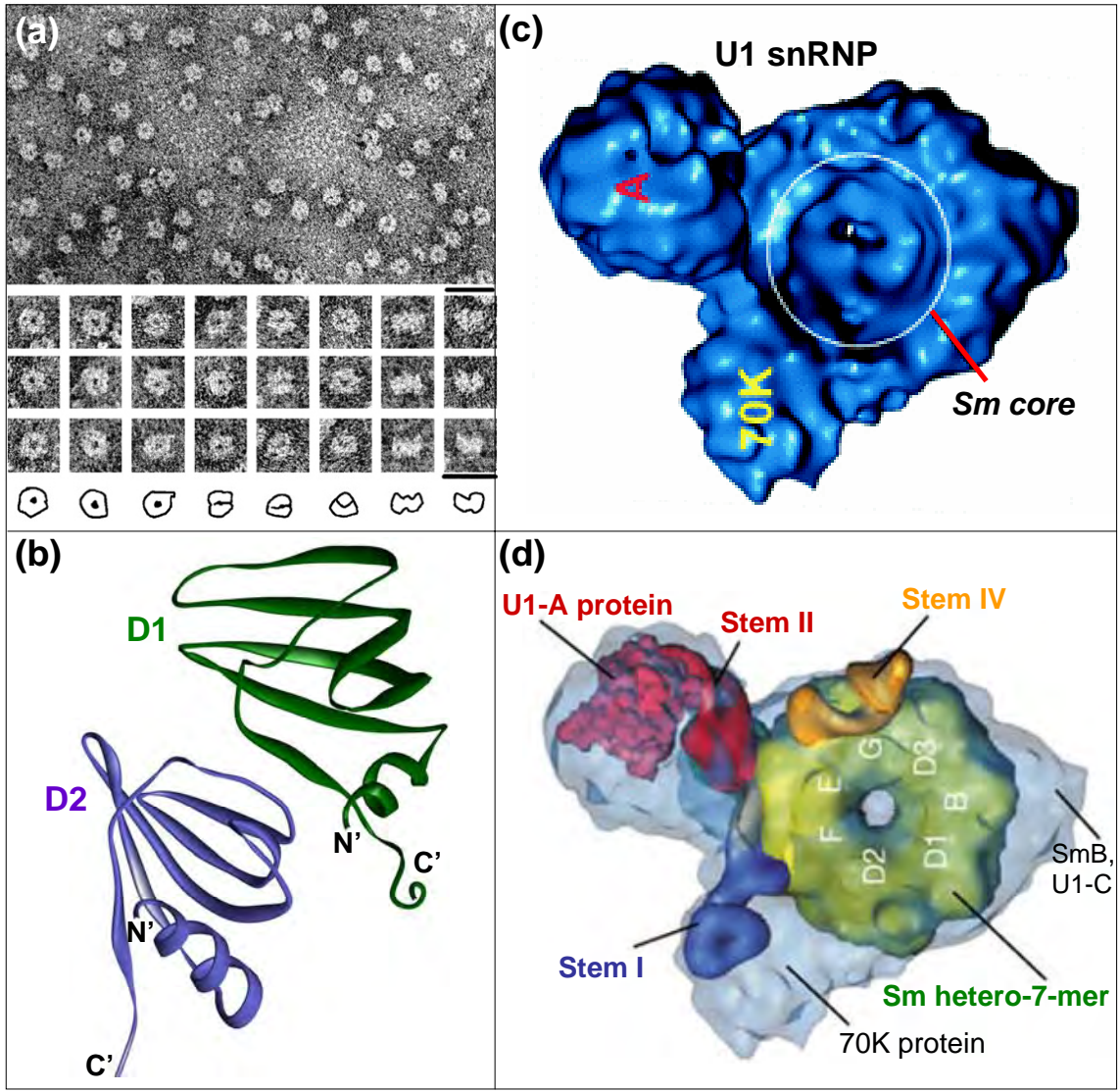

**Table 1.1: A summary of all known Sm and SmAP structures.**

| Protein | Structure determination / asymmetric unit contents | | Significance / Motivation | Reference |
|---|---|---|---|---|
| *Human* **D3•B, D1•D2** | 2.0 Å (D3•B) (PDB `1D3B`) | 6 heterodimers arranged as hexamers ($P2_12_12_1$) | The first Sm protein structures: the human D3•B and D1•D2 heterodimers are nearly identical, and the monomers fold as strongly-bent, five-stranded antiparallel β-sheets | Kambach *et al.* (1999) |
| | 2.5 Å (D1•D2) (PDB `1B34`) | 1 heterodimer ($P6_2$) | | |
| *Afu* **SmAP1** | 2.5 Å (PDB `1I4K`) | 4 heptamers ($P2_1$) | this first structure of an Sm heptamer revealed the binding site of oligouridine ($U_3$) | Törö *et al.* (2001) |
| | 2.75 Å (PDB `1I5L`) | 2 heptamers ($P2_1$) + 2 $U_3$'s | | |
| *Mth* **SmAP1** | 2.0 Å (PDB `1I81`) | 1 heptamer ($P2_1$) | the first structure of an Sm heptamer, model for RNA binding? | Collins *et al.* (2001) |
| | 1.85 Å (PDB `1JBM`) | 1 heptamer ($P1$) | another crystalline packing, obtained concurrent with (but independent of) the work of Collins *et al.* | Mura *et al.* (Chapter 3) |
| | 2.80 Å (PDB `1JRI`) | 2 heptamers ($P2_12_12_1$) | yet another crystalline packing, but one that reveals the likely structure of the fibers observed by EM | |
| | 1.90 Å (PDB `1LOJ`) | 2 heptamers ($P2_1$) + 14 UMPs + 14 MPDs | yet another packing, but one that (i) reveals the binding site of uridine, and (ii) reveals the structure of a 14-mer with pseudo-72 point group symmetry (thus corroborating the *Pae* SmAP1 14-mer) | |
| *Pae* **SmAP1** | 1.75 Å (PDB `1I8F`) | 1 heptamer ($C2$) | the first structure of an Sm heptamer suggested a model for RNA binding (which is probably incorrect) | Mura *et al.* (2001) |
| | 2.05 Å (PDB `1LNX`) | 1 heptamer ($C222_1$) + 7 UMPs | *Pae* 14-mers (with 72 symmetry) exist in this crystal form, comparison of the *Pae* uridine binding site with the conserved ligand-binding sites found in *Afu* and *Mth* SmAP1 | Mura *et al.* (Chapter 3) |
| *Pae* **SmAP2** | *in progress* (data to 2.0 Å) | 2 heptamers ($P42_12$) | to get a complete picture of the SmAP content of a single archaeal species, and to compare to the other two *Pae* SmAPs | — |
| *Pae* **SmAP3** | *in progress* (data to 2.0 Å) | 4 heptamers ($P2_1$) | to get a complete picture of the SmAP content of a single archaeal species, and because *the SmAP3 sequence resembles certain eukaryotic Sm proteins more than does any other SmAP* | — |

**Chapter 2:**

**The crystal structure of a heptameric archaeal Sm protein: Implications for the eukaryotic snRNP core\***

# The crystal structure of a heptameric archaeal Sm protein: Implications for the eukaryotic snRNP core

Cameron Mura*[†], Duilio Cascio*, Michael R. Sawaya*, and David S. Eisenberg*[†‡]

*University of California at Los Angeles–Department of Energy Laboratory of Structural Biology and Molecular Medicine, 201 Boyer Hall/Molecular Biology Institute, Box 951570, Los Angeles, CA 90095-1570; and [†]Department of Chemistry and Biochemistry, University of California, Los Angeles, CA 90095

Sm proteins form the core of small nuclear ribonucleoprotein particles (snRNPs), making them key components of several mRNA-processing assemblies, including the spliceosome. We report the 1.75-Å crystal structure of SmAP, an Sm-like archaeal protein that forms a heptameric ring perforated by a cationic pore. In addition to providing direct evidence for such an assembly in eukaryotic snRNPs, this structure (i) shows that SmAP homodimers are structurally similar to human Sm heterodimers, (ii) supports a gene duplication model of Sm protein evolution, and (iii) offers a model of SmAP bound to single-stranded RNA (ssRNA) that explains Sm binding-site specificity. The pronounced electrostatic asymmetry of the SmAP surface imparts directionality to putative SmAP–RNA interactions.

**E**ukaryotic pre-mRNA processing is an intricate cellular task whose many steps include intron excision. This final maturation step occurs in the spliceosome, a large (~60 S), transiently stable ribonucleoprotein particle that ligates two exons and releases an intron lariat. The major spliceosome contains several small nuclear ribonucleoprotein particles (snRNPs; e.g., U1, U2, and U4/U6•U5), and each U snRNP consists of a respective small nuclear RNA (snRNA; e.g., U1, U2) and many proteins (reviewed in refs. 1 and 2). The subset of proteins common to all spliceosomal U snRNPs is the Sm proteins. Discovered as a group of eight small antigens involved in autoimmune diseases such as systemic lupus erythematosus (3), these core snRNP Sm proteins have been found in most eukaryotes (4) and, recently, in a few archaeal species (ref. 5 and this report). Also, several sets of Sm-like (Lsm) proteins have been discovered in organisms with Sm proteins (4, 6). Biochemical characterization of Lsm proteins has verified their similarity to canonical Sm proteins (reviewed in ref. 7), and this report treats them as equivalent.

Together with U snRNAs, canonical Sm proteins form snRNP core complexes. The molecular structure and function(s) of Sm assemblies and Sm–RNA interactions within these core complexes are unknown, but are presumably more generic than those of snRNP-specific proteins. Cytoplasmic Sm proteins associate with exported snRNAs at short single-stranded regions that are usually flanked by stem-loop structures (8, 9). The consensus sequence for the Sm binding site is RAU$_{4-6}$GR (R = purine), although this selectivity is not very stringent (10, 11). The core complex is thought to be a heteroheptamer of Sm proteins, and electron microscopic investigations of U snRNP core particles suggest that the Sm (12) and Lsm (13) cores are composed of a doughnut-shaped heteromer. The current paradigm is that seven Sm proteins (e.g., human B/B′, D$_1$, D$_2$, D$_3$, E, F, and G) assemble stepwise into a heteroheptameric ring with snRNA through various intermediates, such as D$_1$•D$_2$ and E•F•G heteromers. This snRNP core complex is then imported to the nucleus for completion of spliceosome assembly. Functional complexes of RNA and homologous Sm protein septets, such as the Lsm 1–7 and Lsm 2–8 sets of yeast, are thought to assemble in a similar manner.

There are no atomic resolution structures of snRNP cores, although a recent electron cryomicroscopic study by Stark et al. (14) underscores the importance of Sm proteins in forming the core of the U1 snRNP. Crystal structure determinations of D$_1$•D$_2$ and D$_3$•B heterodimers by Kambach et al. (15, 16) provide the only known Sm structures, and show that each monomer folds as a strongly bent, five-stranded antiparallel β-sheet. These monomers have nearly identical three-dimensional structures, as do the heterodimers. Kambach et al. used their heterodimer structures to model an Sm heteroheptamer with a positively charged central hole. We now report the 1.75-Å crystal structure of Sm-like archaeal protein (SmAP)—a heptameric Sm protein from the hyperthermophilic archaeon *Pyrobaculum aerophilum*—and describe the implications of this structure for the Sm core of eukaryotic snRNPs.

## Materials and Methods

**Protein Preparation and Crystallization.** *P. aerophilum* SmAP was cloned and over-expressed in *Escherichia coli* cells. Heat treatment of lysed cells (80°C) was followed by standard chromatographic steps and proteolytic removal of a C-terminal His$_6$ tag. Final purification steps yielded full-length, wild-type SmAP with an appended glycine. Monoclinic crystals of SmAP (space group C2; $a = 100.26$ Å, $b = 95.74$ Å, $c = 62.16$ Å, $\beta = 92.69°$; $V_M = 2.33$ Å$^3$/Da for seven monomers per asymmetric unit) were used to solve the structure reported here by multiwavelength anomalous diffraction (MAD) phasing of an iridium derivative (Na$_3$IrCl$_6$).

**Data Collection, Phasing, and Refinement.** Synchrotron data were collected at 105 K. Data were processed with DENZO/SCALEPACK (17), and MAD phasing proceeded by the usual methods of heavy atom location [SHELXD (http://shelx.uni-ac.gwdg.de/SHELX/) and SOLVE (18)], maximum likelihood phase refinement [MLPHARE (19)], and density modification [DM (20)]. Phase extension to 1.75 Å permitted automated model building for most of the protein with WARP/ARP (21). Model building and refinement were done in O (22) and CNS (23), respectively. Averaging with sevenfold noncrystallographic symmetry led us to conclude that the only significant asymmetry within the heptameric ring is attributable to side-chain rotameric variation (each monomer of the final model was refined independently). Refinement rounds ended with inspection of the model and $\sigma_A$-weighted $2F_o-F_c$, $F_o-F_c$, and simulated annealing omit

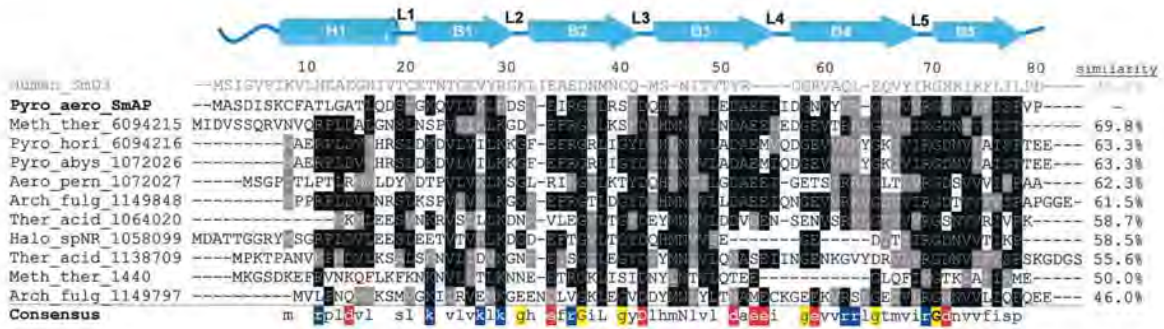

Pyro_aero_SmAP    --MASDISKCFATLGAT...DS...RS...KF...ICN...VP---    -
Meth_ther_6094215 MIDVSSQRVNVC...NSP...I.GD--FP...KS...EDT...TEE--- 69.8%
Pyro_hori_6094216 -----AE...HP...DS--FP...KD...ICN...TEE--- 63.3%
Pyro_abys_1072026 -----HP...DS--FP...KD...IC...AS.TEE--- 63.3%
Aero_pern_1072027 -----MSGP-TLPTL...LDY.DTP...SL-R1...KT...-GETS...AI...AA---- 62.3%
Arch_fulg_1149848 -----PP...NR.KSP...L...KT...KT...A.APGGE- 61.5%
Ther_acid_1064020 ----------K...DN--VLE...GY...VN-SENS...K---- 58.7%
Halo_spNR_1058099 MDATTGGRY.SG...EE...T...GH...SY...IM...K---- 58.5%
Ther_acid_1138709 ----MPKTFANV...KE...SE...M...ESS.INENKGVYDRG...SKGDGS 55.6%
Meth_ther_1440     ----MKGSDKEF.VNKQFLKFKN...NNE...KIS.INNV.QTE...LQF...TK.A...ME---- 50.0%
Arch_fulg_1149797 -----MVL.NQ...KSM.G...HR.KG.GEENLLV.GS.VGDYN...CKGE...KI...QEE---- 46.0%
Consensus          m    p1.v1   s1    v1v...g.gh   .L gy.lhmNlv1 g .  i  g.vv..lgtmvi g .nvvfisp maps (the latter only as necessary). The final model consists of a SmAP heptamer (seven chains, labeled A–G), 130 waters, and five glycerols. Each monomer is complete except for the (*i*) absence of 4 (monomer F) to 13 (monomer C) N-terminal residues per monomer, and (*ii*) an average of 3 residues per monomer that are truncated to either alanine or glycine. Ramachandran plots (PROCHECK, ref. 24) and ERRAT (25) were used for model validation. Structure factors and atomic coordinates have been deposited in the Protein Data Bank (ID code 1I8F).

**Fig. 1.** Sequence analysis of SmAPs. A multiple sequence alignment of SmAPs is shown, along with the sequence of the most similar known Sm structure (human D₃) as a reference point (residue numbering is for the *P. aerophilum* sequence). Pairwise sequence similarity scores between SmAP and its homologs are provided in the last column. The top diagram depicts the SmAP monomer in terms of secondary structure elements. Glycine (yellow), acidic (red), and basic (blue) residues are highlighted in the consensus sequence and are referred to in the text and in Fig. 2c.

**Sequence and Structure Analyses.** PSI-BLAST (26) and CLUSTALW (27) were used for database searches and multiple sequence alignments, respectively. Pairwise alignments were calculated by the Smith–Waterman algorithm. Similar protein structures [rms deviation (rmsd) < ~2.5 Å, e.g., human D₂, D₃ proteins to SmAP] were easily superimposed with ALIGN (28), whereas more dissimilar structures (rmsd > ~2.5 Å by ALIGN, e.g., human B, D₁ proteins) were optimally aligned by combinatorial extension (29). Electrostatic and surface area calculations were performed with GRASP (30). Figs. 2b, 2c, 3a, and 5 were produced with WEBLAB VIEWERPRO 3.7 (Molecular Simulations); Figs. 3 b and c were produced with SETOR (31, 32); and Fig. 4 was created in GRASP.

**The SmAP·ssRNA (Single-Stranded RNA) Model.** The SmAP·ssRNA model was constructed by manual docking of an ssRNA con-taining an Sm consensus sequence (GAU₄GA) through the pore of the refined SmAP heptamer, and did not require any alteration of the SmAP structure. Adjustment of only four phosphate backbone α torsion angles near the poly (U) tract sufficiently extended the RNA so that it was easily accommodated in the SmAP pore; moreover, the values were not adjusted drastically—from α ≈ −50° in A-form RNA to α ≈ −30° in the extended form shown in Fig. 5. The planes of the first and last uracil bases threaded through the pore are separated by roughly 17 Å.

## Results

**A Family of SmAPs.** The occurrence of Sm and Sm-like proteins is not limited to eukaryotes. Along with *P. aerophilum* SmAP, we have uncovered several archaeal Sm sequences within the genomes of *Pyrococcus abyssi*, *Aeropyrum pernix*, *Thermoplasma acidophilum*, and a halobacterium (Fig. 1). Added to the initial list of five SmAPs reported by Salgado-Garrido *et al.* in 1999 (5), these archaeal sequences clearly form a well defined protein family that may be ancestral to modern eukaryotic Sm proteins. On the basis of the structure described here, we propose that *P. aerophilum* SmAP is a representative member of such a family of Sm-like archaeal proteins (SmAPs).

**SmAP Monomer and Dimer Structures.** The crystal structure of SmAP was determined by MAD phasing (Table 1), and reveals

**Table 1. Crystallographic statistics**

| | Data collection & MAD phasing | | | | Model refinement | |
|---|---|---|---|---|---|---|
| | Native | Inflection | Peak | Low-λ remote | | |
| Wavelength, Å | 0.9794 | 1.1058 | 1.1055 | 1.0960 | Resolution range, Å | 20–1.75 |
| Resolution range, Å | 100–1.71 | 100–1.95 | 100–1.95 | 100–1.95 | No. of protein atoms | 3,796 |
| Completeness, % | 98.5 | 97.3 | 97.0 | 95.0 | No. of solvent molecules (H₂O/glycerol) | 130/5 |
| $I/\sigma(I)$ | 44.4 | 24.3 | 25.2 | 22.1 | $R_{cryst}/R_{free}$, %[§] | 23.5/26.6 |
| $R_{merge}$, %[*] | 4.0 | 4.8 | 4.8 | 4.0 | (B-factor) protein, Å² | 38.5 |
| # Ir sites/asymmetric unit | — | — | 8 | — | rmsds: bonds, Å | 0.018 |
| Phasing resolution range, Å | | 20–2.0 | 20–2.0 | 20–2.0 | angles, ° | 1.90 |
| $R_{cullis}$[†]: acentric | | — | 0.71/0.42 | 0.83/0.70 | | |
| centric | | — | 0.64 | 0.82 | | |
| Figure of merit[‡] | | 0.65/0.82 | | | | |

[*] $R_{merge}(I) = \Sigma_{hkl}(\Sigma_i |I_{hkl,i} - \langle I_{hkl}\rangle|)/\Sigma_i I_{hkl,i}$.
[†] $R_{cullis} = (\Sigma_{hkl} \|F_{PH} \pm F_P| - F_{H,calc}|)/\Sigma_{hkl} |F_{PH} \pm F_P|$. Statistics for acentric reflections are given as isomorphous/anomalous.
[‡] Values are given before/after density modification.
[§] $R_{cryst} = \Sigma_{hkl} |F_{obsl} - |F_{calc}||/\Sigma_{hkl} |F_{obsl}|$. $R_{free}$ was computed identically, except that 5% of the reflections were omitted as a test set.

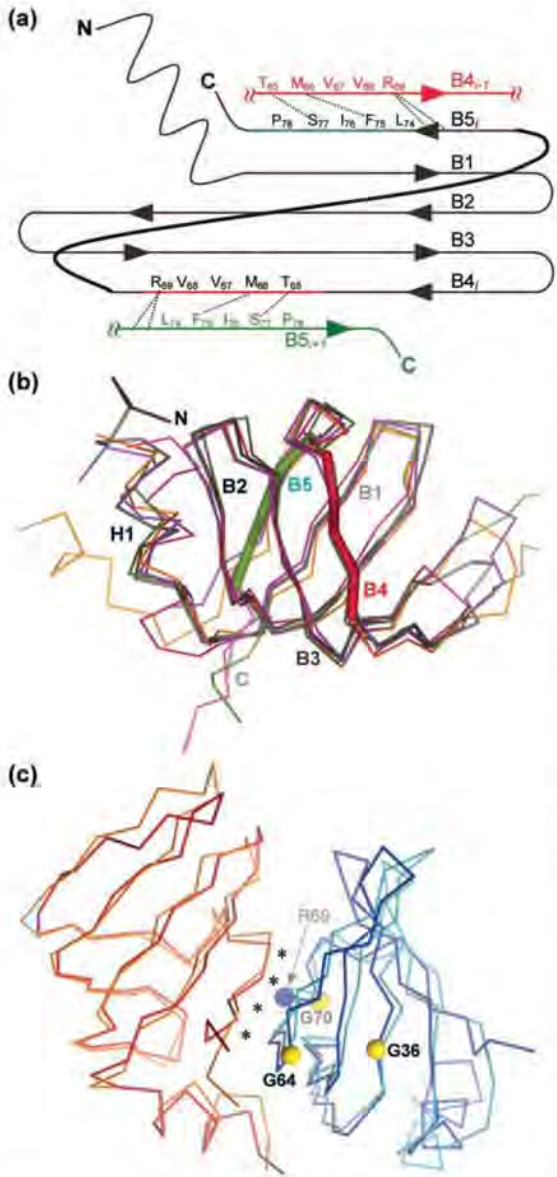

that this archaeal protein—like the human Sm B, $D_1$, $D_2$, and $D_3$ monomers—folds as a strongly bent, five-stranded, antiparallel β-sheet capped by an N-terminal α-helix (Fig. 2 *a* and *b*). This protein fold strongly resembles members of the oligonucleotide/oligosaccharide-binding fold family (33). The β-strands **B2**, **B3**, and **B4** create this peculiar structure by twisting back upon themselves to form a U-shaped core that is roughly elliptical in cross section. The amount of curvature may be measured as the distance between the N and C termini of a strand. As an example of the SmAP curvature, the distance between the **B2** strand termini ($C_\alpha^{Ser31}$ and $C_\alpha^{Gln43}$) is 23.7 Å, whereas this distance would be 39.7 Å in an un-bent, fully-extended conformation. The structural plasticity needed for such a high degree of β-strand curvature is apparently provided by several strictly conserved glycines that occur near the pivot points (Gly36, Gly64, Gly70 in the archaeal Sm sequences shown in Fig. 1). The segment linking **B4** and **B5** lies at the top of the U-shaped trough to close the protein into a β-barrel-like structure (Fig. 2*b*); it also positions **B5** for antiparallel hydrogen bonding to **B1** within the same monomer and the **B4** strand of an adjacent monomer (Fig. 2*a*).

This latter interaction—hydrogen bonding between strands **B4**$_i$··**B5**$_{i+1}$ for each monomer *i*—creates the seven dimer interfaces that orient the monomers head-to-tail around the ring. Specifically, each monomer contributes five residues from each **B4** and **B5** strand (Fig. 2*a*) to create an interface that occludes $1,682 \pm 78$ Å$^2$ of surface area per dimer (Fig. 2*c*). The interfacial residues are predominantly apolar, and a few interesting side-chain interactions supplement the standard hydrogen bonds between the β-strand backbones. Such interactions include: (*i*) the guanidinium group of Arg-69$_{B4}$ in most of the monomers engages in several hydrogen bonds with main-chain and side-chain atoms from the adjacent **B5** strand, and (*ii*) van der Waals contact of the sulfur of Met-66$_{B4}$ with the aromatic ring of Phe-75$_{B5}$ suggests that a favorable S···π aromatic interaction (34, 35) may stabilize the interface.

The structural superimpositions shown in Fig. 2 illustrate the strong similarity between SmAP and human Sm monomers and dimers. SmAP aligns best with Sm $D_3$ (31% sequence similarity, 1.0 Å rmsd over backbone atoms), and least well with $D_1$ (40% sequence similarity, 1.7 Å rmsd). The twisted β-sheet core of each Sm monomer is nearly identical (Fig. 2*b*), the main difference being the shorter **L4** loop of SmAP versus three of the human structures (B, $D_1$, and $D_3$). A similar trend toward shorter loops has also been found in other pairs of thermophilic-mesophilic proteins (see ref. 35 and references therein). Both $D_3$·B and $D_1$·$D_2$ human heterodimer structures align very closely with a SmAP homodimer (1.4 Å and 1.6 Å rmsd, respectively). Superimposition of $D_3$·B on SmAP shows that the dimer interface is conserved (Fig. 2*c*), although the amino acids in this region (**B4** residues 65–69 and **B5** residues 74–78) show greater phylogenetic variation than in the rest of the sequence (Fig. 1). This variation is explained by the fact that the dimer interface consists mainly of hydrogen bonds between the backbones of the β-strands. In addition to showing that human and archaeal Sm proteins belong to the same fold family, the SmAP structure is apparently an example of an ancestral homodimeric interface that evolved into several distinct, functional heterodimeric interfaces ($D_3$·B, $D_1$·$D_2$, etc. interfaces in human).

**The SmAP Heptamer.** The heptameric organization of the SmAP shown in Fig. 3 was revealed by the binding of iridium ions (used for phasing purposes, Fig. 3*b*) and by *in vitro* biophysical characterization (C.M. and D.E., unpublished results). The refined crystallographic model reveals a disk-shaped homoheptameric ring that measures ≈65 Å in diameter and ≈38 Å in height (Fig. 3*c*), which is consistent with the dimensions from electron microscopy of human Sm cores (12). One of the most notable features of the heptamer is the propagation of the β-sheet core of individual monomers across the ring to give a circular, 35-stranded β-sheet. This extended β-sheet does not lie in a single plane because of the strong curvature of the constituent strands. Rather, the U-shaped monomers are arranged like the blades of a turbine, rotated by ≈45° out of the horizontal plane of the heptamer. In contrast to the model proposed for the human heptamer (15), the remarkable asymmetry in the electrostatic surface of the SmAP heptamer gives the disk a large dipole moment (Fig. 4). This feature may be of functional relevance because it imparts directionality to putative SmAP–RNA interactions.

**Fig. 2.** SmAP monomer and dimer structures. (*a*) A cartoon of the SmAP fold, along with the B4 and B5 strands of the two neighboring monomers. Dashed lines indicate side-chain interactions that supplement the backbone hydrogen bonding of the B4–B5 pairs. (*b*) A depth-cued illustration of the structure of one of the SmAP monomers (gray) is superimposed on the $C_\alpha$ traces of four aligned structures: human Sm $D_3$ (violet, 1.0 Å rmsd), B (green, 1.1 Å rmsd), $D_2$ (magenta, 1.2 Å rmsd), and $D_1$ (orange, 1.7 Å rmsd). The extensive L4 loop of Sm B has been truncated for clarity, and the segments of strands B4 and B5 that bind adjacent monomers are colored as thick green and red lines as in *a*. This orientation illustrates the strongly bent five-stranded antiparallel β-sheet that forms the Sm structures. (*c*) The human $D_3$·B heterodimer (cyan-orange) is superimposed on an SmAP homodimer (red-blue). Asterisks mark the conserved dimer interface, and colored balls give the positions of conserved residues shown in Fig. 1. Note that the archaeal homodimer, taken directly from the SmAP heptamer crystal structure, has essentially the same structure as the human heterodimer (1.4 Å rmsd over main-chain atoms).

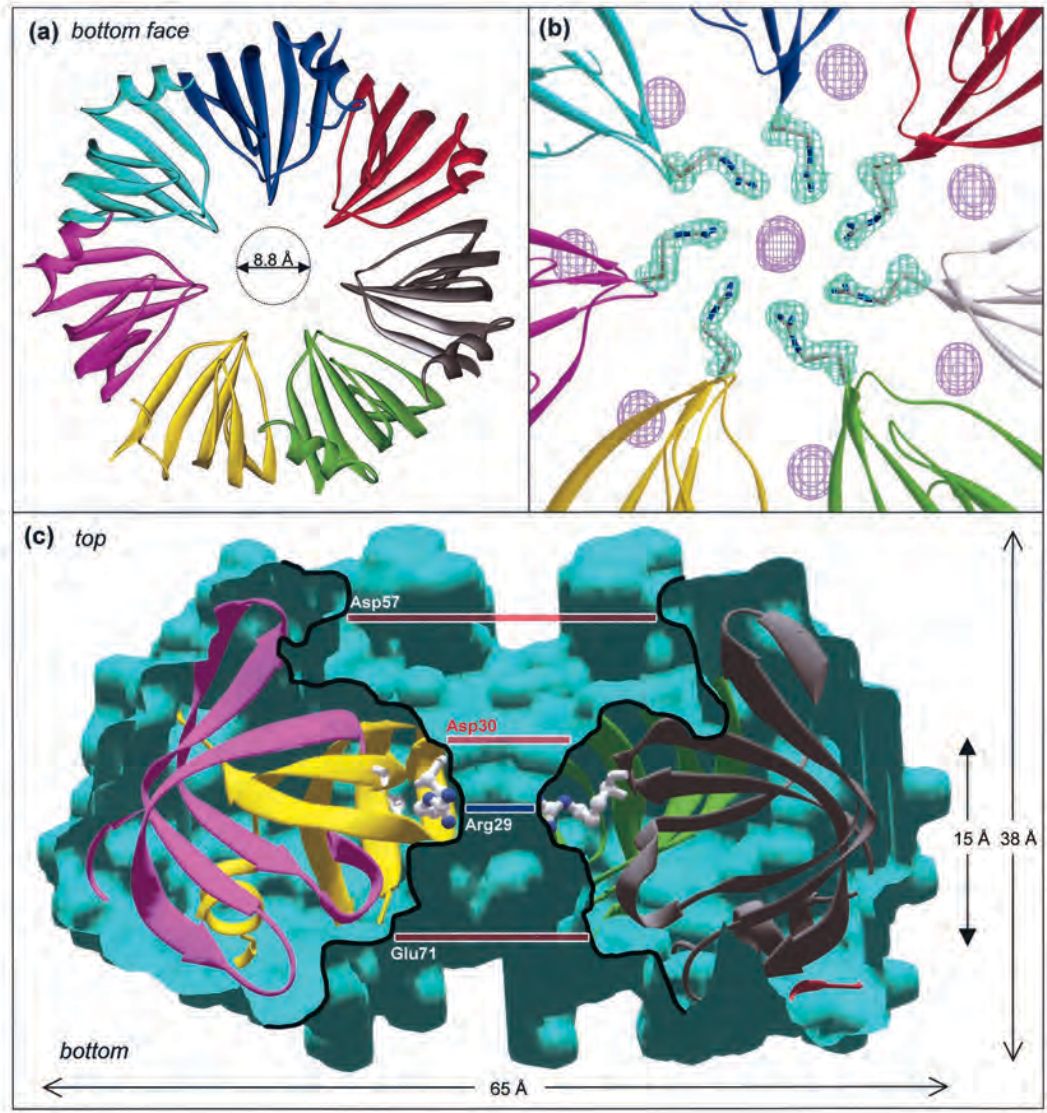

**Fig. 3.** Structure of the SmAP heptamer and the cationic pore. (*a*) A ribbon diagram of the SmAP heptamer, viewed from the "bottom" face, parallel to the sevenfold axis. (*b*) The enlarged view of experimental $2F_o - F_c$ electron density for the Arg-29 side chains that line the pore (green, contoured at 1.2 $\sigma$). The anomalous difference density marks the sites of the eight bound iridium ions (indigo, contoured at 3.5 $\sigma$). (*c*) Illustration of the molecular surface for a sagittal section of the heptamer, with the sevenfold axis vertical. Arg-29 side chains are rendered as ball-and-stick models, and reveal that the most constricted section of the hourglass-shaped pore (traced in black) has a diameter of 8.8 Å [heavy, closed arrows (↕) are pore dimensions; light, open arrows (↕) are for the entire heptamer]. Horizontal bars indicate the concentric charged rings that are discussed in the text (red = anionic; blue = cationic).

## Discussion

**The Cationic Pore.** Interactions of SmAP with RNA (or DNA) are likely to occur within the shallow hourglass-shaped pore that runs along the sevenfold axis (Fig. 3 *a* and *c*). This cationic pore is ≈15 Å in depth and has a minimum diameter of 8.8 Å, the most constricted region being formed by the Arg-29 side chains of loop **L2**. The pore tapers steeply away on both sides of the narrow ring of arginines so that no other residues can be said to *line* the pore. Fig. 3c shows that four layers of charged heptagonal rings lie near to or within the pore. Starting at the bottom face of the heptamer, there is a negatively charged ring of glutamic acid residues (Glu-71, 18-Å diameter) followed by the narrow cationic ring of arginine residues (Arg-29, 8.8-Å diameter); a larger ring of aspartic acid residues (Asp-30, 11.5-Å diameter) is slightly above this; finally, there is a much wider ring of aspartic acid residues at the top face of the heptamer (Asp-57, 28.8-Å diameter). Note that unproductive SmAP·ssRNA bonding (i.e., interactions outside of the pore) would be diminished by such an arrangement of anionic rings.

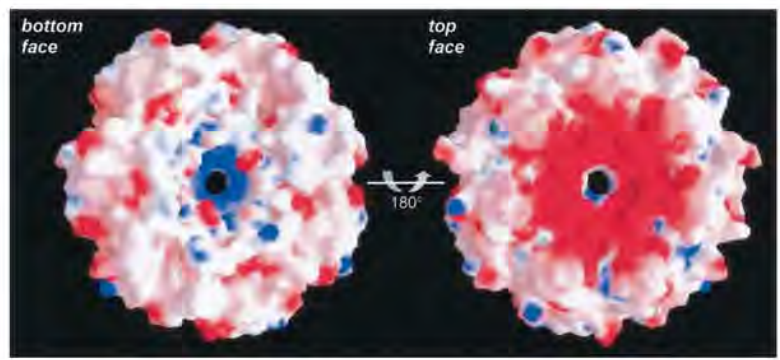

**Fig. 4.** Electrostatic properties of the SmAP heptamer surface. The molecular surfaces of both faces of the SmAP heptamer are shown color-coded by electrostatic potential (red = −9.3 $kT$; blue = +10.9 $kT$). Note that there are two distinctive features: a ring of strong positive potential that lines the heptamer pore toward the bottom face, and a diffuse zone of intense negative potential across most of the top face. Such pronounced charge asymmetry gives the heptameric disk a large dipole moment (calculated to be 553 debye).

Strict conservation of Arg-29 in archaeal Sm homologs (Fig. 1) suggests that a cationic pore with similar properties exists in all archaeal Sm heptamers (this arginine is also conserved in the human $D_3$ protein). By superimposing the human heterodimers on the SmAP heptamer, we find that the cationic character of the SmAP pore is probably conserved in the human Sm heptamer. Several positively charged residues from the human proteins would lie near the SmAP pore (although not in exactly the same location as the Arg-29 ring), including Arg-51 and Lys-67 from $D_3$, and Lys-54 from the B monomer.

The aforementioned residues that form the pore and charged heptagonal rings all lie in loops **L2**, **L4**, and **L5**. However, recent experiments by Urlaub *et al.* (37) demonstrate that residues from loop **L3** in human Sm B and G proteins can crosslink with an RNA nonanucleotide containing the Sm binding site (AAU$_5$GA). Assuming that snRNA binds in the pore *in vivo*, these results are not consistent with the relative locations of the pore and loop **L3** in SmAP, and suggest that either (*i*) the human Sm monomers adopt altered conformations in the heteroheptamer to position the **L3** loops more proximal to the pore, or (*ii*) the RNA nonanucleotide used by Urlaub *et al.* binds near the pore, but not within it.

**Elucidation of Sm Binding Site Specificity.** Notably, the SmAP heptamer structure offers a simple explanation for the specificity of the snRNA Sm binding site (RAU$_{4-6}$GR). The depth of the pore varies depending on exactly how it is defined, and a lower bound estimate of the pore depth (15 Å) is shown in Fig. 3c. More importantly, the aperture of the cationic Arg-29 ring is clearly defined in electron density (Fig. 3b) and gives the pore a minimum diameter of 8.8 Å. Fig. 5 shows a model of the SmAP heptamer complexed with a 20-nucleotide ssRNA that contains the consensus Sm binding site. The minimal diameter of the pore is ideally suited for binding of a single strand of polypyrimidyl RNA, but is too narrow to accommodate a polypurine, thereby affording steric discrimination between polypyrimidine and polypurine tracts. At least four or five nucleotides are required to span the $\approx 16$ Å length of the pore, thus explaining the minimum requirement of four to five uridines in the Sm binding site. A more difficult matter to explain is the selection of poly(U) versus the other possible polypyrimidine, poly(C). It is possible that the cationic pore favors poly(U) because it is a better hydrogen-bond acceptor than the slightly more electropositive poly(C). Most importantly, the eighth tritium peak in anomalous difference Fourier maps is located in the pore, $\approx 3$ Å below the cationic Arg-29 plane (Fig. 3b). This peak proves that the pore is accessible to polyvalent anions (e.g., octahedral $IrCl_6^{3-}$) by an approach from the bottom face of the heptamer.

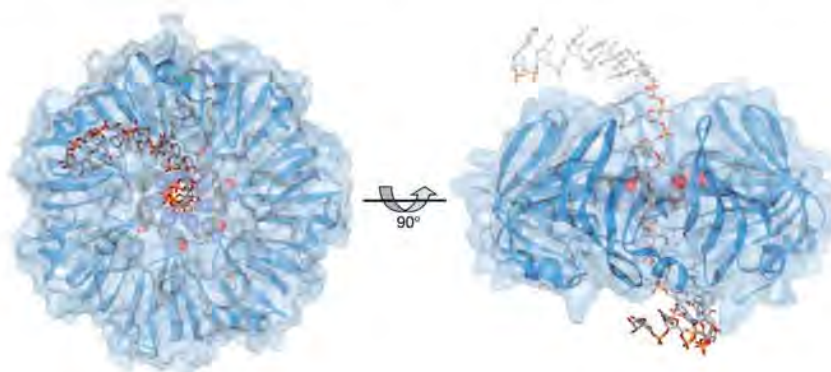

**Fig. 5.** Model of a SmAP·ssRNA complex. Orthogonal views are shown for a model of SmAP bound to a hypothetical 20-nucleotide ssRNA (e.g., eukaryotic snRNA) that consists of three segments from 5′ to 3′: a random string of 6 nucleotides in the A-form conformation (ACGAUC), followed by a minimal consensus Sm binding site (GAU$_4$GA), and ending with 6 more nucleotides in A-form geometry (ACGAUC). The SmAP heptamer is depicted as a ribbon diagram along with the solvent-accessible surface. The Arg-29 ring that forms the pore is colored by atom type and rendered in space-filling form, with the ssRNA drawn as a stick model. The steric and electrostatic environment of the pore is ideally suited to accommodate a single-stranded polypyrimidyl nucleic acid.

**The Sm Core of Eukaryotic snRNPs.** The crystal structure of SmAP and the model of it bound to an snRNA Sm binding site have several implications for the Sm core of eukaryotic snRNPs. We note that there are two possible paths by which SmAP may bind snRNA. Either the RNA is threaded through the pore (from the bottom face), or the heptamer assembles onto the snRNA binding site. The SmAP structure shows that the latter scenario is more plausible because purine bases cannot fit in the Sm pore. In vitro evidence for stable, subheptamers of human Sm proteins (e.g., an E·F·G heterotrimer, ref. 8), along with the fact that a minimal complex of five Sm proteins will bind RNA, lends credence to this stepwise core snRNP assembly pathway.

The regulation of such an assembly pathway could be achieved by noting the recent finding by Brahms et al. (32) that the C-terminal Arg-Gly dipeptides of human Sm proteins $D_1$ and $D_3$ are symmetrically dimethylated on the arginine $\eta^1$ and $\eta^2$ nitrogens in vivo. Fig. 1 shows that a C-terminal Arg-69–Gly-70 dipeptide of SmAP is strictly conserved among archaeal Sm proteins and aligns with one of the Arg-Gly dipeptides of human $D_3$ protein. Inspection of the structural environment of Arg-69 in the human and SmAP structures shows that it lies at the Sm dimer interface, where it makes many contacts (Fig. 2c). Dimethylation of Arg-69 guanidinium groups by a protein arginine methyltransferase is predicted to interfere sterically with this interface, preventing Sm heptamer formation. Therefore, the structure of SmAP leads us to propose that regulation of the snRNP core assembly may be achieved by arginine dimethylation.

Because SmAP is a homoheptamer, our results support a mechanism of modern eukaryotic Sm protein evolution by early gene duplication events. Presumably, an archaeal SmAP gene was duplicated and gradually accrued neutral point mutations, with the structural restraint that the independently evolving monomers still form a heteroheptamer. Such a scheme would give rise to modified/modern Sm heteroheptamers, such as the human B·D₁·D₂·D₃·E·F·G heptamer, with modified/modern biochemical activities, such as U snRNA binding. This model implies asymmetric heteroheptamers composed of paralogous Sm proteins, and explains the diversification of Sm protein function. Our preliminary crystallographic results with another archaeal Sm protein suggest that it also assembles into a homoheptamer, thereby supporting this model. Finally, we note that these results raise several interesting questions regarding the potential RNA that P. aerophilum SmAP interacts with in vivo, the cellular role of this putative RNA binding, and the possibility of snRNP-like particles in archaeal species.

We thank Dr. Kym Faull for help with mass spectrometry, and Adam Frankel, Linda Columbus, Drs. Dan Anderson, Doug Black, Jim Bowie, Guillaume Chanfreau, Sorel Fitz-Gibbon, Celia Goulding, and Ioannis Xenarios for useful discussions. We also thank Brookhaven National Lab for the use of beamline X8C of the National Synchrotron Light Source. We gratefully acknowledge financial support from the National Institutes of Health, the Department of Energy, and a National Science Foundation predoctoral fellowship (to C.M.).

**Chapter 3:**

**The oligomerization and ligand-binding properties of**

**Sm-like archaeal proteins (SmAPs)**

# Abstract

Intron splicing is one example of the many types of RNA processing that directly utilize small nuclear ribonucleoprotein (snRNP) complexes. Sm proteins form the cores of most snRNPs, so to further elucidate structural principles of snRNP assembly, we have characterized the oligomerization and ligand-binding properties of Sm-like archaeal proteins (SmAPs) from *Pyrobaculum aerophilum* (*Pae*) and *Methanobacterium thermautotrophicum* (*Mth*). Ultracentrifugation shows that *Mth* SmAP1 is exclusively heptameric in solution, whereas *Pae* SmAP1 forms either disulfide-bonded 14-mers or sub-heptameric states (depending on the redox potential). By electron microscopy, we show that *Pae* and *Mth* SmAP1 polymerize into sheets composed of well-ordered polar fibers that are formed by head-to-tail stacking of heptamers. The crystallographic results reported here corroborate these findings by showing heptamers and tetradecamers of both *Mth* and *Pae* SmAP1 in several new crystal forms. The 1.9-Å resolution structure of *Mth* SmAP1 bound to uridine-5'-monophosphate (UMP) reveals conserved ligand-binding sites. The likely RNA binding site in *Mth* agrees with that determined for *Archaeoglobus fulgidus* (*Afu*) SmAP. Finally, we find that both *Pae* and *Mth* SmAP1 gel-shift negatively supercoiled DNA. These results distinguish SmAPs from eukaryotic Sm proteins, and suggest possible differences in their functions.

---

*Abbreviations*: snRNP, small nuclear ribonucleoprotein; SmAP, Sm-like archaeal protein; *Pae*, *Pyrobaculum aerophilum*; *Mth*, *Methanobacterium thermautotrophicum*; *Afu*, *Archaeoglobus fulgidus*; UMP, uridine-5'-monophosphate; MPD, 2-methyl-2,4-pentanediol; EM, electron microscopy; NCS, non-crystallographic symmetry; *wt*, wild type; nt, nucleotide; DTT, dithiothreitol; ss(D/R)NA, single-stranded (D/R)NA; OB-fold, oligosaccharide/oligonucleotide-binding fold

## Introduction

Excision of non-coding regions (introns) is one of the most vital steps in the maturation of precursor mRNAs. Most eukaryotic protein-coding genes contain multiple introns,[1] so high-fidelity pre-mRNA processing is essential to ensure a mature mRNA with correctly registered exons. A transiently stable assembly of five small nuclear ribonucleoproteins (snRNPs) catalyzes the simultaneous excision of introns and splicing of exons in eukaryotic pre-mRNA. This large assembly of uridine-rich snRNPs (U snRNPs) is known as the spliceosome. It contains five small nuclear RNAs (snRNAs) and at least 80 proteins,[2] making it roughly the same size as the ribosome (sedimentation coefficient of $\approx$ 60S).[3] At various stages in its catalytic cycle, the spliceosome consists of the U1, U2, U4/U6, and U5 snRNPs.[4] The recent isolation of a novel U1•U2•U4/U6•U5 "penta-snRNP" devoid of mRNA suggests needed modification of the long-held belief that spliceosome assembly requires pre-mRNA.[5]

Extensive biochemical and genetic data have shown that stepwise binding of seven cytoplasmic Sm proteins to exported snRNAs is a key step in snRNP biogenesis (recently reviewed in ref. 6). Each U snRNP complex is composed of a $\approx$ 110-180 nucleotide uridine-rich snRNA and two classes of proteins: *(i)* snRNP-specific proteins that provide snRNP-specific functions (*e.g.*, U1A protein of U1 snRNPs), and *(ii)* a set of Sm or Sm-like (Lsm) proteins that are common to each snRNP core.[7] The snRNA component contains a single Sm or Lsm binding site with the uridine-rich consensus sequence PuAU$_{\approx 4\text{-}6}$GPu (Pu = purine). However, specificity for this sequence is not very stringent and there can be redundancy in Sm•snRNA binding.[8] The Sm sites are predicted to be single-stranded RNA (ssRNA) regions flanked by stem-loop structures (reviewed in refs. 2, 4). Sm binding is highly sensitive to modifications of the flanking stem-loops and the Sm site of a given snRNA, and varies from one snRNA to another.[9] Sm•snRNA binding may be modulated by recently discovered interactions between some Sm proteins and the survival of motor neurons (SMN) protein complex,[10] and by symmetric dimethylation of arginine residues in some of the RG dipeptide repeats of Sm[11,12,13] and Lsm[14] proteins by a putative "methylosome".[15] In eukaryotes, the Sm D1•D2 and E•F•G heteromers simultaneously bind to snRNA to yield a "subcore" snRNP complex.[6,16,17] The final component to join the Sm complex (the B/B'•D3 heterodimer) triggers hypermethylation of the 5' $m^7G$ cap of snRNA to a trimethylated cap ($m_3G$). The $m_3G$ cap and the snRNA•Sm core complex form a bipartite nuclear localization signal that results in transit of the snRNP core to the nucleus, where association of various snRNP-specific proteins completes the assembly process.

Aside from their roles in assembly and nuclear import of snRNP cores, the primary structural or catalytic function of Sm proteins is not known. Forming the core of snRNPs, Sm proteins probably mediate critical RNA-RNA, RNA-protein, and protein-protein interactions, and may recruit both snRNP and non-snRNP proteins to the assembling spliceosome. The vast network of protein-protein and protein-RNA interactions in which Sm proteins probably engage was recently reviewed by Will and Lührmann[7,6] and experimentally verified by genome-wide two-hybrid screens of yeast Lsm proteins.[18] An example of such interactions is the discovery that the extended and highly charged C-terminal tails of Sm B, D1, and D3 are involved in sequence-independent binding to pre-mRNA substrate, and may stabilize U1 snRNP•pre-mRNA interactions.[19,20] Chemical cross-linking of the U1 snRNP 70K protein to the Sm B and D2 proteins provides suggests a possible role of snRNP Sm cores in recruiting other proteins to the assembling spliceosome.[21] The importance of Sm proteins in RNP assemblies is underscored by their phylogenetic distribution: in addition to the canonical Sm and Lsm proteins found in eukaryotes ranging from yeast to humans, the Sm-like archaeal protein ("SmAP") family has been discovered.[22,23] The recent demonstration that the *E. coli* bacteriophage host factor *Hfq* is an Sm-like protein provides the first example of a eubacterial Sm protein.[24,25] These results imply fundamental roles for Sm proteins in the early evolution of RNA metabolism.

Sm proteins have a remarkable tendency to associate into ring-shaped oligomers. Prompted by biochemical and genetic data, electron microscopic (EM) investigations of U snRNP particles revealed the "doughnut-shaped" ultrastructure of Sm and Lsm cores.[26,27] The realization that Sm and Lsm proteins occur in groups of at least seven paralogs within the genome of a given organism suggests that snRNP cores are formed from Sm heteroheptamers, and two recent results verify this. First, Stark *et al.* reconstructed a 10-Å resolution map of the U1 snRNP by cryo-EM and found that a model of the Sm heptamer could be docked into the ring-shaped body of the snRNP.[28] Next, the *in vivo* stoichiometry of Sm proteins in yeast spliceosomal snRNPs was determined by a differential tag/pull-down assay, showing that the snRNP core domain contains a single copy of each of the seven Sm proteins.[29] Intriguingly, stable sub-heptameric Sm complexes have been suggested as intermediates along the snRNP core assembly pathway (*e.g.*, a D1•D2•E•F•G complex capable of binding snRNA),[17] and ultracentrifugation and EM show that some of these oligomers (*e.g.*, a (E•F•G)$_2$ hexamer) can form ring-like structures that resemble intact, heptameric snRNP cores.[30] Such findings emphasize the importance of the Sm heptamer at the snRNP core, and suggest the possibility of other oligomeric states.

There is no atomic-resolution structure of a eukaryotic snRNP core; however, the crystal structures of Sm-like archaeal proteins from *Afu*,[31] *Pae*,[23] and *Mth*[32] reveal a ring-shaped Sm heptamer and provide a model for snRNA binding in the snRNP core. Sm monomers fold as strongly bent, five-stranded antiparallel β-sheets,[33] and form toroidal heptamers that are perforated by a conserverd cationic pore. The inner surface of this pore appears to be the oligouridine binding site. The structural similarity between the SmAP1 monomers and dimers and the nearly identical human Sm D1•D2 and B•D3 heterodimers[34] justifies SmAP1-based models for the heptameric snRNP core. In order to elucidate further structural principles of snRNP assembly, we have characterized the oligomerization and ligand-binding properties of *Pae* and *Mth* SmAP1. Many of our results distinguish these two Sm-like archaeal proteins from eukaryotic Sm proteins, and suggest that the functions of archaeal Sm proteins may be quite different from the snRNP-based roles of eukaryotic Sm proteins.

## Materials and Methods

### *Cloning, expression, and purification of* **Pae** *and* **Mth** *SmAP1s*

A genomic phosmid clone that contains the *Pae* SmAP1 open reading frame (ORF) was kindly provided by the laboratory of Jeffrey H. Miller (UCLA), and genomic DNA containing the *Mth* (strain ΔH) SmAP1 ORF was kindly provided by the laboratory of John Reeve (Ohio State University). Based on the ORF DNA sequences, we used these primers for PCR amplification with Deep Vent$_R$ polymerase (New England Biolabs):

| (*Pae* sense) | 5' CCATATGGCCTCGGATATATCT 3' |
| (*Pae* antisense) | 5' AAGCTTTCCCCGTCCTGGTACT 3' |
| (*Mth* sense) | 5' CCATATGATAGATGTGAGTTCAC 3' |
| (*Mth* antisense) | 5' AAGCTTTCCCCGGGATATGTA 3' |

Blunt-end PCR products were cloned into a pET-22b(+) expression vector (Novagen) *via* intermediate subcloning into the pCR-Blunt vector (Invitrogen). Ligation products were directly transformed into chemically competent NovaBlue *E. coli* (Novagen), and plasmids from overnight cultures of positive transformants (as assayed by PCR screening of colonies) were mini-prepped (Qiagen). DNA sequencing (David Sequencing) of these plasmids verified that the expressed proteins would contain a C-terminal His-tag after a 10-residue serine protease-sensitive linker. That is, the constructs were designed as: wild type (*wt*) SmAP1 + G**R**\*GKLAAALEHHHHHH (single letter amino acid codes, \* indicates intended protease site). Recombinant proteins were over-expressed in BL21(DE3) *E. coli* at 37°C by standard protocols using 1 mM isopropyl-β-D-thiogalactoside induction of the T7*lac*-based promoter. At least 120 mg of soluble protein was expressed per liter of cell culture. The Cys8→Ser mutant of *Pae* SmAP1 was created in a similar manner, except that site-directed mutagenesis was achieved *via* overlap-extension PCR with an additional pair of primers that contained the mutant site.

Harvested cells (stored at –20°C overnight) were thawed and re-suspended in a high salt concentration buffer (20 mM NaHEPES pH 7.8, 1.5 M NaCl, 0.5% v/v Triton X100, 30 mM PMSF). Cells were lysed by a combination of lysozyme treatment (0.3 mg/ml chicken egg white lysozyme) and French-press (1000 psig). Initial purification of the thermostable proteins was achieved by heating the cleared supernatant to ≈ 80°C, followed by high-speed centrifugation (37,000$g$) to remove the bulk of denatured *E. coli* proteins. The SmAP1-His6x proteins were further purified by affinity chromatography on a Ni$^{2+}$-charged iminodiacetic acid-sepharose column (both proteins eluted as broad peaks over the range 170-400 mM imidazole). Both *Mth* and *Pae* SmAP1 were >99% pure by this point (as determined by SDS-PAGE and MALDI-TOF mass spectrometry). Since the His6x tag prevents heptamer formation for some SmAPs (unpublished data, Mura & Eisenberg), the next step was proteolytic removal of the C-terminal tag for both *Pae* and *Mth* SmAP1 (*wt Mth* SmAP1 is 81-amino acid residues, and has a MW of 9,029 Da; *wt Pae* SmAP1 is 80-amino acid residues, and has a MW of 8,800 Da).

The His-tag and most of the linker were removed by limited proteolysis with trypsin (since thrombin was ineffective). The peak fractions from the Ni$^{2+}$-column that were judged as pure by SDS-PAGE were pooled and dialyzed at room temperature into phosphate-buffered saline (PBS) supplemented with 15 mM EDTA (to prevent His-tag mediated aggregation). The EDTA concentration was gradually reduced to zero over 2-3 buffer exchanges. Porcine trypsin was added to the SmAP1 (at ≈ 1 mg trypsin per 100 mg

SmAP1), and complete removal of the tag occurred after ≈ 4 h at 37°C (the extent of proteolytic digestion was assayed by MALDI-TOF spectra of time points). Transfer of the protein to 4°C and addition of a protease inhibitor (50 mM PMSF) terminated the reaction. The amino acid composition of *Mth* and *Pae* SmAP1 led to calculated isoelectric points of ≈ 5.2 and 5.8, resepectively; therefore, anion exchange chromatography was used to separate cut (*i.e.*, *wt*) SmAP1 from trypsin, uncut protein, and any other contaminants. In preparation for anion exchange chromatography on a quaternary ammonium matrix (UNO-Q6, BioRad), *Pae* SmAP1 was dialyzed against 20 mM Tris, pH 8.55. *Mth* SmAP1 was insoluble at 4°C or in the Tris-alone buffer, and had to be dialyzed *versus* 20 mM Tris pH 8.55, 30 mM EDTA pH 8.0 at room temperature (EDTA did not interfere with chromatography). Both SmAP1s eluted at ≈ 80 mM NaCl in the salt concentration gradient. Pure fractions (assayed by SDS-PAGE and MALDI-TOF) were pooled and dialyzed into a buffer for crystallization.

### Crystallization of *Pae* *SmAP1 and* Mth *SmAP1*

For *Pae* SmAP1 crystallization, the protein buffer was "XB" (10 mM Tris pH 7.8, 5 mM EDTA pH 8.0), and for *Mth* (which requires higher ionic strength buffers for solubility) it was "XB6β" (10 mM Tris pH 7.8, 5 mM EDTA pH 8.0, and 0.1 M NaCl). Protein concentrations in these buffers were increased to various values for crystallization (noted below) by using 3 kDa molecular weight cutoff Centripreps to reduce sample volume. Initial sparse matrix screening of crystallization conditions utilized the commercially available kits from Hampton Research and Emerald Biosystems, Inc. Final, optimized *Pae* SmAP1 crystals of the $C222_1$ form were grown by the hanging-drop vapor diffusion method in 24-well Linbro trays. An 11 μl drop [4 μl well buffer + 5 μl wt 29.6 mg/ml *Pae* SmAP1 + 1 μl 0.1 M dithiothreitol (DTT) + 1 ml 0.1 M uridine-5'-monophosphate (UMP)] was equilibriated against a 800 μl well [0.1 M sodium acetate pH 8.20, 0.1 M ammonium acetate, 8.6% w/v PEG-4000, and 23.8% v/v glycerol] at room temperature ($\approx$ 19.8˚C). Orthorhombic crystals reached maximum dimensions of $0.1 \times 0.1 \times 0.3$ mm within 5 days. Hanging drops contained a mixture of the new $C222_1$ crystals and the previously reported $C2$ form (used to solve the original *Pae* SmAP1 structure[23]).

Three forms of *Mth* SmAP1 crystals were obtained under three conditions by hanging-drop vapor diffusion at room temperature. For the $P1$ form, *Mth* SmAP1 was at 56 mg/ml in buffer XB6β. The drop was 4 μl of protein + 4 μl of well buffer. The well was 600 μl of [0.1 M sodium citrate pH 5.60, 15% w/v PEG-4000, 0.2 M ammonium acetate]. Crystals grew to maximum dimensions of $\approx 0.1 \times 0.1 \times 0.25$ mm within 7 days. For the $P2_12_12_1$ form, *Mth* SmAP1 was at 42 mg/ml in buffer XB6β. The drop was 3 μl of protein + 3 μl of well buffer. The well was 600 μl of [0.1 M Tris pH 8.50, 10% v/v isopropanol]. Crystals grew to maximum dimensions of $\approx 0.3 \times 0.3 \times 0.6$ mm within 3 days. For the $P2_1$ form, *Mth* SmAP1 was at 30.3 mg/ml in a modified form of buffer XB6β that contained a 26-nt single-stranded DNA [10 mM Tris pH $\approx$ 7.7, 3 mM EDTA pH 8.0, 55 mM NaCl, 0.6 mM ssDNA]. The drop was 2.5 μl of protein/ssDNA + 2.5 μl of well buffer + 1 μl of 0.1 M UMP. The 600 μl well contained 55 μl of 1.0 M sodium citrate (pH 5.6), 5 μl of 1.0 M sodium citrate (pH 8.0), 60 μl of 2.0 M ammonium acetate, 180 µl of neat MPD and 300 µl of sterile dH$_2$O (interestingly, 2.5 M 1,6-hexanediol could be substituted for MPD). Crystals grew to maximum dimensions of ≈ 0.15 × 0.15 × 0.25 mm within 7 days.

### *Cryoprotection and data collection*

The *C*222$_1$ *Pae* SmAP1 and *P*2$_1$ *Mth* SmAP1 crystals did not require the addition of a cryosolvent, due to the 23.8% v/v glycerol or 30% v/v MPD in those drops, respectively. The other two *Mth* SmAP1 crystal forms had to be cryoprotected as follows: *(i)* for the *P*1 form, ethylene glycol was added directly to the drop to a final concentration of ≈ 20% v/v, and crystals were allowed to soak for 20 sec prior to mounting in a cryo loop; *(ii)* for the fragile *P*2$_1$2$_1$2$_1$ crystals, the cryoprotectant was ethylene glycol (mixed with well buffer), and had to be introduced gradually over several hours (in ≈ 5% v/v increments). The *P*2$_1$2$_1$2$_1$ crystals were allowed to soak for only a very short time (2-3 sec) at the final ethylene glycol concentration (30% v/v). Diffraction data were collected either at the synchrotron (*P*1 and *P*2$_1$ form *Mth* xtals) or in-house (*P*2$_1$2$_1$2$_1$ *Mth* and *C*222$_1$ *Pae* crystals) on an ADSC Quantum-4 charge-coupled device (CCD) detector. All crystals were mounted in a cryogenic nitrogen stream at -168°C for data collection.

After autoindexing, all images were indexed/integrated/reduced in DENZO, and reflections were scaled and merged in SCALEPACK.[35] Complete data sets were collected from single crystals (Table 3.1). Unit cell dimensions for the *Pae C*222$_1$ form are *a* = 91.83, *b* = 113.76, *c* = 126.59 Å; for the *Mth* crystals they are: *a* = 45.07, *b* = 54.08, *c* = 62.35 Å, α = 87.58°, β = 72.86°, γ = 81.45° (*P*1); *a* = 65.25, *b* = 109.96, *c* = 83.76 Å, β = 95.81° (*P*2$_1$); *a* = 40.37, *b* = 114.70, *c* = 238.60 Å (*P*2$_1$2$_1$2$_1$). The large unit cell edge of

the *Mth* $P2_12_12_1$ crystals ($c = 238.60$ Å) led to spot overlap for high-resolution reflections ($d < 3$ Å), so multiple data sets were collected at two $2\theta$ values ($0°$, $-12°$) for two crystal alignments (related by a $45°$ azimuthal rotation).

### *Structure determination, refinement, and validation*

Initial phases for the $C222_1$ *Pae* SmAP1 structure were determined by the evolutionarily-programmed molecular replacement algorithm (EPMR).[36] The most reasonable Matthews coefficient ($V_M = 2.58$ Å$^3$/Da) corresponded to a heptamer in the asymmetric unit (a.u.), so the search model was the identical *Pae* SmAP1 heptamer from the $C2$ crystal form.[23] The EPMR solution was used for manual model building in the program O,[37] and model refinement in CNS.[38] Refinement in CNS proceeded by standard protocols, using the maximum-likelihood target function for amplitudes (mlf), bulk solvent correction, and anisotropic B-factor correction terms. Seven-fold non-crystallographic symmetry (NCS) was determined by calculation of a locked self-rotation function, but NCS restraints were not imposed at any time. Solvent molecules were added as necessary (water, glycerol, acetate). Refinement of individual atomic positions, isotropic temperature factors, and simulated annealing torsion angle dynamics was performed in most rounds. Each refinement round ended with inspection of the agreement between the model and $\sigma_A$-weighted $2F_o - F_c$, $F_o - F_c$, and simulated annealing omit maps (the latter only as necessary).

Determination of the *Mth* $P1$ structure proceeded in two steps. First, a homology model of the *Mth* SmAP1 heptamer was built from the *Pae* SmAP1 structure using an in-house script (unpublished, Mura & Eisenberg), and was used as a search model for molecular replacement with EPMR ($V_M$ = 2.29 Å$^3$/Da for a single heptamer in the $P1$ cell). Next, the unambiguous EPMR solution was converted to a polyalanine model and subjected to free-atom model refinement with the ARP/wARP program in the "molrep" mode (side chains from the *Mth* sequence were built in the final wARP stage). This *Mth* $P1$ structure was refined in the usual manner with CNS, as described above for the *Pae* $C222_1$ structure. The $P2_1$ and $P2_12_12_1$ *Mth* structures were solved by molecular replacement (EPMR) with the refined $P1$ *Mth* model. Self-rotation functions and $|F_o|^2$ Patterson maps were calculated to deduce the NCS between heptamers in the $P2_1$ and $P2_12_12_1$ forms (each of which contains 14 monomers per a.u.). Solvent was added as necessary for all structures (see Table 3.1), and no NCS restraints were enforced at any point in the refinements.

Refinement statistics for the single *Pae* and three *Mth* structures are shown in Table 3.1. Each of the four protein models is complete, except for anywhere from 6-11 missing N-terminal residues in various models (see PDB files). The stereochemistry and geometry of each SmAP1 monomer was validated with the programs PROCHECK[39] and ERRAT,[40] and found to be acceptable (*e.g.*, no residues in the disallowed region of φ,ψ space for the *Pae* $C222_1$ model). Final model coordinates and diffraction intensity data were submitted to the PDB, with ID codes 1JBM, 1LOJ, 1JRI, and 1LNX (see Table 3.1).

### *Analytical ultracentrifugation*

The *wt Pae* protein in 75 mM NaCl, 10 mM Tris, pH 7.8, was examined by sedimentation velocity in a Beckman Optima XL-A analytical ultracentrifuge at 52,000 rpm and 20°C using absorption optics at 273 nm and a 12 mm pathlength double sector cell. The sedimentation coefficient distribution was determined from a $g$(s) plot using the Beckman Origin-based software (Version 3.01). The peak sedimentation coefficient was corrected for density and viscosity to an $S_{20,wat}$ value by using a value for the partial specific volume at 20°C of 0.743 (calculated from the amino acid composition[41] and corrected to 20°C[42]).

Sedimentation equilibrium runs were performed on all three proteins – *wt Mth*, *wt Pae*, and the *Pae* C8S mutant – in 150 mM NaCl, 10 mM Tris, pH 7.8, again using a Beckman Optima XL-A analytical ultracentrifuge. Each protein was examined at three different concentrations and four speeds, using 12 mm pathlength six-sector cells. Protein concentrations used were 3.4, 0.69 and 0.19 mg/ml for *wt Pae*; 5.9, 1.26 and 0.32 mg/ml for the C8S mutant of *Pae*; and 4.1, 0.85 and 0.22 mg/ml for *wt Mth*. Rotor speeds were 8,000, 10,000, 12,500 and 14,500 rpm. Protein concentration was monitored by absorption at 280 nm and, for the lowest protein concentrations, at 232 nm. A partial specific volume of 0.743, calculated as described above, was used for all three proteins. Individual scans were analyzed using the Beckman Origin-based software (Version 3.01) to perform a nonlinear least-squares exponential fit for a single ideal species, thus giving the weight-averaged molecular weight for each protein.

### *Transmission electron microscopy*

The following protein samples were prepared for electron microscopy: (1) 0.5 mg/ml wild-type *Mth* SmAP1 in 10 mM Tris pH 7.5, 60 mM NaCl, (2) 1.2 mg/ml wild-type *Pae* SmAP1 in 25 mM Tris pH 7.5, 30 mM NaCl, (3) 1.1 mg/ml C8S mutant *Pae* SmAP1 in the same buffer as the *wt* protein, and (4) 1.2 mg/ml wild-type *Pae* SmAP1 in reductant buffer (25 mM Tris pH 7.5, 30 mM NaCl, 10 mM DTT). Carbon-coated parlodion support films mounted on copper grids were made hydrophilic immediately before use by high voltage, alternating current glow-discharge. Protein samples were applied directly onto the grids and allowed to adhere for 2 min. Grids were rinsed with distilled water and negatively stained with 1% w/v uranyl acetate. Specimens were examined in a Hitachi H-7000 electron microscope at an accelerating voltage of 75 kV.

### Gel-shift assays

For gel-shift experiments, negatively supercoiled plasmid DNA was prepared by transforming the plasmid into *E. coli* BL21(DE3) cells and mini-prepping (Qiagen) it from spun down cells that had reached stationary phase. Several different plasmids were tested, including ones derived from pUC18, pACYC, pET-22b(+) (Novagen), and pCR-Blunt (Invitrogen). Titration of plasmids with ethidium bromide was used to verify the negative superhelicity of the DNA *via* electrophoretic mobility changes in agarose gels. Single-stranded DNAs of various lengths and sequences were synthesized by Integrated DNA Technologies, Inc., and were re-hydrated in 10 mM Tris pH 7.8 (*e.g.*, the 26-mer in Fig. 3.8(b) with the following sequence: $^{5'}$CGGATCCTCAGTAAAAAGTGCGGAAA$^{3'}$). Stock solutions of protein were *wt Pae* at 5.6 mg/ml in buffer XB (see above) or *wt Mth* at 5.6 mg/ml in buffer XB6β (see above). Except as noted, buffer, DNA, and protein samples were mixed to produce 25- or 50-µl reactions that were incubated at room temperature (generally for 30-60 min). Gel-shift of the DNA was assayed by electrophoresis at a constant voltage (120V) in 1.3% or 1.5% w/v TAE/agarose gels. Examples of typical reactions are shown in Fig. 3.8.

## Results

### *Crystallization and determination of the* **Pae** *and* **Mth** *SmAP1 structures*

As with many proteins, crystallization of *Pae* SmAP1 was not straightforward. Wild-type (*wt*) *Pae* SmAP1 crystallized only in the presence of dithiothreitol (DTT), as described in the *Methods* section. Identical crystallization buffers that lacked DTT failed to produce crystals, and presumably this additive is required because it reduces the seven disulfide bonds that form between Cys8 residues in the *Pae* tetradecamer (which can therefore be thought of as a dimer of heptamers rather than as a heptamer of dimers). We found that other reductants (*e.g.*, β-mercaptoethanol) can substitute for DTT to yield crystals, although such crystals are of poorer quality than the DTT-based condition. Apparently, reduction of the disulfides frees heptamers to crystallize independently in orientations that relax crystal lattice strain, even when the 14-mer persists in the crystal (as in the $C222_1$ form reported here). The only other notable (but unnecessary for crystallization) additive to the *Pae* crystallization condition was uridine-5'-monophosphate (UMP).

As shown in Table 3.1, diffraction data extended to at least 2.05-Å resolution for the $C222_1$-form *Pae* SmAP1 crystals. Previously, we determined the crystal structure of *Pae* SmAP1 in spacegroup *C*2 by multiple-wavelength anomalous dispersion phasing.[23] Thus, the $C222_1$ structure reported here was solved by the stochastic molecular replacement method in the EPMR program, using the *C*2 heptamer as a search model (the Matthews coefficient and 7-fold NCS in the locked self-rotation function suggested a heptamer in the $C222_1$ asymmetric unit). However, NCS restraints were not imposed during crystallographic refinement. As discussed in detail below, only the uridine fragment of UMP was built into the final refined model, and atomic occupancies ($q$) were refined only for uridine (not for any other ligand or protein atoms). Partial occupancies for uridine atoms were restricted to a reasonable range ($0.2 < q < 1.5$). The final structure was refined to an $R/R_{\text{free}}$ of 18.2%/22.6%, with reasonable model geometry (Table 3.1) and no outliers in a Ramachandran plot.

Crystallization of *Mth* SmAP1 was relatively simple, and in fact this protein could be crystallized in three forms ($P1$, $P2_1$, $P2_12_12_1$) under three dissimilar conditions (also, a fourth form was crystallized by Collins *et al.*[32]). The most intriguing result is that *Mth* SmAP1 crystallized in the $P2_1$ form only in the presence of a single-stranded DNA (ssDNA) to which it was thought to bind, even though ssDNA was not found in the crystal structure. Since diffraction data were obtained from the $P1$ form before the first *Mth* SmAP1 structure was reported by Collins *et al.*, we solved the $P1$ *Mth* structure by a combination of molecular replacement and free-atom model refinement (in ARP/wARP). Briefly, a homology model of *Mth* SmAP1 was built from the *Pae* SmAP1 structure. An unambiguous molecular replacement solution was found for this search model against the *Mth* $P1$ data (using EPMR). In order to reduce *Pae* model bias, this solution was converted to polyalanine and phases from this initial model were used to autobuild a completely new model with the ARP/wARP program. Initial phases for the $P2_1$ and $P2_12_12_1$ *Mth* data were obtained by molecular replacement with the refined $P1$ model (as summarized in Table 3.1). No NCS restraints were applied in refinement of any of the *Mth* structures, and various non-protein molecules were built into electron density as appropriate (based on the crystallization condition and $F_o - F_c$ density $> +3\sigma$ in strength). Electron density for the UMP-binding sites was more interpretable in *Mth* SmAP1 than in the *Pae* structure, and permitted model building of six complete UMPs (only uridine fragments were built for the other eight UMPs in the *Mth* 14-mer). As with the *Pae*•UMP model, partial occupancies of UMP atoms were refined ($0.2 < q < 1.5$). All three *Mth* structures were refined to reasonable values of $R/R_{\text{free}}$ and model geometries (Table 3.1).

### *Comparisons of known SmAP monomer, dimer, and heptamer structures*

Several structures of Sm proteins and SmAPs are now available, and make possible the comparative structural analyses of these proteins. The previously reported *Mth* SmAP1 heptamer structure[32] is virtually identical to the *Mth* structures reported here (0.65 Å RMSD for superimposition of the *P*1 heptamer using mainchain atoms). The results from pairwise comparisons of the *Pae*, *Mth*, and *Afu* SmAP1s are shown in Fig. 3.1 and Table 3.2, and show that the compact, $\approx$ 80-amino acid SmAP monomer structures are nearly identical. The most similar monomer structures are the *Afu/Mth* pair (0.51 Å RMSD), and the most dissimilar are *Mth/Pae* SmAP1 (1.02 Å RMSD). These values do not correlate to pairwise sequence similarities. The increase in pairwise RMSDs in going from monomer alignments to dimers and heptamer alignments (Table 3.2) suggests that there are slight rigid-body variations in the monomer orientations in the higher-order oligomers (*i.e.*, slight variations in interfaces cause the RMSDs to propagate when comparing heptamers to dimers and dimers to monomers).

The absolute conservation of the dimer interface in three different SmAP1s is emphasized by the view in Fig. 3.1. Sequence conservation is low for interfacial residues relative to the rest of the SmAP monomer sequence; this is probably because the interface is largely formed by hydrogen bonding between mainchain atoms of the β4 strand of one monomer and the β5 strand of an adjacent monomer. The interface structure is also conserved between SmAPs and human Sm heterodimers.[23] The main structural difference in the three SmAP1 heptamers is the width of the pore: ≈ 8-9 Å diameter for *Pae versus* ≈ 12-15 Å in *Mth* and *Afu*. Such variation largely arises from differences in the structures of the pore-forming loops L2 and L4 (Fig. 3.1). Mapping of the phylogenetic conservation of SmAP residues onto the *Pae*, *Mth*, or *Afu* heptamer structures shows that most of the conserved residues cluster about the pore region (data not shown, Mura & Eisenberg). One of the least conserved features of the SmAP1 heptamer structures is the calculated electrostatic potential of the surfaces: the *Pae* and *Mth* heptamers display a strongly acidic L4 face, while the surface of the *Afu* heptamer is much more basic (ref. 23 for *Pae*, ref. 31 for *Afu*, and unpublished data for *Mth*, Mura & Eisenberg).

***Various oligomeric states of SmAP1, including sub-heptamers and 14-mers***

Biophysical characterization of *Pae* and *Mth* SmAP1 by a variety of methods reveals peculiar oligomerization properties. These methods include: mass spectrometry, size exclusion HPLC, native polyacrylamide gel electrophoresis (PAGE), and analytical ultracentrifugation. Sedimentation velocity ultracentrifugation revealed that *wt Pae*: *(i)* is monodisperse in solution; *(ii)* has a symmetric and narrow Gaussian-shaped distribution of sedimentation coefficients, with a coefficient at 20°C of $S_{20,w} = 6.49$ S; and *(iii)* has a frictional coefficient ratio close to one ($f/f_o = 1.2$, where $f$ = experimentally derived frictional coefficient and $f_o$ = ideal frictional coefficient for a sphere with the MW of SmAP1). These preliminary results suggested a roughly spherical, high-order *Pae* oligomer (SmAP1)$_n$ with $n \approx 12 \pm 2$ (data not shown, Mura, Phillips, & Eisenberg).

The results of equilibrium sedimentation analyses of *wt Mth*, *wt Pae*, and the C8S mutant of *Pae* SmAP1 are shown in Fig. 3.2. Molecular weights were estimated by fitting experimental curves to single exponential models, and include a roughly 2-3% error (residuals are shown in the top panels). The calculated molecular weight of *wt Pae* suggests that it exists as a tetradecamer. Other data suggested a disulfide-bonded 14-mer (see *Discussion*), so the single cysteine of *Pae* SmAP1 was mutated to serine to give the C8S mutant of *Pae* SmAP1. Sedimentation results with this mutant can be fit only by species with molecular weights much less than that of a heptamer (*e.g.*, the 46.7 kDa species shown in Fig. 3.2*(b)*), suggesting pentameric or hexameric states (*n* = 5 gives a MW of $\approx$ 45 kDa). The monodispersity of the data in Fig. 3.2*(b)* suggests a single, stable sub-heptameric complex. In contrast to *Pae*, sedimentation equilibrium data for *Mth* SmAP1 show that it only forms a stable, monodisperse heptamer (Fig. 3.2*(c)*). The concentration dependence of the experimentally calculated MWs (not shown), as well as the slight upward concavity of the residuals in Fig. 3.2*(b)* and 2*(c)*, provide additional evidence for *Pae* and *Mth* SmAP1 monomer ↔ oligomer association reactions.

### Polymerization of SmAP1 into polar fibers

The polymerization of both *Pae* and *Mth* SmAP1 into well-ordered fibers was an unexpected result, and is shown in the transmission electron micrographs (EM) of Fig. 3.3. Protein samples were in standard buffers (*e.g.*, 25 mM Tris pH 7.5, 30 mM NaCl for *Pae* SmAP1), and reproducibly formed the striated sheets of fibers seen in these EMs.

Measurement of the sheet and fiber dimensions, together with the diameters of SmAP1 heptamers from crystal structures ($\approx$ 70-75 Å), suggests a model in which the fibers are formed by head-to-tail stacking of heptamers, with the SmAP1 7-fold axis roughly parallel to the fiber axis (see white arrows in Fig. 3.3*(b)*). Several fibers may associate laterally to form sheets, such as those seen most clearly in Figs. 3*(a)* and *(b)*.

In order to test this head-to-tail stacking model, we assayed fiber formation by *wt Pae* and the C8S mutant. Under oxidative conditions, *wt Pae* SmAP1 forms disulfide-bonded 14-mers in which the highly acidic L4 faces are exposed at either end of the barrel-shaped structure (see the *Pae* 14-mer in Fig. 3.4*(c)*). Such a 14-mer would be constrained to form only head-to-head interfaces (*i.e.*, loop L4 face-to-loop L4 face) in a fiber, and would probably not do so because of the unfavorable electrostatic cost of closely apposing these anionic faces (at least not at the neutral pHs or low ionic strength conditions in which the SmAP1s were buffered). As expected, *wt Pae* forms only ring-shaped structures under oxidative conditions (Fig. 3.3*(c)*). However, when the seven disulfide bonds that link heptamers into 14-mers are eliminated, *Pae* SmAP1 assembles into fibers with roughly similar morphologies as *Mth* fibers. Polymerization can be achieved either by addition of a reducing agent (as in Fig. 3.3*(d)*) or by mutation of the cysteine (C8S mutant in Fig. 3.3*(e)*). Such fiber formation has been hitherto unreported for Sm proteins.

### *Packing of* **Mth** *and* **Pae** *SmAP1 heptamers in four crystal forms*

Crystallization of *Mth* and *Pae* SmAP1 in several forms is a fortuitous result, since different packing geometries of SmAP1 heptamers in these various crystal forms shed light on the oligomerization results described above. The *Pae* SmAP1 $C222_1$ structure differs from the original $C2$ form in that heptamers pack face-face in the orthorhombic lattice to give a 14-mer with 72-point group symmetry. The crystal packing is shown in Fig. 3.4*(a)* and interacting surfaces are shown in Fig. 3.4*(b)*. This 14-mer is likely to be significant because: *(i)* it is consistent with the oligomerization results described above from biophysical characterization, *(ii)* it persists in the $C222_1$ lattice despite the requirement of DTT for crystallization (the sulfhydryls in Fig. 3.4*(c)* are separated by >8-9 Å), *(iii)* the heptamer-heptamer interface occludes 7,550 $Å^2$ of surface area, and *(iv)* it is corroborated by an *Mth* 14-mer in the asymmetric unit of the $P2_1$ form. The total buried surface area in the heptamer interface of the $P2_1$ *Mth* 14-mer is probably significant (3,000 $Å^2$), although less than half as much as the *Pae* interface.

In the *Mth* $P1$ and $P2_12_12_1$ lattices, SmAP1 heptamers form quasi-hexagonal layers that stack upon one another to give a crystal. In the $P1$ form these layers are staggered; however, in the $P2_12_12_1$ form these layers are in register. Fig. 3.4*(c)* shows how the head-to-tail stacking of SmAP1 heptamers in this crystal form produces cylindrical tubes. A slight tilt of each heptamer ($\approx15°$) with respect to the tube axis results in the SmAP1 7-fold axes being parallel, but not coaxial. Since they are formed by head-to-tail stacking of asymmetric heptamers, these tubes have a defined polarity, and, when rendered as molecular surfaces, they bear a striking resemblance to the EM fibers shown in Fig. 3.3. The tubes are also consistent with EM fiber dimensions. Therefore, the $P2_12_12_1$ crystal structure provides a model for the atomic structure of SmAP1 fibers. In addition to providing insights into oligomerization states, two of the crystal forms (*Pae C*222$_1$ and *Mth P*2$_1$) were used to investigate the ligand-binding properties of SmAP1s.

***Crystal structures of* Mth *and* Pae *SmAP1 bound to various ligands***

The 1.90-Å resolution crystal structure of *Mth* SmAP1 bound to uridine-5'-monophosphate (UMP) is shown in Fig. 3.5. The protein was co-crystallized with this ribonucleotide in an effort to determine its likely RNA-binding site (co-crystallization efforts were unsuccessful with single-stranded DNA or RNA oligonucleotides). As shown in Fig. 3.5*(a)*, SmAP1 binds UMP with a 1:1 stoichiometry, so that 14 UMPs are bound to the 14-mer near the pore region. The orthogonal view in Fig. 3.5*(a)* shows that the UMPs are bound near the flat face of the heptamer, opposite the highly acidic loop L4 face. The structure of the SmAP1•UMP complex is shown in more detail in Fig. 3.5*(b)*, where it can be seen that the binding site is well defined by electron density. The uracil ring intercalates between the guanidinium group of Arg72 and the imidazole ring of His46 (both of these residues are highly conserved in SmAPs). The planes of these three moieties are spaced $\approx$ 3.6 Å apart, as expected for energetically favorable stacking interactions between these conjugated $\pi$-systems. Individual protein-UMP contacts are discussed in greater detail below.

In addition to the expected UMP binding site, we found that each *Mth* SmAP1 monomer binds a molecule of MPD. The MPD binding site is somewhat solvent-exposed, near the periphery of the SmAP1 ring (Fig. 3.5*(a)*). Protein-MPD recognition is the same in each of the 14-monomers, and is shown in detail in Fig. 3.5*(c)*. The primary contact is hydrogen bonding between the Ser21 hydroxyl and MPD, and there are several water-mediated SmAP···$H_2O$···MPD contacts. The cryoprotectant in the $P1$ and $P2_12_12_1$ *Mth* SmAP1 structures was ethylene glycol (Table 3.1), and in these structures some of the SmAP1 monomers bind ethylene glycol in the same site as MPD.

A UMP binding site was found in the *Pae* SmAP1•UMP co-crystal structure as well, but it is not as clearly defined in electron density as for *Mth* SmAP1•UMP. Fig. 3.6 shows the *Pae*•UMP structure, which was refined to a resolution of 2.05-Å. UMPs bind to the same face of the heptamer as in *Mth* (*i.e.*, the "flat face" opposite L4), but are much more distant from the pore. As shown by the $2F_o - F_c$ maps in Fig. 3.6*(b)*, only the planar uracil fragment of UMP is clearly defined in electron density. Protein-UMP contacts are scarce in this binding site. Asn46···UMP distances are shown by dashed lines in Fig. 3.6*(b)* only for the sake of completeness – the geometries of these interactions do not satisfy standard hydrogen bond criteria (in terms of both distances and angles), and favorable interactions probably do not exist between the UMP O4 oxygen and the amide nitrogen of the Asn46 side chain or between the UMP N3 nitrogen and the amide oxygen of Asn46. Also, there are no aromatic side chains in this region to participate in π-stacking interactions with the uracil base. As with *Mth* SmAP1, additional small-molecule binding sites exist in *Pae* SmAP1: many of the modeled glycerol molecules are bound identically near the loop L4 faces (see Fig. 3.6*(a)*). The significance of such binding sites is unknown.

The structure of an *Afu* SmAP1•$U_3$ complex was determined recently by Törö *et al.*,[31] and permits a comparison of the mode of uridine recognition in *Afu* and *Mth* SmAP1. The UMP binding site and SmAP1···UMP interactions clearly differ in *Mth* and *Pae* SmAP1, and, since the binding site was poorly resolved in the *Pae*•UMP complex, this structure was not included in the comparative analysis shown in Fig. 3.7. In the *Mth* and *Afu* structures, the aromatic pyrimidine ring intercalates between the side chains of the highly conserved Arg/His pair, and specific uracil recognition is achieved by hydrogen bonding of the uracil ring to the side chain of a strictly conserved asparagine residue (Asn48$_{Mth}$). The main chain amide nitrogen of a highly conserved aspartate (Asp74$_{Mth}$) also participates in hydrogen bonding to a uracil carbonyl oxygen. The pattern of hydrogen bond donors/acceptors in the Asn48/Asp74$_{Mth}$ pair makes binding specific for a uracil (if RNA) or thymine (if DNA) base. Additional specificity for uracil may be achieved by two means: *(i)* recognition of the 2' hydroxyl of the ribose (RNA *versus* DNA discrimination) and *(ii)* the C5 carbon of the pyrimidine ring of uracil is only 3.8 Å from the backbone carbonyl oxygen of Leu45$_{Mth}$ from an adjacent monomer, thus providing steric and polar discrimination against the methyl on the C5 carbon of thymine. We crystallized *Pae* and *Mth* SmAP1 in the presence of various other nucleoside monophosphates (*e.g.*, AMP, CMP, GMP), but there was no evidence for binding of these non-uridine NMPs (data not shown, Mura & Eisenberg). The only significant differences in uridine recognition by *Mth* and *Afu* SmAP1 are highlighted by green arrows in Fig. 3.7*(b)*. These are: *(i)* hydrogen bonding of an *Mth* Arg72 side chain from an adjacent monomer to the 2' hydroxyl of the ribose, and *(ii)* hydrogen bonding between a phosphate oxygen and an imidazole nitrogen from the His46 residue of an adjacent monomer. Overall, it appears that the mode of uridine recognition is conserved in the SmAP family.

**Pae *and* Mth *SmAP1 gel-shift negatively supercoiled DNA***

In our initial attempts to determine the biochemical function of *Pae* SmAP1, we inadvertently found that this protein gel-shifts supercoiled plasmid DNA. This activity was further investigated for both the *Mth* and *Pae* SmAP1s, and examples of it are shown in Fig. 3.8. Migration of the negatively supercoiled plasmid "p5L1c1" is severely retarded by incubation with μM amounts of *Mth* heptamer in Fig. 3.8*(a)*. Interestingly, the extent of gel-shift increases at higher concentrations of *Mth* SmAP1, until saturation of the effect occurs at ≈ 60 μM (compare lanes 7 and 8). A similar gel-shift occurs to supercoiled DNA when it is incubated with *wt Pae* SmAP1, as shown in lane 4 of Fig. 3.8*(b)*. This experiment also shows that the gel-shift can be eliminated by incubation with a 26-nucleotide single stranded DNA (ssDNA). Inhibition of the gel-shift activity is titratable, and at higher concentrations of ssDNA there is no gel-shift (lane 8).

Similar DNA gel shift assays and control experiments have revealed that: *(i)* the *Pae* activity is specific for supercoiled (*sup*) plasmid DNA, whereas *Mth* SmAP1 gel-shifts both *sup* and linearized (*lin*) plasmids; *(ii) Pae* activity is eliminated by $MgSO_4$, whereas the dependence of *Mth* activity on divalent metals such as $Ca^{2+}$, $Mg^{2+}$, and $Mn^{2+}$ is not as straightforward; *(iii)* ssDNA of any sequence and length >≈ 20-nt inhibits the gel-shift activity of *Pae* and *Mth* in a concentration dependent manner; *(iv)* both *Pae* and *Mth* activities are nonspecific with respect to the *sup* DNA; *(v) Pae* and *Mth* are not linearizing or otherwise cutting both strands of the sup DNA; *(vi) Mth* gel shift activity is not temperature-dependent at and above room temperature, whereas the extent of *Pae*-induced gel shift abruptly increases at ≈ 55-60°C. All of these results come from experiments in which the migration of a large (>4,000 nt) plasmid DNA is assayed in agarose gels. Binding of *Mth* SmAP1 to any one of the ssDNAs that inhibit the *sup* DNA gel shift (*e.g.*, Fig 8*(b)*) has been assayed in preliminary native PAGE experiments; these results suggest that ssDNA inhibits the *sup* DNA gel shift by directly binding to SmAP1.

## Discussion

### *Comparative structural analysis of Sm proteins and SmAPs*

Recent work has shown that SmAPs form a phylogenetically well-conserved family of proteins whose sequences are similar to eukaryotic Sm and Lsm proteins, and whose structures are nearly identical to the human Sm D3•B and D1•D2 heterodimer structures.[32,23,31] The Sm and SmAP monomers form antiparallel, five-stranded β-sheets capped by a short N-terminal α-helix. The Sm β-sheet is highly bent, into a β-barrel like structure that closely resembles proteins of the oligosaccharide/oligonucleotide binding (OB) fold family.[43] As shown by Fig. 3.1 and Table 3.2, the *Afu*, *Mth*, and *Pae* SmAP1 monomers and homodimers are nearly identical to one another. Besides the N- and C-termini, the only significant deviations in Sm and SmAP monomer structures occur in the loops: the L2 and L4 loops are the most structurally variant regions in Fig. 3.1, and several eukaryotic Sm proteins have insertions of up to 30 amino acids in loop L4. Work in progress with another SmAP homolog that contains a similar L4 insertion (*Pae* SmAP3) shows that it too forms heptamers. The recent solution structure of the SMN Tudor domain, which interacts with Sm proteins to form snRNP cores, has provided an unexpected result: the SMN and Sm monomers have the same fold, and nearly identical structures.[10] This raises the intriguing possibility that the SMN protein interacts with the Sm complex by forming mixed heteromers. Overall, there is a high degree of phylogenetic and structural conservation of the SmAP domain from archaea to eukaryotes

Afu, Mth, and Pae SmAP1s form heptamers with remarkably similar structures, primarily because the structure of the homodimer interface is extremely well conserved (Fig. 3.1). Greater RMSDs for heptamers compared to dimers (and dimers compared to monomers) shows that a large fraction of the structural variation in higher-order SmAP oligomers (Table 3.2) is due to rigid-body displacements of monomers with respect to one another. A feature of the SmAP1 heptamers that is highly conserved in terms of sequence and overall structure is the central cationic pore. The largest difference between SmAP monomers (L2, L4 loops) results in the largest difference between SmAP heptamers: variation in the width of the pore in *Pae* (≈ 8-9 Å diameter) *versus Afu* and *Mth* (≈ 12-15 Å) is due to main chain and side chain rotamer variations in the L2 and L4 loops. The other significant difference between *Afu*, *Mth*, and *Pae* SmAP1s is the calculated electrostatic potential of the heptamer surface: the L4 face of the *Afu* surface is very basic, while these *Pae* and *Mth* faces are intensely acidic. Such differences are likely to be important for modulating putative SmAP-RNA interactions. Overall, the near identity of the human Sm D3•B and D1•D2 heterodimers to *Afu*, *Mth*, and *Pae* SmAP homodimers qualifies the SmAP1 *homo*heptamer as an accurate model for the Sm *hetero*heptamer of eukaryotic snRNP cores.

### The oligomerization properties of SmAPs

Like the Lsm (but not Sm) proteins, *Pae*, *Mth*, and *Afu* SmAP1 form heptamers in the absence of RNA. We also found that SmAP1 exhibits complex self-association properties that result in 14-mers and sub-heptameric oligomers, in addition to the expected heptamers. Various oligomeric states were characterized *in vitro* (primarily by ultracentrifugation, Fig. 3.2), revealing roughly spherical disulfide-bonded *Pae* SmAP1 14-mers and a monodisperse population of *Mth* SmAP1 heptamers. Additionally, we created a cysteine-free point mutant of *Pae* SmAP (C8S), and found that it forms sub-heptameric states (most likely pentamers). Interestingly, similar plasticity of oligomerization behavior has been reported for human Sm proteins. Lührmann *et al.* found that a human Sm E•F•G complex forms a stable oligomer – most likely a (E•F•G)$_2$ hexamer – whose ring-shaped structure resembles intact Sm heteroheptamers by EM.[17,30] One of these studies also found that stable, sub-heptameric complexes of human Sm proteins (*e.g.*, a D1•D2•E•F•G pentamer) may be intermediates in the Sm-RNA assembly pathway.[17] Recently, another *Afu* SmAP was reported to form hexamers (personal communication cited in ref. 29), and in the structure of the human Sm D3•B the heterodimers pack as (D3•B)$_3$ hexamers in the asymmetric unit of the crystal.

We found that *Pae* and *Mth* SmAP1 reproducibly oligomerize into 14-mers, either *in vitro* (*Pae*) or in various crystal forms (*Pae* and *Mth*). The highly acidic L4 faces are exposed in the barrel-shaped 14-mers (Fig. 3.4), as expected from electrostatics. The heptamer-heptamer interface buries a large amount of surface area in both *Pae* (7,550 Å$^2$) and *Mth* (3,005 Å$^2$), suggesting the significance of these oligomers. Preliminary crystallographic data from another SmAP homolog (*Pae* SmAP3) show that it also forms 14-mers in the asymmetric unit (unpublished results, Mura & Eisenberg). The propensity of ring-shaped SmAPs to crystallize as head-head oligomers with dihedral symmetry is shared by another single-stranded RNA binding protein: the *trp* RNA-binding attenuation protein (TRAP) forms a toroidal 11-mer that forms both head-head and head-tail 22-mers in the crystal[44] (in fact, the structures of Sm and TRAP monomers are quite similar).

Perhaps the most novel property of SmAP1s is their polymerization into extremely well-ordered fibers. *Pae* and *Mth* SmAP1 samples at physiological conditions form these fibers, which we observe by EM. Three lines of evidence suggest that these fibers form by the head-to-tail stacking of heptamers (Fig. 3.3): differential fiber formation by C8S and *wt Pae* SmAP1, comparison of measured fiber dimensions with SmAP1 heptamer dimensions, and electrostatic considerations for the packing of highly charged heptameric disks. The packing of *Mth* SmAP1 heptamers in the $P2_12_12_1$ lattice supports our head-to-tail polymerization model, and provides an atomic-resolution structure for the fibers (Fig. 3.4). Such peculiar oligomerization properties have not been reported for eukaryotic Sm proteins, and the biological significance of SmAP1 14-mers and homogeneous, fibrillar polymers is not yet known.

***Comparison and analysis of the ligand-binding properties of SmAPs***

Comparison of the structures of *Mth* SmAP1 bound to UMP and *Afu* SmAP1 bound to oligouridine ($U_3$) suggest a highly conserved mode of RNA recognition in SmAPs. UMP binds near the 7-fold axis, suggesting the pore as a putative RNA binding site. Diagrams of SmAP1⋯UMP interactions show that these proteins specifically bind the uracil base by a combination of π-stacking and hydrogen bond interactions with strictly conserved SmAP residues (Fig. 3.7). Differences between UMP binding in *Mth* and *Afu* are limited to interactions with the ribophosphate moiety, and may not be significant since *Mth* SmAP1 was co-crystallized with free UMP nucleotide, whereas *Afu*

SmAP1 was crystallized with a U$_3$ oligouridine. The oligo(U) specificity of RNA binding to *Afu* SmAP1 mimics the substrate specificity of eukaryotic Sm proteins.[45,31] The geometry of binding of several of the UMPs in *Mth* SmAP1 allows them to be strung together into a hypothetical oligouridine that may mimic biologically relevant RNA binding in the Sm core of snRNPs. Failure of other NMPs to co-crystallize with *Mth* or *Pae* SmAP1 supports the specificity of uridine binding that we infer from the crystal structures.

In addition to the *Mth* UMP-binding site, several other ligand-binding sites exist in *Mth* and *Pae* SmAP1. Unlike the well-defined UMP site in *Mth*, the uridine-binding site in *Pae* is distant from the pore and not easily interpretable in electron density maps, suggesting low affinity binding at this site (Fig. 3.6). Also, the *Pae* SmAP1 residues in the region of this uridine are not very conserved. If all SmAPs specifically bind to an oligouridine site in RNA *in vivo*, then geometric considerations require any such RNA to bind near the 7-fold symmetry axis (*i.e.*, the pore), and therefore the *Pae* binding pocket described here cannot be biologically relevant. Presumably, breaking of 7-fold symmetry in eukaryotic Sm *hetero*heptamers is reconcilable with RNA binding away from the pore (although there is no evidence for this). We note that the UMP-binding site in *Afu* and *Mth* exists in *Pae* SmAP1, and that UMP can be docked into this putative *Pae* binding site with only minimal changes required for side chain rotamers. We also found other sites in *Mth* and *Pae* occupied in each monomer by MPD, ethylene glycol, or glycerol. Though these ligands are clearly defined by electron density and many of the residues in these binding sites are phylogenetically conserved, any biological significance of these additional sites is not yet clear.

The gel-shift activity of *Mth* and *Pae* SmAP1 on negatively supercoiled DNA substrates (Fig. 3.8) is especially interesting given the similarity of SmAP monomers to the OB fold. We found that SmAP1s non-specifically gel-shift a variety of supercoiled DNA substrates and that ssDNA oligonucleotides of >20-nt inhibit the gel-shift (Fig. 3.8*(b)*), possibly by direct binding to SmAP1. Since eukaryotic Sm proteins bind to ssRNA, and since SmAP *homo*heptamers probably do not function *identically* to Sm *hetero*heptamers, we propose that SmAPs may have a generic single-stranded nucleic acid binding activity (*e.g.*, as a nucleic acid chaperone). The striking resemblance of the SmAP and OB folds corroborates this idea, given that several OB-fold proteins bind to ssDNA non-specifically. The following recently determined structures are all very similar (and in some cases nearly identical) to the Sm fold: the single-stranded DNA-binding domain of replication factor A,[46] the S1 RNA-binding domain,[47] the single-stranded telomeric DNA binding protein,[48] and the *Streptococcus pneumoniae* SP14.3 protein (which is fused to a domain that is homologous to ribosomal protein S3).[49]

***Emerging differences between SmAPs and canonical Sm proteins***

Eukaryotic Sm and Lsm proteins and their archaeal homologs, which we term Sm-like archaeal proteins, share a number of structural and functional features. Most significant is the similarity in Sm and SmAP 3D and quaternary structures: the monomers are nearly identical, and the SmAP *homo*heptamer parallels the Sm *hetero*heptamer that forms snRNP cores. Also, both sets of proteins apparently bind specifically to oligouridine-containing RNA. However, several differences are emerging between SmAPs and the canonical Sm proteins. The results presented here show that SmAPs associate into many oligomeric states besides the standard heptamer (*e.g.*, 14-mers and sub-heptamers), and can polymerize into homogeneous fibers. Such behavior is unreported for eukaryotic Sm proteins. No structural information is available for Sm proteins bound to RNA (or any other ligand), so it is difficult to evaluate the similarity of uridine binding by eukaryotic Sm proteins and SmAPs. Cross-linking experiments corroborate RNA binding near the pore in human Sm heptamers.[50] The near identity of the Sm and SmAP dimer structures, as well as the strictly conserved mode of uridine recognition between *Afu* and *Mth* SmAP1, suggest that the SmAP1 UMP-binding site is an accurate model for RNA binding in the snRNP core. In this model, snRNA wraps around the circumfrence of the pore, but does not thread through it. Further elucidation of the similarities and differences between archaeal SmAP complexes and the Sm cores of eukaryotic snRNPs will be the aim of future experiments, and will provide insight into the structure and assembly of snRNPs.

A.J., & Horton, J.C., ed.), pp. 90-125. The Royal Society of Chemistry, Cambridge, England.

**Figure and Table legends**

**Table 3.1: Statistics for several crystal forms.** Crystallographic statistics are given for the *Mth* and *Pae* SmAP1 structures in different spacegroups (with various packing geometries) and with bound ligands (UMP, MPD, *etc.*). Data were collected either in-house ($\lambda = 1.54$ Å) or at the synchrotron ($\lambda = 1.10$ Å). Statistics for the highest resolution shell are given in [square brackets]. $R_{cryst} = \Sigma_{hkl} ||F_{obs}| - |F_{calc}|| / \Sigma_{hkl} |F_{obs}|$, and $R_{free}$ was computed identically, except that 5% of the reflections were omitted as a test set. Non-protein molecules were added based on the chemical composition of the crystallization condition and sufficiently strong $F_o - F_c$ density ($>3\sigma$).

**Table 3.2: Pairwise RMSDs between *Pae*, *Mth*, and *Afu* SmAP1.** RMSDs are shown for pairwise 3D alignments of *Pae*, *Mth*, and *Afu* SmAP1 monomers, dimers, and heptamers (using mainchain atoms only). The *Afu-Mth* pair superimposes best, while the *Pae-Mth* monomer structures are most dissimilar.

**Figure 3.1: 3D structural alignment of *Pae*, *Mth*, and *Afu* SmAP1 dimers.** A depth-cued stereoview is shown of the Cα trace for aligned *Pae* (red hues), *Mth* (blue hues), and *Afu* (green hues) SmAP1 dimers. N- and C-termini, as well as loops L2 and L4 are indicated. The greatest structural variation is in the positions of these two pore-forming loops, and the dimer interface is strictly conserved (asterisks). The large difference in the width of the heptameric pores in *Pae* ($\approx$8-9 Å diameter) *versus Afu* and *Mth* ($\approx$12-15 Å diameter) is due to two structural features: *(i)* side chain variation: the position of the positively-charged, pore-lining side chain R29 is extended *into* the pore in *Pae*, but K31 extends *along* the direction of the pore in *Mth*, and *(ii)* backbone variation: the distance

between identical Cα atoms in loop L2 of the seven monomers is greater in *Mth* than in *Pae*, *i.e*., the backbone protrudes further into the pore in the *Pae* heptamer.

**Figure 3.2: The oligomeric states of *Pae* and *Mth* SmAP1s in solution.** Representative sedimentation results for analytical ultracentrifugation of *wt Pae* SmAP1 *(a)*, the C8S mutant of *Pae* SmAP1 *(b)*, and *wt Mth* SmAP1 *(c)* are shown. Data were collected at 20°C, at a rotor speed of 12,500 rpm, with absorbance measured at 280 nm. Protein concentrations were 0.69 mg/ml *(a)*, 1.26 mg/ml *(b)*, and 0.85 mg/ml *(c)*. Weight-average molecular weights (given in kDa) were determined by fitting experimental data (circles) with a single exponential (solid line), and include roughly 2-3% error (residuals are in top panels); note that the protein samples are monodisperse. The molecular weight of the *wt Pae* protein suggests that it exists as a 14-mer, while the *wt Mth* data closely fits a heptamer. The molecular weight of the C8S mutant is significantly less than that of a heptamer, suggesting that it may exist in lower oligomerization states (4-, 5-, or 6-mers). Such heptamer "subcomplexes" have been detected for eukaryotic Sm proteins (see text for details).

**Figure 3.3: Polymerization of SmAP1s into polar fibers.** Transmission EMs are shown for *wt Mth* (*a*, *b*), *wt Pae* (*c* oxidized, *d* reduced), and the C8S mutant of *Pae* SmAP1 (*e*). The scale bar represents 10 nm for panel *(c)*, and 50 nm for all other panels. The striated sheets formed by *Mth* SmAP1 (*a, b*) and non-disulfide bonded *Pae* SmAP1 (*d, e*) are extremely well ordered. The distance between the inner arrow tips in *(b)* corresponds to ≈ 8.3 nm (in agreement with the heptamer diameters from crystal structures), and suggests that the fiber axis is parallel to the heptameric 7-fold. The ≈ 50 nm distance between the outer white arrows in *(b)* corresponds closely to six heptamer widths. Together with heptamer packings in various crystal forms, these EMs suggest that SmAP1 fibers form by head-to-tail stacking of heptamers (see Fig. 3.4). Doughnut-shaped SmAP1s are visible in the background of these EMs (most clearly for the *wt Pae* sample in panel *(c)*).

**Figure 3.4: Various crystalline oligomers of *Pae* and *Mth* SmAP1.** Panel *(a)* provides orthogonal views of the quasi-hexagonal packing of *Mth* SmAP1 heptamers in the $P2_12_12_1$ crystal form. Heptamers stack upon one another to form cylindrical tubes, thus providing a model for the structure of the EM fibrils (see text for explanation). The head-to-tail association of heptamers gives the tubes a defined polarity (colored arrows). Molecular surfaces show that the lateral packing of tubes in the crystal may generate the striated sheets seen by EM. A unit cell of the *Pae* SmAP1 $C222_1$ crystal form is shown in *(b)*, along with examples of crystallographic 2-fold and $2_1$ screw axes. The asymmetric unit is a heptamer (shown as Cα traces in red or blue), and a *Pae* SmAP1 14-mer is formed from adjacent asymmetric units as shown in *(c)*. Interaction surfaces and cysteines are illustrated. The 14-mer has 72-point group symmetry, with a 2-fold axis coinciding with a crystallographic 2-fold, and buries 7,547 $\text{Å}^2$ of surface area at the heptamer-heptamer interface.

**Figure 3.5: Ligand-binding sites in the structure of the *Mth* 14-mer bound to UMP.** Orthogonal views are shown for the two *Mth* heptamers (red, blue) in the asymmetric unit of the $P2_1$ form *(a)*. A single molecule of MPD binds identically to each monomer, and is shown in space-filling (colored by atom type, yellow carbons). Space-filling models of the 14 UMP ligands show that they bind in the pore region (colored by atom type, gray

carbons). Electron densities for the UMP and MPD binding sites are shown in *(b)* and *(c)*, respectively. The $2F_o - F_c$ density is contoured at $+1.2\sigma$ (green) and $F_o - F_c$ maps are contoured at $-3.2\sigma$ (red) or $+3.2\sigma$ (blue). Conserved residues that form these ligand-binding sites are labeled, and residues from different monomers are distinguished by primes. Hydrogen-bond distances are not shown in *(b)* for the sake of clarity (see Fig. 3.7 and the text for details of the SmAP1-ligand interactions).

**Figure 3.6: Ligand-binding sites in the structure of the *Pae* 14-mer bound to UMP.** Orthogonal views are shown in *(a)* for the *Pae* SmAP1 14-mer that is found in the $C222_1$ lattice (heptamer per asymmetric unit). Ten glycerol molecules bind to each heptamer (shown in space-filling, green-colored carbons), and seven of them occupy identical sites. The uridine fragments of UMP were modeled, and are shown in space-filling (gray-colored carbons). Electron density for one of the UMP-binding sites is shown in *(b)*, contoured at $+1.2\sigma$ for $2F_o - F_c$ density (green) and at $+/-3.2\sigma$ for $F_o - F_c$ density (blue/red). While electron density for the uracil moiety is clearly defined, this ligand-binding site may not be a biologically relevant RNA-binding site (see text for discussion).

**Figure 3.7: Conserved mode of uridine recognition by *Mth* and *Afu* SmAP1.** Interactions between SmAP1 and uridine are diagrammed for *Afu (a)* and *Mth (b)*. The remainder of the $U_3$ oligouridine from the *Afu* structure (indicated by a U2~~) has been omitted in *(a)* for the sake of clarity.[31] Parenthesized letters after residue labels denote individual monomers. In both structures, the aromatic uracil base intercalates between a highly conserved pair of Arg/His side chains – *e.g.*, the guanidinium of Arg72 and imidazole of His46 for *Mth* SmAP1. Specific interactions and differences between *Afu* and *Mth* are discussed in the text. This figure was derived from a LIGPLOT[51] output.

**Figure 3.8: Gel-shift of supercoiled DNA by *Mth* and *Pae* SmAP1.** The ability of *Mth* and *Pae* SmAP1 to shift the electrophoretic mobility of supercoiled plasmid DNA is shown in the agarose gels of *(a)* and *(b)*, respectively. In *(a)*, increasing concentrations of *Mth* SmAP1 were incubated with a negatively supercoiled plasmid ("p5L1c1"). The first onset of gel-shift is apparent at the lowest concentration of *Mth* (1.1 μM heptamer, arrow in lane 3), and saturates by the highest concentration (60 μM, lane 8). The ability of a 26-nucleotide ssDNA to inhibit the gel-shift induced by *Pae* SmAP1 is shown in *(b)*. Lane 1 provides a DNA ladder, lanes 2 and 3 serve as negative controls, and the arrow in lane 4 shows the maximal gel-shift in the absence of ssDNA (which may inhibit the gel-shift by binding to SmAP1).

**Table 3.1: Statistics for several crystal forms.**

| Crystal form | $P1$ (*Mth*) | | $P2_1$ (*Mth* with UMP) | | $P2_12_12_1$ (*Mth*) | | $C222_1$ (*Pae* with UMP) | |
|---|---|---|---|---|---|---|---|---|
| **Data collection** | | | | | | | | |
| X-ray wavelength (Å) | 1.1000 | | 1.1000 | | 1.5418 | | 1.5418 | |
| Resolution range (Å) | 90.0 – 1.85 | | 100.0 – 1.90 | | 100.0 – 2.80 | | 100.0 – 2.05 | |
| # reflections (total / unique) | 145,416 / 44,472 | | 337,336 / 89,378 | | 329,838 / 28,487 | | 330,687 / 40,722 | |
| Completeness (%) | 93.8 [67.4] | | 97.0 [92.4] | | 99.0 [92.5] | | 97.4 [95.5] | |
| I / σ(I) | 20.4 [4.1] | | 25.8 [2.3] | | 19.3 [3.3] | | 17.9 [4.1] | |
| $R_{merge}$ (%) | 5.4 [25.5] | | 4.8 [56.5] | | 11.2 [38.7] | | 11.5 [50.9] | |
| **Molecular replacement** | | | | | | | | |
| Search model | Homology model of *Pae* SmAP1 structure (1I8F) | | Refined $P1$ *Mth* SmAP1 structure (1JBM) | | Refined $P1$ *Mth* SmAP1 structure (1JBM) | | Refined $C2$ *Pae* SmAP1 structure (1I8F) | |
| Search method | Search w/ EPMR, then ARP/ wARP of polyalanine-ized version (to reduce model bias) | | EPMR | | EPMR | | EPMR | |
| Crystal packing | Heptamer per a.u. | | *Face-face* tetradecamer per a.u. (pseudo-72 point group symmetry) | | *Edge-edge* tetradecamer per a.u. | | Heptamer per a.u.; *face-face* tetradecamer in crystal (72 point group symmetry) | |
| **Model refinement** | | | | | | | | |
| Resolution range (Å) | 20.0 – 1.85 | | 20.0 – 1.90 | | 15.0 – 2.80 | | 20.0 – 2.05 | |
| <B-factor> (prot / water, Å²) | 27.9 / 37.2 | | 41.7 / 48.6 | | 50.9 / 40.1 | | 25.1 / 34.8 | |
| Number of solvent or ligand molecules included in model: | water | 273 | water | 387 | water | 86 | water | 325 |
| | ethylene glycol | 11 | MPD | 14 | ethylene glycol | 13 | glycerol | 10 |
| | acetate | 5 | UMP | 14 | chloride | 3 | acetate | 2 |
| | | | | | | | UMP | 7 |
| RMSDs: bonds (Å) | 0.021 | | 0.014 | | 0.007 | | 0.013 | |
| angles (°) | 2.06 | | 1.85 | | 1.44 | | 1.61 | |
| PDB code for submission | 1JBM | | 1LOJ | | 1JRI | | 1LNX | |
| $R_{cryst}$ / $R_{free}$, (%) | 19.6 / 23.8 | | 20.7 / 25.0 | | 19.9 / 29.0 | | 18.2 / 22.6 | |

**Figure 3.1: 3D structural alignment of *Pae*, *Mth*, and *Afu* SmAP1 dimers.**

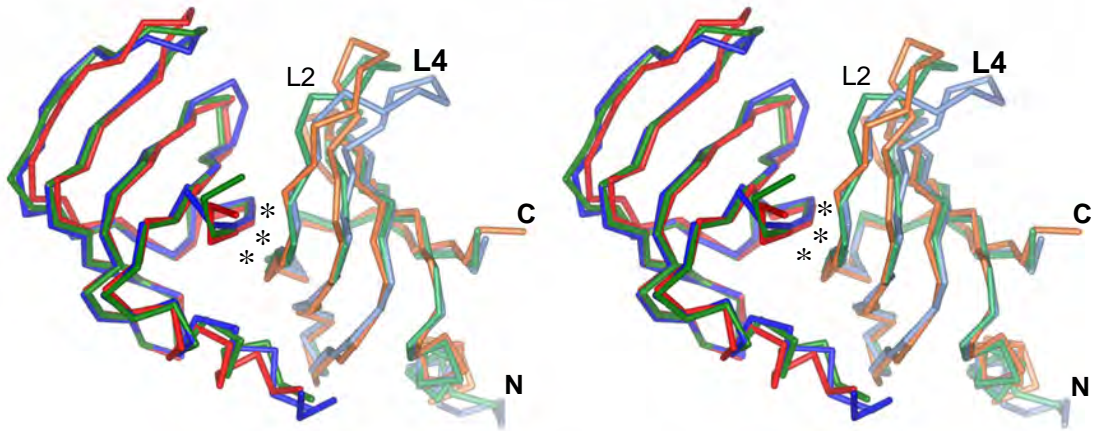

**Table 3.2: Pairwise RMSDs between *Pae*, *Mth*, and *Afu* SmAP1.**

|  | *Afu* | *Mth* | *Pae* |
|---|---|---|---|
| *Afu* | — | 0.51 Å (monomer)<br>0.61 Å (dimer)<br>0.81 Å (heptamer) | 0.90 Å (monomer)<br>1.02 Å (dimer)<br>1.96 Å (heptamer) |
| *Mth* |  | — | 1.02 Å (monomer)<br>1.19 Å (dimer)<br>1.90 Å (heptamer) |
| *Pae* |  |  | — |

**Figure 3.2: The oligomeric states of *Pae* and *Mth* SmAP1 in solution.**

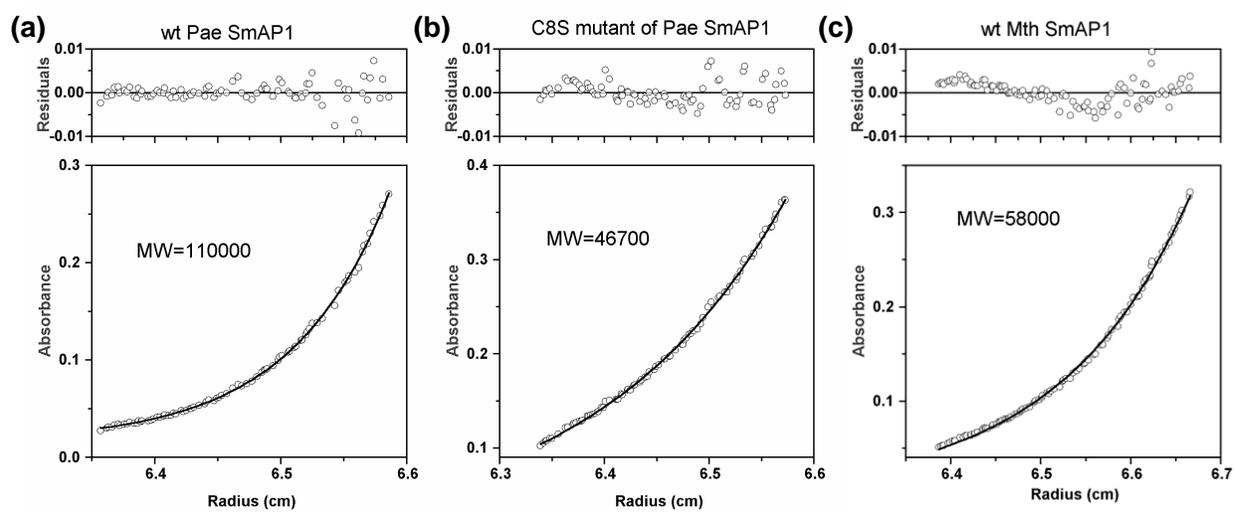

**Figure 3.3: Polymerization of SmAP1s into polar fibers.**

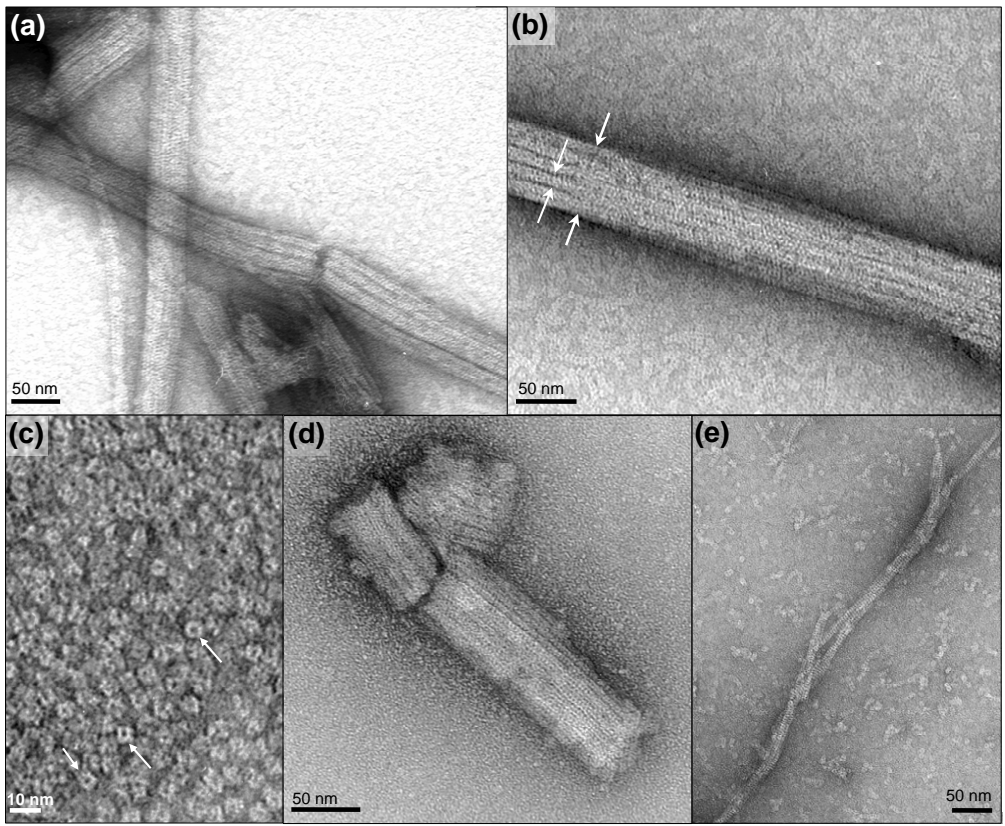

**Figure 3.4: Various crystalline oligomers of *Pae* and *Mth* SmAP1.**

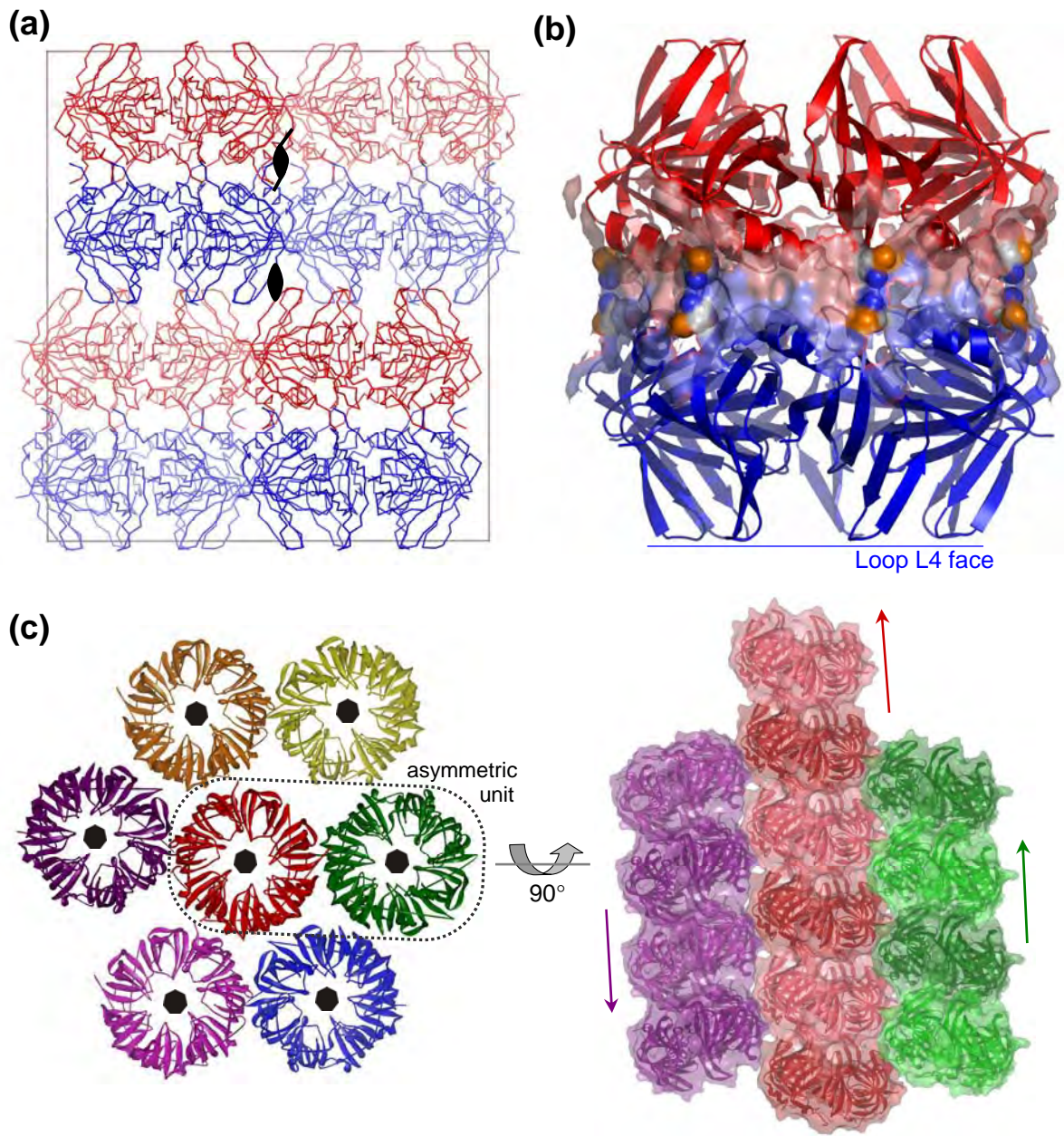

**Figure 3.5: Ligand-binding sites in the structure of the *Mth* 14-mer bound to UMP.**

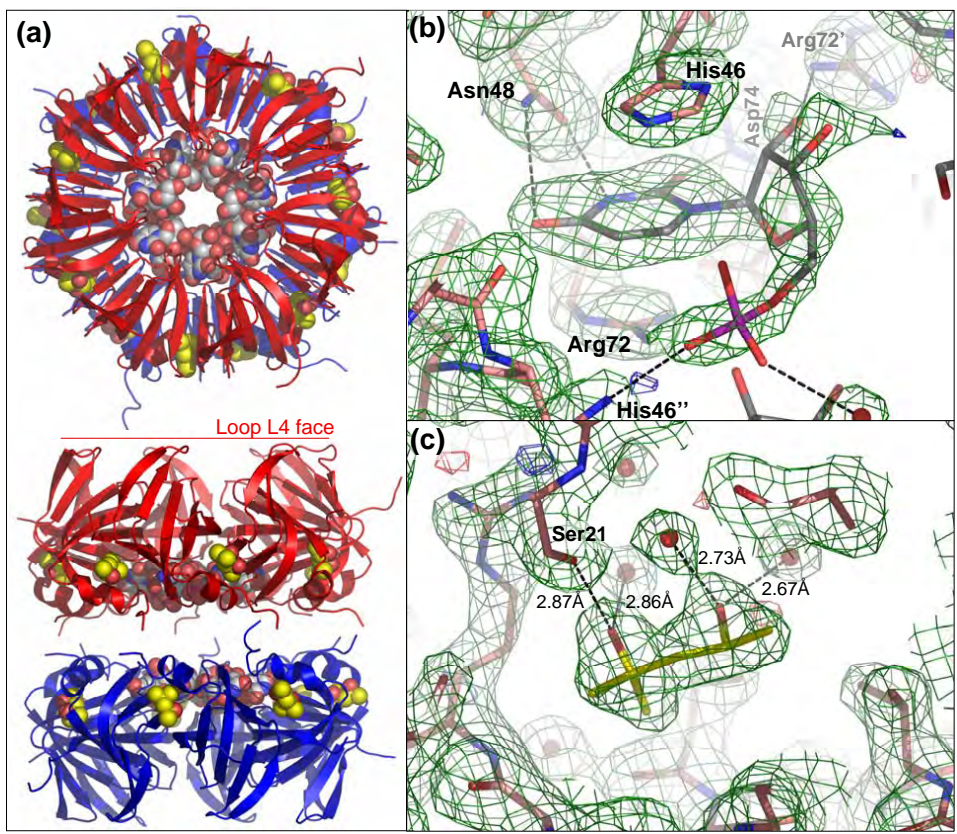

**Figure 3.6: Ligand-binding sites in the structure of the *Pae* 14-mer bound to UMP.**

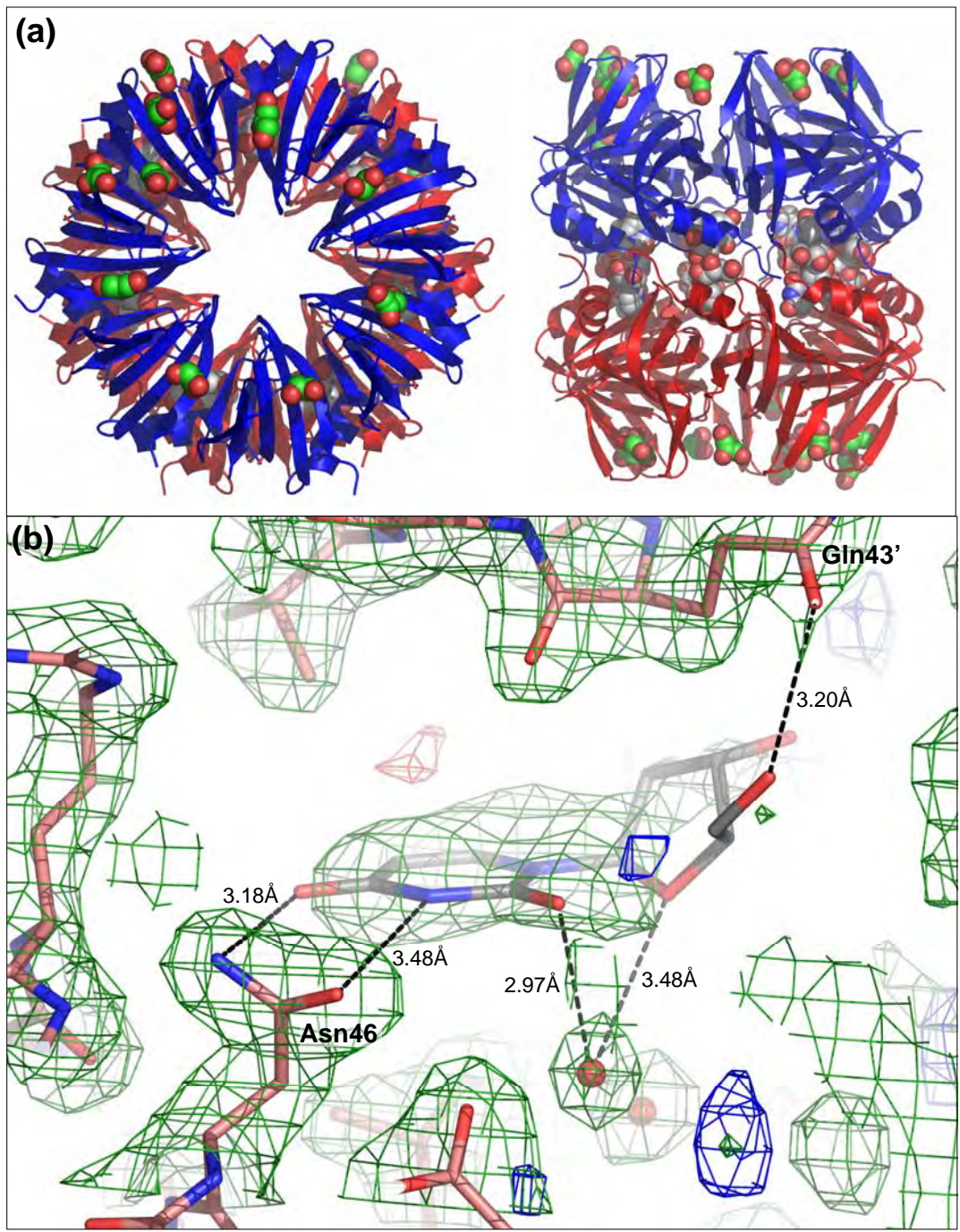

**Figure 3.7: Conserved mode of uridine recognition by *Mth* and *Afu* SmAP1.**

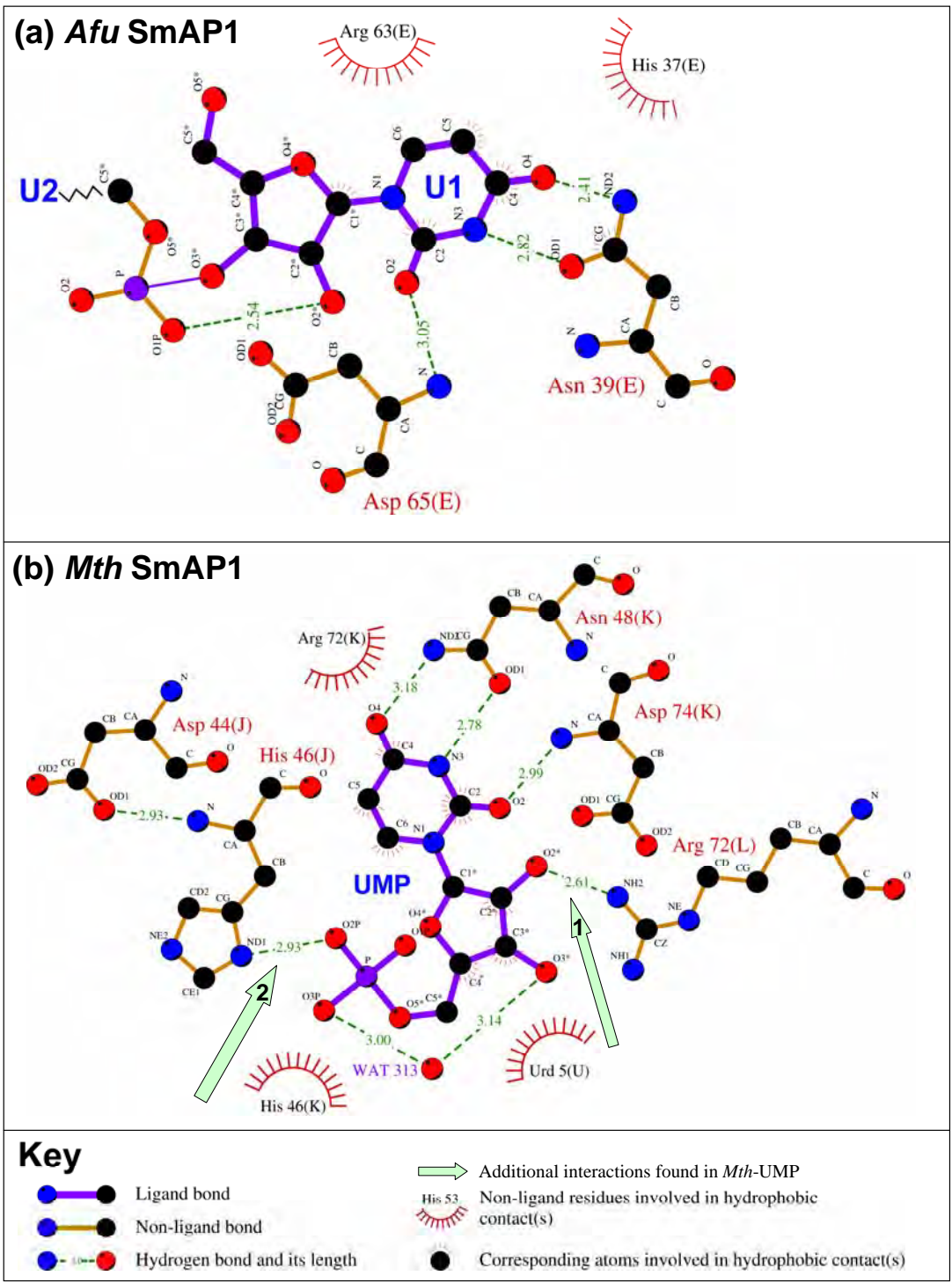

**Figure 3.8: Gel-shift of supercoiled DNA by *Mth* and *Pae* SmAP1.**

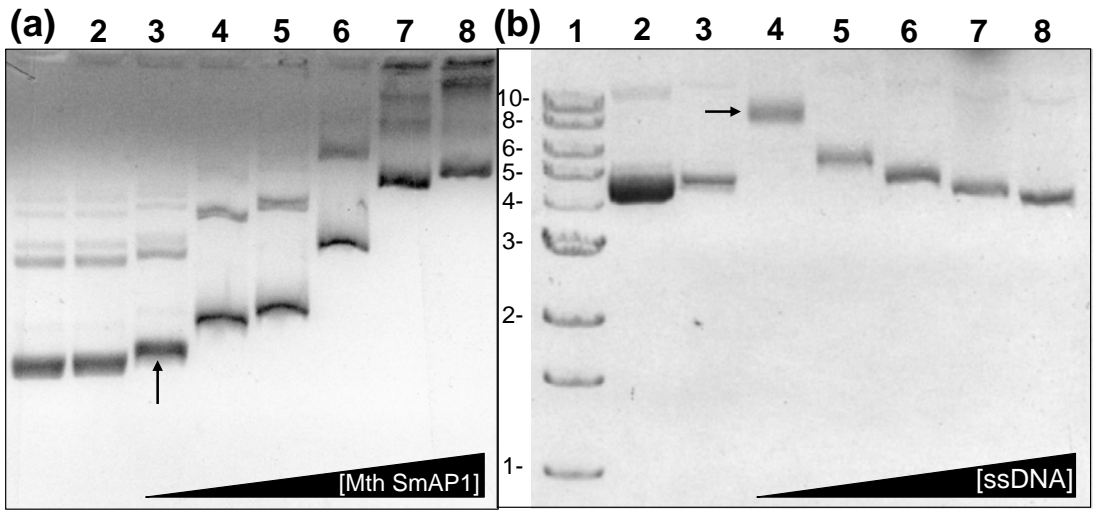

| lane | 1 | 2 | 3 | 4 | 5 | 6 | 7 | 8 |
|---|---|---|---|---|---|---|---|---|
| p5L1c1 | + | + | + | + | + | + | + | + |
| *wt Mth* | – | – | | | | | | |

| lane | 1 | 2 | 3 | 4 | 5 | 6 | 7 | 8 |
|---|---|---|---|---|---|---|---|---|
| p9L2c1 | – | + | + | + | + | + | + | + |
| *wt Pae* | – | – | – | + | + | + | + | + |
| ssDNA | – | – | – | – | | | | |

**Chapter 4:**

**The structure and potential function of an archaeal homolog of**

**survival protein E (SurEα)**

# Abstract

The survival protein E (SurE) family was discovered by its correlation to stationary phase survival of *Escherichia coli* and various repair proteins involved in creating this and other stress-response phenotypes. In order to better understand the structures and functions of this ancient and well-conserved protein family, we have determined the 2.0 Å-resolution crystal structure of SurEα from the hyperthermophilic crenarchaeon *Pyrobaculum aerophilum* (*Pae*). This first structure of an archaeal SurE reveals significant similarities and differences to the only other known SurE structure, that from the eubacterium *Thermatoga maritima* (*Tma*). Both SurE monomers adopt similar folds; however, unlike the *Tma* SurE dimer, crystalline *Pae* SurEα is predominantly non-domain swapped. Comparative structural analyses of *Tma* and *Pae* SurE monomers suggest conformationally variant regions, such as a hinge loop that may be involved in domain swapping. The putative SurE active site is highly conserved, and implies a model for SurE bound to one of its substrates, Guanosine-5'-monophosphate (GMP). Analyses of the sequences, phylogenetic distribution, and genomic organization of the SurE family reveal examples of genomes encoding multiple *surE* genes, and suggest that SurE homologs may constitute a broad family of enzymes with phosphatase-like activities.

---

## Introduction

The survival protein E (SurE) family was discovered nearly ten years ago by Clarke and colleagues by its correlation to stationary phase survival of *E. coli* and various repair proteins thought to be involved in creating this stress-response phenotype, *e.g.*, protein-L-isoaspartate(D-aspartate)-*O*-methyltransferase (pcm, EC 2.1.1.77).[1] The *E. coli surE* gene lies immediately upstream of the *pcm* gene, overlapping it by four nucleotides; together, these genes are thought to form a bicistronic operon that is essential for *E. coli* viability under stressful conditions, such as elevated temperatures, osmotic stress, or high cell density. The *surE* and *pcm* genes are co-transcribed as detected by *in vitro* transcription assays,[1] although each gene may also be transcribed independently from its own promoter.[2] Several bacteria contain an additional conserved gene of unknown function (*ORF0*) directly upstream of *surE*. Taken together, these several genes are thought to cluster into a stationary phase stress-survival operon, *surE-pcm-nlpD-rpoS*, where *nlpD* is an outer-membrane lipoprotein gene and *rpoS* encodes an alternative RNA polymerase σ factor (σ$^s$) that plays a regulatory role by inducing the transcription of several other stationary phase survival genes.[3]

The results of Visick *et al*. (1998) showed that both the *surE* and *pcm* genes are ancient and well-conserved, with orthologous genes being found in several eubacterial and archaeal species.[4] The only prokaryotes in which a *surE* gene is not found are gram-positive bacteria and mycobacteria. The phylogenetic distribution of *surE* genes is apparently more extensive than was initially thought, with SurE homologs having been found in eukaryotes ranging from simple protozoa (*e.g.*, the yeast *Saccharomyces* *cerevisiae*) to metazoa (*e.g., Arabidopsis thaliana*). The results reported here emphasize the distribution and genomic organization of *surE* genes in the archaea.

The increased accumulation of isoaspartyl damage and diminished viability of stationary phase *E. coli* that have various combinations of *surE* and *pcm* mutations further supports the idea that these two proteins interact either directly or indirectly (or in parallel pathways) to provide a stress-survival phenotype: *pcm/surE* double mutants accumulated much higher levels of isoaspartyl residues than did the parent strain or either single mutant, and a *surE* null mutation was able to suppress stress-survival defects in a *pcm* mutant strain.[4] Recently, it was shown that the stress-survival operon noted above (*surE•••rpoS*) was duplicated by several *E. coli* strains that were evolved over 2000 generations at high temperatures.[5] Moreover, the few *E. coli* strains which adapted to high temperatures without duplicating this *surE*-including region did display elevated expression from their single *surE* gene. A conclusion of these results is that *surE* plays a significant physiological role in stress-response.

The earliest hint about the biochemical function of SurE came from genetic experiments with a protein from the yeast *Yarrowia lipolytica*. This *Y. lipolytica* protein (P30887, or *PHO2*) bears weak sequence similarity to the N-terminal domain of the SurE family, and was found to complement mutations in two of the major acid phosphatases of *S. cerevisiae*.[6] Because of its lack of sequence similarity to known phosphatases (or any other functionally characterized protein), *PHO2* was described as a novel acid phosphatase. Most recently, the crystallographic and biochemical work of two groups has illuminated both the structure and function of SurE in greater detail. Lee *et al.* (2001) and Zhang *et al*. (2001) independently determined the crystal structure of a SurE homolog from the eubacterium *Thermatoga maritima* (*Tma*).[7,8] Their primary findings were that: (i) the *Tma* SurE monomer consists of an N-terminal globular domain of ≈ 180 residues that resembles a Rossmann fold and a novel, extended C-terminal region of ≈ 70 residues; (ii) monomers assemble into dimers (and possibly tetramers) with extended C-terminal α-helices that domain swap; (iii) *Tma* SurE exhibits a divalent cation-dependent acid phosphatase activity that is inhibited by vanadate or tungstate; (iv) divalent metal ions bind in a putative conserved active site; and (v) the *Tma* enzyme shows no protease or nuclease activity, but has a slight preference for Guanosine-5'-monophosphate (GMP) or Adenosine-5'-monophosphate substrates. These results suggested for the first time that SurE may be a novel acid phosphatase.

In order to better understand the structures and functions of members of this ancient and well-conserved protein family, we determined the 2.0 Å-resolution crystal structure of SurEα from the hyperthermophilic crenarchaeon *P. aerophilum* (*Pae*, $T_{max}$ = 104°C).[9] Comparison to the eubacterial *Tma* SurE structure reveals several significant similarities and differences between these SurEs; such comparison are justified because the *Pae* structure was determined independently of *Tma*, by multi-wavelength anomalous dispersion (MAD) phasing. An analysis of the phylogenetic distribution and genomic organization of *surE* genes is also provided, and these results are discussed in terms of the SurEα structure and its likely biochemical function.

## Materials and Methods

### *Cloning, expression, and purification of* **Pae** *SurEα*

A genomic phosmid clone that contained the *Pae* SurEα open reading frame was kindly provided by the laboratory of Jeffrey H. Miller (UCLA). Using its DNA sequence, we designed the following primers (Invitrogen/Life Technologies, Inc.) for polymerase chain reaction (PCR) amplification with Deep Vent$_R$ polymerase (New England Biolabs):

```
5'GGCCATATGAAGATCTTGGTCACTAATG 3'  (sense)
5'AAGCTTTGATAGCGACGCGTTAATGTAC 3'  (antisense)
```

Blunt-end PCR products were cloned into a pET-22b(+) expression vector (Novagen) *via* intermediate subcloning into the pCR-Blunt vector (Invitrogen). DNA sequencing (Davis sequencing, Inc.) of transformants into chemically-competent NovaBlue *E. coli* (Novagen) verified that the cloned protein would be identical to wild-type (*wt*) SurEα, except for the addition of the following 14-residue His-tag to the C-terminus: `SKLAAALEHHHHHH`. Due to rare usage of the arginine codons AGG and AGA in *E. coli*, *Pae* SurEα was over-expressed only after co-transformation with a tRNA$^{Arg}$-encoding vector (see the rare codon calculator at http://www.doe-mbi.ucla.edu/cgi/cam/racc.html). Otherwise, the recombinant protein was over-expressed in BL21(DE3) *E. coli* at 37°C by standard protocols using 1 mM isopropyl-β-D-thiogalactoside (IPTG) induction of the T7*lac*-based promoter. Approximately 10 mg of soluble protein were expressed per liter of cell culture. Selenomethionine (SeMet)-substituted SurEα was prepared for MAD phasing in exactly the same way as the wild-type protein, except that the expression was performed in M9 minimal media supplemented with SeMet (as described in ref. [10]).

Harvested cells (stored at –20˙C overnight) were thawed and re-suspended in a high-salt concentration buffer (20 mM NaHEPES pH 7.8, 1.5 M NaCl, 0.5% v/v Triton X100, 30 mM PMSF). Cells were lysed by a combination of lysozyme treatment (0.3 mg/ml chicken egg white lysozyme) and French-press (1000 psig). Initial protein purification was achieved by heating the cleared supernatant to ≈ 80˙C (>85% purity as estimated by density scans of SDS-PAGE lanes), followed by high-speed centrifugation to remove the bulk of denatured *E. coli* proteins. The SurEα-His6x was further purified by chromatography on a $Ni^{2+}$-charged iminodiacetic acid-sepharose column (Pharmacia); the protein was eluted in an imidazole gradient, and it may be significant that yellow-colored fractions from an earlier point in the gradient reproducibly contained a single protein of ≈ 20 kDa. Affinity chromatography resulted in >99% pure protein as estimated by several independent techniques (SDS-PAGE, MALDI-TOF and electrospray mass spectrometries, and gel-filtration chromatography). These methods also verified that the final, purified protein consists of full-length wild-type SurEα with the appended 14-residue His-tag, and mass spectrometry was used to verify the incorporation of SeMet. After affinity chromatography, all attempts to exchange the protein into a buffer incapable of chelating divalent metal ions (*e.g.*, any buffer lacking imidazole or EDTA) were unsuccessful – the protein would invariably precipitate out of solution, presumably due to His-tag mediated polymerization in the presence of divalent cations. Therefore, for crystallization efforts SeMet-labelled *Pae* SurEα was exchanged by dialysis into the following buffer: 10 mM Tris-Cl, 5 mM EDTA pH 8.0.

***Crystallization and x-ray data collection***

SeMet-labelled SurEα was concentrated to ≈ 11 mg/ml in Amicon ultrafiltration devices (10 kDa molecular weight cutoff), and several initial crystallization leads with different habits were obtained by hanging-drop vapor diffusion and sparse matrix screening (Hampton Research; Emerald BioSystems). Initial leads had to be extensively optimized by systematic variation of crystallization parameters, particularly protein and precipitant (PEG) concentrations. In terms of a crystallization response surface,[11] the most critical parameter was the sampling of a different region of crystallization space – in this case, *via* the addition of a reductant (dithiothreitol, DTT). The efficacy of adding 10 mM DTT to the crystallization condition is easily rationalized, because biophysical characterization of *Pae* SurEα shows that it forms disulfide-bonded dimers *in vitro* (unpublished data, Mura and Eisenberg). The final, optimized crystallization condition for the trigonal SurEα crystals grown in hanging-drops at 19.8°C follows: 5 μL drops (2.5 μL well + 2.5 μL 11.4 mg/ml SeMet SurEα) over 600 μL wells (0.083 M Tris pH 8.55, 21.7% v/v PEG-4000, 0.17 M NaOAc, 15% v/v glycerol). A single SeMet crystal of reasonable diffraction quality appeared within one year, and grew as trigonal prisms of maximum size ≈ 0.2 mm x 0.4 mm (crystals of the native protein were never obtained).

Initial auto-indexing of diffraction patterns with DENZO revealed that the SeMet crystal was of a hexagonal Bravais lattice, and further scaling of data revealed the space group to be either $P3_121$ or $P3_221$, with unit cell dimensions $a$ = 90.5 Å, $c$ = 129.95 Å. The molecular mass of *Pae* SurEα-His6x (30,733.1 Da), together with these cell dimensions, suggested a dimer in the asymmetric unit. For $Z$ = 12 monomer/cell, the calculated Matthew's coefficient ($V_M = 2.50$ Å$^3$/Da) corresponds to a solvent content of 50.8% by volume.

After initial screening of crystals on a RAXIS-IV++ image plate detector (UCLA), final SeMet MAD data sets were collected on an ADSC Quantum-4 charge-coupled device (CCD) detector at ALS beamline 5.0.2. A crystal was transferred directly from the hanging drop in which it grew to a cryocane stored in liquid nitrogen. Data were collected on this crystal in a cryogenic nitrogen stream at –168°C (105 K). All images were indexed/integrated/reduced in DENZO, and reflections were scaled and merged in SCALEPACK.[12] X-ray fluorescence scans about the K absorption edge of the SeMet crystal (≈12.6578 keV) were used to select appropriate wavelengths for *inflection*, *peak*, and *high-energy remote* data sets. Complete data sets were collected from the single SeMet crystal (Table 4.1).

***Crystallographic MAD phasing and initial refinement***

There were two indications that the SeMet crystal would be suitable for multi-wavelength anomalous dispersion (MAD) phasing: (i) $\chi^2$ values > 1 for the merging of I$^+$ and I$^-$ reflections indicated the presence of a reasonable anomalous signal (Table 4.1), and (ii) there were large anomalous difference ($\Delta F_{ano}^2$) Patterson peaks for the data set collected at the selenium *peak* wavelength. The integrated Patterson and direct methods program SHELXD (http://shelx.uni-ac.gwdg.de/SHELX/) was used to locate 8 Se sites per asymmetric unit (out of an expected 12 sites), using a single-wavelength anomalous scattering approach with the *peak* data set. The sites were verified by comparing their predicted self- and cross-vectors with observed peaks in the anomalous difference

Patterson maps. The program MLPHARE[13] was used for maximum likelihood heavy atom and MAD phase refinement to 2.8 Å (that being the high resolution limit of the *high-energy remote* data set); the *inflection point* data set was treated as a native data set for refinement of Se occupancies, as it has the smallest dispersive scattering component of the three wavelengths used ($f' = -10.1e$). Phases for the centrosymmetric solution were also calculated and refined – *i.e.*, the inverted hand of the Se positions in the enantiomorphic space group ($P3_221$).

Next, density modification using solvent flattening and histogram matching was performed with the program DM[14] in order to distinguish the correct enantiomorph of Se sites and to improve electron density map quality. Maps calculated from these experimental phases were of excellent quality (Fig. 4.1), with protein secondary structure elements clearly identifiable. Phases were extended from 2.8 to 2.0 Å with DM (including 2-fold NCS averaging). Rigid secondary structure elements were initially fit into 2.5 Å-resolution maps automatically with the program MAID[15], and this served as a useful starting point for automated model building of $\approx 87\%$ of the protein backbone (485 out of 560 residues/dimer in 13 chains) with the program ARP/wARP.[16]

Manual model building was done with the program O[17], and the program CNS[18] was used for model refinement. Refinement in CNS proceeded by standard protocols, using the maximum-likelihood target function for amplitudes (mlf), bulk solvent correction, and anisotropic B-factor correction terms. Initially, the two monomers in the asymmetric unit were refined with only weak NCS restraints imposed, and for the final rounds of refinement the two monomers were refined independently. Solvent molecules (water, glycerol, and acetate) were added as necessary. Refinement of individual atomic positions, isotropic temperature factors, and simulated annealing torsion angle dynamics was performed in most rounds. Also, the occupancies of the Se atoms were refined. Each refinement round ended with inspection of the agreement between the model and $\sigma_A$-weighted $2F_o - F_c$, $F_o - F_c$, and simulated annealing omit maps (the latter only as necessary).

### *Final occupancy refinement and model validation*

For the purpose of distinguishing the domain swapped (DS) from the non-domain swapped (non-DS) conformation of the hinge loop region (residues ≈ 244-248), the occupancies of atoms in these residues were refined as groups in the program CNS. Calculation of $2F_o - F_c$, and $F_o - F_c$ maps from models that used these refined occupancies, together with the values of these occupancies, suggested that the crystal consists of a mixture of DS and non-DS states (see Fig. 4.5 and the *Results* section). Further efforts led to a final, refined model consisting exclusively of the non-DS conformer, with occupancies for the hinge loop residues (and all other atoms except for selenium) set to one. This final model contains 276/280 residues for one monomer, and 278/280 for the other. A total of 287 water molecules, 7 glycerols, and 2 acetates were modeled as solvent. Final $R/R_{free}$ values are 18.5%/22.3%, with reasonable root mean square deviations (RMSDs, Table 4.1). The programs ERRAT[19], PROCHECK[20], and Verify3D[21] were used in model validation. Experimental structure factors and the refined *Pae* SurEα structure have been submitted to the PDB (code 1L5X).

### *Sequence and structure analyses*

Homologs of *Pae* SurEα were found *via* iterative PSI-BLAST searches of the most current non-redundant database of deposited protein sequences at NCBI. This final list of 43 SurE homologs is shown in Table 4.3. Multiple sequence alignments over the entire list, as well as just the seven archaeal sequences, were performed with CLUSTALW.[22] Pairwise sequence similarity scores were calculated by the Smith-Waterman algorithm, as implemented in the GCG software package.[23] An unrooted phylogenetic tree for all 43 SurEs was inferred from the distance matrix methods in the Phylogeny inference program PHYLIP.[24]

Structural alignments were created with various programs as necessary. For example, active site regions (which are similar in structure) were aligned with the Kabsch least squares method in the program ALIGN (Cohen, unpublished result cited in ref. [25]), whereas entire monomeric or dimeric *Pae* and *Tma* structures (which are more dissimilar) were optimally aligned by the combinatorial extension algorithm.[26] Calculations of the electrostatic potentials at surfaces were performed in GRASP,[27] and buried surface areas were calculated by the Lee & Richards approach[28] as implemented in CNS. Comparative structural analyses of several SurE models was performed *via* error-scaled difference distance matrices[29] in the program ESCET.[30] To this end, the ESCET analysis was performed twice: (i) using an ensemble of the 8 crystallographically independent *Tma* SurE models refined by Lee *et al*[7] and Zhang *et al*[8] or (ii) using a single *Tma/Pae* pair of structures (*e.g*., a single *Tma* monomer from PDB code 1J9J and a single *Pae* monomer). In the latter case, the ESCET analysis was restricted to portions of the two chains that aligned in 3D (as determined by combinatorial extension). The programs GRASP and PyMOL[31] were used for electron density figures and other structural illustrations.

## Results

### *Structure determination, refinement, and validation*

Since molecular replacement efforts with *Tma* SurE failed, the *Pae* SurEα crystal structure was determined by MAD phasing of data that were collected from a single crystal (Table 4.1). As described in the *Methods* section, native crystals were never obtained, so the final model was refined against the best data from the SeMet-substituted crystal. Despite the poor phasing statistics (Table 4.1), electron density maps calculated with experimental MAD phases were of excellent quality (Fig. 4.1). The high-resolution limit of the data (2.0 Å) at the *peak* wavelength permitted automatic model building for much of the structure by successive steps of secondary structure fragment matching (MAID) and free-atom model refinement (wARP). No non-crystallographic symmetry restraints were imposed on the SurEα dimer after the first few rounds of model refinement. The final model was refined to an $R/R_{free}$ of 18.5% / 22.3%, with reasonable geometry (Table 4.1).

As discussed in further detail below, crystalline *Pae* SurEα is a mixture of domain swapped (DS) and non-domain swapped (non-DS) states. Residues ≈ 244-248 in *Pae* SurEα correspond to the putative hinge loop region in the DS *Tma* SurE structures.[7,8] Electron density for this region was poorer than in any other part of *Pae* SurEα. Because of the ambiguity in building these residues, and to substantiate a non-DS model for the SurEα dimer, the final *Pae* structure underwent extensive model validation with ERRAT, PROCHECK, and Verify3D. Results from some of these verification methods are shown in Fig. 4.1*a* and *b*, along with a representative example of the agreement between the final model and experimental (Fig. 4.1$c$) and $2F_o - F_c$ electron densities (Fig. 4.1$d$). The Ramachandran plot shows a single residue (Ser99) in a strictly disallowed region of $\phi$, $\psi$ space (Fig. 4.1$a$); however, the electron densities shown in Fig. 4.1 illustrate that this residue is modeled correctly.

### *The topology and fold of SurEs*

Like *Tma* SurE, *Pae* SurEα is a Rossmann-like fold with an extended C-terminal domain; the topology of this mixed α/β protein is shown schematically in Figs. 2 and 3. The N-terminal core domain of ≈ 170 residues adopts a Rossmann-like fold, and the C-terminal region of ≈ 90 residues forms an irregular structure that is dominated by a 40-residue β-hairpin. This hairpin protrudes from the body of the protein, and mediates possible tetramerization of both *Pae* and *Tma* SurE (discussed below). Despite its Rossmann fold, no structures in the PDB are significantly similar to *Pae* (or *Tma*) SurEα. Three homology searches were performed with the DALI program: using the entire SurEα monomer, using the highly conserved N-terminal domain alone, or using the C-terminal region alone. The closest match was the Rossmann fold of phosphofructokinase against the N-terminal domain of SurEα, but this has a *Z*-score of only 6. Similar results were found by Lee *et al.* and Zhang *et al.* for *Tma* SurE.[7,8] As has been observed for other nucleotide-binding proteins that utilize Rossmann folds, the most highly conserved residues in the SurE family (Fig. 4.2) map to the C-terminal loops of the β-strands that form the core of this fold (Fig. 4.3). Two of the largest structural differences between *Tma* and *Pae* SurEα are also indicated in this figure: (i) the more extended β-sheet core of *Tma* SurE contains 7 strands rather than 5, and (ii) the C-terminal α-helices exchange in *Tma* to form a domain swapped dimer.

### *Comparative structural analysis of SurE monomers*

In order to dissect more quantitatively the structural differences between the *Tma* and *Pae* structures, we utilized error-scaled difference distance matrices (DDMs). This is a recent structure comparison approach that explicitly takes into account the crystallographic data for the two models under comparison (*e.g*, resolution, *B*-factors, $R_{free}$) *via* a diffraction precision index.[29] The output from a pairwise comparison of models is a DDM, which is a symmetric matrix with the common core of the two models represented by the rows and columns. The entries of this matrix are the statistical signficance of the deviation from the mean of the two structures (expressed as a standard deviation). Recently, a genetic algorithm has been devised to allow simultaneous comparison of all pairwise DDMs of an ensemble of structures.[30] Two applications of this method were made for SurE: (i) the ensemble of 8 crystallographically-independent *Tma* SurE models was analyzed (two monomers in the asymmetric unit of each of the PDB entries 1J9J, 1J9K, 1J9L, 1ILV) and (ii) various *Tma-Pae* SurEα pairs were compared.

The results of this analysis, along with a standard 3D structure superimposition, are shown in Fig. 4.4. The superimposition in Fig. 4.4*a* shows that the N-terminal β-sheet core is highly conserved (1.1 Å RMSD over all atoms). Minor structural differences occur in the β-turn between strands B3 and B4 (green arrow) and the N-terminal region of the interrupted helix H4 (purple arrow). The largest differences occur in the C-terminal β-hairpin (orange arrow, Fig. 4.4*a*) and C-terminal α-helix, which is either swapped (*Tma*) or mostly non-swapped (*Pae*) depending on the hinge loop (red arrow, Fig.4.4*a, c*). For an ensemble of 8 structures, there are 28 unique DDMs. An example of one of these difference distance matrices as applied to the ensemble of 8 *Tma* SurEs is shown in Fig. 4.4*b*, and an example of the difference distance matrices between *Tma* and *Pae* SurEα is shown in Fig. 4.4*c*. In the latter figure, the lower diagonal represents the statistical significance of the scaled structural differences and the upper diagonal gives the actual difference in measured distances. Regions of relatively minor structural variation between *Pae* and *Tma* are apparent from the DDM analysis (see arrows in Fig. 4.4*c* and corresponding differences in Fig. 4.4*a*). For example, β-strand B4 hydrogen bonds to the adjacent strand (B2) in *Tma* to extend the β-sheet core, whereas in *Pae* SurEα this strand curves away from the sheet (see Fig. 4.3 and green arrows in Fig. 4.4*a, c*).

### Biochemical characterization of **Pae** *SurEα*

Biochemical characterization of the phosphatase activity of *Pae* SurEα is currently in progress. In preliminary results, SurEα exhibits acid phosphatase activity on some substrates, *e.g.*, *p*-nitrophenyl phosphate (unpublished data, Katz and Clarke). The SurEα activity is ≈ 20-fold less than that of the *Tma* SurE; however, the *Pae* SurEα was assayed in solution conditions that were determined to be optimal for the *Tma* enzyme, and which may not be optimal for the *Pae* enzyme (*e.g., Pae* SurEα apparently has a higher $T_{opt}$ than the *Tma* enzyme). In addition to determining the optimal condition for activity, future biochemical assays will characterize the phosphatase function of *Pae* SurEα with both DNA and RNA substrates (particularly rRNA and tRNA).

## Discussion

### *Significance of domain swapping in SurEs*

Throughout the crystallographic refinement, inspection of $2F_o - F_c$, $F_o - F_c$, and $F_o - F_c$ simulated annealing omit maps made it apparent that the crystal consists neither entirely of domain swapped (DS) SurEα dimers nor non-domain swapped dimers (non-DS). Evidence for such an inhomogeneous crystalline mixture of DS and non-DS states is two-fold: (i) neither conformation alone could be satisfactorily fit into the various maps mentioned above (Fig. 4.5) and (ii) refinement of occupancies for residues in the putative hinge loop region invariably led to values significantly less than 1 (but greater than 0.5). Attempts to refine alternate conformations of the hinge loop in a mixed DS/non-DS model provided little improvement over the non-DS model, suggesting that the majority of the crystal contains non-DS dimers. Also, the non-DS model agrees more closely with relatively unbiased $F_o - F_c$ simulated annealing omit maps (compare Figs. 5*c* and *d*). Parallel refinement of a DS and a non-DS model of *Pae* SurEα reinforced the conclusion that SurEα is (mostly) non-domain swapped (compare Figs. 5*a* and *b*). We note that stereochemically reasonable models of SurEα with no φ, ψ dihedral violations can be built for both the DS and non-DS conformations of the hinge loop region (data not shown).

Another example of a crystalline mixture of domain swapped states was recently reported by Zhang and coworkers for the 64-residue B1 domain of protein L.[32] However, in that case the non-DS protein is naturally monomeric, and the homogeneity of the mixture allowed both DS and non-DS states to be distinguished in the crystal. Both of these examples support the hypothesis that protein oligomers may have evolved from monomers by passing through a DS stage,[33] and the possibility of mixed DS/non-DS states in a single crystal seems plausible for other domain swapped oligomers.

A consideration of the various dimerization states of *Pae* and *Tma* SurE is required in order to decide if domain swapping is of general significance for SurE proteins. Because there is no evidence for an independently stable, closed monomer form of *Tma* SurE *in vitro*, the tertiary packing of the exchanged C-terminal α-helix leads us to classify the *Tma* dimer as a *candidate* for 3D domain swapping, adopting the nomenclature of Schlunegger *et al.*[34] The *Pae* SurEα structure reported here provides strong evidence for both non-DS dimers (composed of "closed" monomers) and DS dimers (composed of "open" monomers) in the same crystal. Although crystallographic evidence for the non-DS conformer is stronger, this mixture of both DS/non-DS states classifies the SurE proteins as a *bona fide* example of domain swapping. Apparently domain swapping is a feature of SurE proteins, but its function is not yet known.

***Similarities and differences between* Pae *and* Tma *SurE monomers***

Comparative structural analyses of *Pae* and *Tma* SurEs *via* error-scaled difference distance matrices suggest that the conformationally variable regions comprise much of the irregular C-terminal region, including the hinge loop that connects the swapped helix to the N-terminal core. The most conformationally invariant region of *Tma* SurE is the strongly conserved N-terminal ≈ 170-residue core (Fig. 4.2), and even within the *Tma* ensemble alone there are significant deviations in the β-hairpin and the hinge loop region that precedes the domain swapped α-helix. Whether such conformational variance is primarily due to high-amplitude fluctuations in certain dynamic regions of *Tma* SurE or discrete, slowly exchanging conformations is unclear. Note that we found similar conformationally invariant regions in the *Tma*-*Pae* comparison (Fig. 4.4*c*) and in the ensemble analysis of *Tma* alone (Fig. 4.4*b*). This consistency lends support to the interpretation of these results and to the generalization of these results to include other SurEs.

### Significant differences between **Pae** and **Tma** *SurE dimers and tetramers*

Several lines of evidence suggest that the biologically relevant oligomerization state of *Pae* and *Tma* SurEs is a dimer, and more tenuous evidence suggests a tetramer. Briefly, Lee *et al*. and Zhang *et al*. found an identical domain swapped dimer in the asymmetric unit of their *Tma* crystal structures, and detected dimers *in vitro* by various biophysical methods.[7,8] Zhang *et al*. further suggested that a *Tma* SurE tetramer exists on the basis of size exclusion chromatography and their crystal packing (the tetramer being generated by a crystallographic 2-fold). Our *in vitro* data for *Pae* SurEα reveal only a dimer, although we observe a crystalline SurEα tetramer that has the same overall structure and 222-point group symmetry as the *Tma* tetramer (data not shown).

The non-DS *Pae* SurEα dimer has a much less extensive dimer interface than the DS *Tma* SurE dimer, largely because of the non-swapped C-terminal helix. These interfaces are compared in Fig. 4.6, which shows the substantial difference ($\approx 1000$ Å$^2$) between the total buried surface area in the *Tma* (open) dimer interface ($6892 \pm 43$ Å$^2$) and the *Pae* (closed) dimer interface ($5835$ Å$^2$). Aside from the swapped α-helix, the two dimer interfaces are structurally similar; there are only slight rigid-body rotations of the monomers with respect to one another in the *Pae vs.* the *Tma* dimer (as evidenced by a 2.1 Å RMSD for alignment of dimers *vs.* 1.1 Å RMSD for alignment of monomers). Table 4.2 provides the buried surface area statistics for the *Pae* and *Tma* SurE dimers as well as tetramers. Notice that much more surface area is buried in the crystallographic *Pae* tetramer (3716 Å$^2$) than in either *Tma* tetramer (2215 ± 208 Å$^2$). The structural basis for this disparity is that a slightly different conformation of the extended β-hairpins in the *Pae* structure (Fig. 4.4) allows a much closer approach and more extensive contacts between the two SurEα dimers than is the case with the *Tma* dimers. Because the two dimers are related by a crystallographic 2-fold, this difference (and, in fact, the entire tetramer) may be an artefact of crystal packing.

Two results support the significance of a SurE tetramer: (i) the tetramer interface is extensive and well-defined (*e.g.*, 3716 Å$^2$ for *Pae*) and (ii) *Pae* and *Tma* SurE form identical tetramers in three independent crystallization efforts (*i.e.*, that of Lee *et al.*,[7] Zhang *et al.*,[8] and this report). However, two data that refute the importance of a tetramer are: (i) no *Pae* SurEα tetramer is found by size exclusion chromatography (unpublished data, Mura and Eisenberg), and (ii) in the tetramer, the conserved active site is largely occluded by the C-terminal β-hairpins of the second dimer (data not shown). Of special interest will be any connection between these large discrepancies in dimerization and tetramerization behavior and the biochemical functions of *Pae*, *Tma*, and other SurEs.

***The conserved active site and a substrate-binding model***

The putative active site is a highly conserved, acidic cleft on the surface of both *Tma* and *Pae* SurEs (Fig. 4.7). Structural and biochemical characterization of *Tma* SurE showed it to have acid phosphatase activity, and allowed a putative active site to be identified based on the crystallographically-located divalent metal ion binding sites.[7,8] The most striking attribute of the electrostatic potential of both *Tma* and *Pae* SurEα surfaces in Fig. 4.7*a* and *b* is also the most conserved feature (Fig. 4.7*c*): there is a relatively large acidic cleft that forms the putative SurE active site and continues as a narrow anionic channel in *Tma* SurE. An interesting feature that is not shared between *Pae* and *Tma* SurE is the weak – but significant – amount of basal level anionic charge that covers most of the *Tma* SurE surface (Fig. 4.7*b*). This property may be related to the fact that the *Tma* protein is an *acid* phosphatase: most of the protein surface would be neutralized only at the acidic pHs in which *Tma* SurE is optimally functional.

In order to gain further insight into substrate binding and the potential functions of SurEs, we created a model of guanosine-5'-monophosphate (GMP) bound to the conserved *Pae* SurEα active site. GMP was chosen as a ligand because it was found to be the best substrate among the phosphate-containing compounds tested by Zhang *et al.*[8] and Lee *et al.*[7] In the complex shown in Fig. 4.8, the GMP was docked so that its phosphate moiety coincided with the binding site of the inhibitory tungstate described by Lee *et al.*[7] In addition, several other favorable protein-GMP interactions are formed in this complex, *e.g.*, the aromatic stacking interaction between the guanine ring and the highly conserved Tyr192 side chain.

In addition to showing that GMP can be easily accommodated in the *Pae* SurEα active site, the hypothetical SurE•GMP complex elucidates the importance of many conserved active site residues, and suggests that the catalytic mechanism may rely on nucleophilc attack by a polarized water. The *Tma* SurE protein binds a divalent metal ion ($Mg^{2+}$ or $Ca^{2+}$) in the active site such that the metal is chelated by several conserved residues, including the strictly conserved Asp8/Asp9 pair. There is also a water molecule that is tightly bound 2.2 Å away from $Mg^{2+}$ in the structure of Lee *et al.*[7], and this polarized water may serve as a nucleophile for attack on the substrate's phosphate center. As *Pae* SurEα could be crystallized only in the presence of EDTA, no divalent metal ions were found at the active site. However, waters were found in the two sites just described: one water occupies the same divalent metal ion binding site as in *Tma*, and the other water is hydrogen bonded to this one. Both of these waters are compatible with the location of the modeled GMP in Fig. 4.8.

### *Phylogenetic distribution of* **surE** *genes*

The only SurE structure known prior to the archaeal *Pae* SurEα structure reported here is that from the thermophilic eubacterium *Tma*. In order to better understand the phylogenetic distribution and possible evolution of other SurEs, we compiled a comprehensive list of all open reading frames (ORFs) with sequence similarity to *Pae* SurEα. These 43 putative SurE proteins are shown in Table 4.3, and were found *via* iterative PSI-BLAST searches of a non-redundant database at NCBI. The SurE database compiled here consists of 32 eubacterial sequences, 4 eukaryotic sequences, and 7 archaeal proteins, represented by a total of 39 species (including the extremely radioresistant eubacterium *Deinococcus radiodurans*). We found examples of organisms with more than one *surE* gene: the genomes of *Synechocystis sp. PCC 6803*, *Nostoc sp. PCC 7120*, *Arabidopsis thaliana*, and *P. aerophilum* each contain pairs of SurE paralogs, which we designate SurEα and SurEβ (Table 4.3 and Fig. 4.9*a*). Our finding of two *surE* genes in the hyperthermophile *Pae* (which grows up to 104°C) is especially interesting given a recent report that the single *surE* gene in *E. coli* is duplicated in strains that are evolved for 2,000 generations at elevated temperatures.[5] The authors speculated that such duplication – along with the *pcm*, *rpoS*, and *nlpD* genes – may facilitate thermal adaptation in *E. coli*. However, duplication of *surE* genes is unlikely to be strongly correlated to extremophile survival, since (i) the three other organisms with duplicate SurE paralogs are mesophiles, and (ii) several thermophilic or halophilic species harbor only a single *surE* homolog (Table 4.3).

An unrooted phylogenetic tree was inferred by the application of distance matrix methods to multiple sequence alignments of the 43 surE sequences, and shows a large dispersion in the SurE lineages. That is, the tree displays very few bifurcated nodes, and a large fraction of the 43 SurE sequences cannot be grouped into clades. Interestingly, the unrooted tree also shows that the SurE sequences do not cluster by kingdom: archaeal SurEs are interspersed with eukaryotic and eubacterial ones in an apparently random way (Fig. 4.9*a*). The phylogenetic relationships of SurE paralog pairs – such as *Pae* SurEα/β or *Nostoc* SurEα/β – suggest that the second member of these SurE pairs may not have arisen by gene duplication and neutral drift within these genomes. If gene duplication led to two SurEs in these genomes, it is likely that the two paralogs would have a greater degree of sequence similarity and would share a stronger phylogenetic similarity than that shown in Fig. 4.9*a*. For example, the two SurEs from two subspecies of *H. pylori* are closely related, as are the *E. coli* and *S. enterica* SurEs; however, members of the four α/β pairs are apparently only distantly related. Thus, duplicate *surE* genes may have arisen by horizontal gene transfer.

### *Genomic organization of archaeal* **surE** *genes*

Phylogenetic analysis of the SurE family illuminates the *inter*-genomic distribution of *surE* genes. What about features of the *intra*-genomic organization of *surE* genes, especially in terms of the possibility that they cluster with archaeal homologs of other stress-survival genes? Also, what are the gene neighbors of *Pae* SurEα and SurEβ, and where do any *pcm*, *rpoS*, or *nlpD*-like genes lie in the archaeal genomes (*e.g.*, is there any operon-like clustering)?

These questions were addressed by: (i) looking at the ORFs encoded in all six reading frames upstream and downstream of all the archaeal *surE* genes (± 2,500 bp) and (ii) using sequence similarity searches to search for archaeal homologs of the *pcm*, *rpoS*, and *nlpD* genes. Notably, we found no strong sequence homologs of the *rpoS* or *nlpD* genes in *Pae* or any other archaea: *nlpD* homologs were found only in the eubacteria, and *rpoS* homologs could be found only in eubacteria and a few eukaryotes (primarily of the plant lineage *Viridiplantae*). Sequence searches revealed one significant *pcm* homolog in *Pae*. However, unlike the case in several eubacterial genomes, this *pcm* is not located near either *surE* gene in *Pae*, but is almost 0.5 Mbp away (Fig. 4.9*b*). The same result was found in other archaeal genomes: in each case the nearest gene neighbors of *surE* were not homologous to the *pcm*, *rpoS*, or *nlpD* genes. In fact, several of the ORFs adjacent to *surE* in archaeal genomes have no strong sequence matches to proteins of known function, and are annotated as *conserved hypothetical proteins*.

The genomic organization of – or even presence of – putative stress-survival genes is clearly not conserved in the eubacteria and archaea. However, in several cases a homolog of known function can be found for *surE* gene neighbors. For example, the nearest gene neighbor of *Pae surEα* encodes a putative purine NTPase (≈1500 nt upstream and in the same reading frame). A homolog of CTP-synthase lies ≈1200 nt upstream of – and in the same reading frame as – *Pae surEβ* (Fig. 4.9*b*). Three other examples of archaeal gene neighbors are: (i) a putative protein tyrosine phosphatase (PTP) upstream of *M. thermautotrophicum surE*; (ii) a homolog of ribose-5'-phosphate isomerase downstream of (and overlapping) the *A. fulgidus surE* gene; and (iii) an adjacent dihydropteroate synthase (DHPS) gene in *A. pernix* (encoded in a reverse reading frame). Some representative reactions catalyzed by these enzymes are shown in Fig. 4.10. In a reaction analogous to that catalyzed by DHPS, the nearest *Pae* SurEβ gene neighbor (CTP-synthase) mediates the condensation of UTP and an amino group (from either glutamine or ammonia) to form Cytosine-5'-triphosphate.

A final example of how the genomic organization of *surE* and its homologs may illuminate its function is the fact that the *S. cerevisiae* SurE ortholog is a large protein (>700 residues) that can be divided into two regions: an N-terminal half with sequence similarity to SurE, and a C-terminal domain of >300 residues that has significant sequence homology to tubulin-tyrosine ligase (TTL). TTL catalyzes the ATP-dependent post-translational addition of a tyrosine residue to the carboxy-terminus of α-tubulin, and is thought to be phosphorylated in its $Mg^{2+}$/ATP-binding domain (reviewed in ref. 35). Note that several of the reactions catalyzed by homologs of the gene neighbors of *surE* either directly or indirectly involve some form of phosphate ester hydrolysis. These results are consistent with the recent discovery of an acid phosphatase activity for *Tma* SurE, as well as our *Pae* SurEα structure.

## Conclusions

Until the recent reports of Lee *et al.*[7] and Zhang *et al.*[8] for *Tma* SurE, there was no structural or biochemical knowledge about the SurE family and its *in vivo* function. In order to extend and generalize their results, we determined the crystal structure of *Pae* SurEα to 2.0 Å-resolution. The *Pae* and *Tma* monomers adopt similar structures, consisting of N-terminal Rossmann-like folds and irregular, C-terminal domains that mediate oligomerization. Crystalline *Pae* SurEα differs from *Tma* SurE in that it forms an inhomogeneous mixture of domain swapped and non-domain swapped dimers, with the non-domain swapped form predominating. This shows that SurE proteins can exist in both monomeric and dimeric forms, and suggests that the transition could be of functional significance. More minor differences in the two structures were revealed by the application of error-scaled difference distance matrices. The considerable structural similarity of the SurE active sites allowed us to model the protein bound to substrate (GMP), and the SurE•GMP model elucidates the importance of conserved active site residues and suggests that phosphoester bond hydrolysis may proceed *via* nucleophilic attack of an active site water molecule. Finally, analyses of the phylogeny and genomic organization of SurE reveal examples of genomes with multiple *surE* genes and suggest a generic phosphatase-like function for other members of the SurE family.

**Figure Legends**

**Figure 4.1: Structure validation and sample electron density for *Pae* SurEα.** A Ramachandran plot for the refined model is shown in *(a)*, and Verify3D scores for each of the two monomers (A in red and B in blue) are plotted in *(b)*. Out of 560 residues per dimer, only 2 are disallowed (Ser99 in chains A, B) and an additional 6 are generously allowed. Similarly, most of the profile scores in *(b)* are significantly greater than zero; the poorest-scoring region (near residue 200) is near the C-terminal β-hairpin, and the positive scores for the putative hinge loop region (near residue 245) suggest that it is correctly modeled (in a non-DS conformation). Representative examples of electron density (contoured at +1.4σ) illustrate the close agreement between the map calculated from experimental MAD phases $\{F_o, \Phi_{MAD}\}$ *(c)* and that calculated from final model phases $\{2F_o - F_c, \Phi_{model}\}$ *(d)*. Together with the lack of significant positive or negative $F_o - F_c$ density in this region, these maps show that Ser99 – the only strong outlier in the Ramachandran plot – is modeled correctly.

**Figure 4.2: Sequence analysis of all archaeal SurEs.** A multiple sequence alignment is shown for all known archaeal SurE sequences. Clusters of conserved residues are shaded in black (stringent) or grey (less stringent), and the consensus sequence is given in the last line (conservative substitutions are italicized, and identities are capitalized). Numbering is for the *Pae* SurEα sequence. Regular secondary structure elements are shown as arrows and cylinders, and pleated lines indicate regions that can only loosely be classified as β-strands or turns. Secondary structures that form the irregular C-terminal region are shaded in a gray box (see Fig. 4.3). The strictest conservation occurs in the N-

terminal ≈ 150-residue core, which is also where all of the putative active site residues are located (underlined).

**Figure 4.3: *Pae* SurEα is a Rossmann-like fold with an extended C-terminal domain.**

A cartoon of the *Pae* SurEα monomer topology is shown, illustrating the N-terminal Rossmann fold and the extended C-terminal region (with gray-shaded background). The intensity of shading conveys the approximate 3D structure: lighter-colored elements are below the plane of the paper and darker elements are above it (asterisks denote positions that are near in 3D space). The most conserved residues in the SurE family (circles) map to the C-terminal loops of the β-strands which form the Rossmann fold. The primary differences between the *Pae* and *Tma* SurE structures (double-headed arrows) and the hinge loop region (dashed line) are indicated.

**Figure 4.4: Structural comparison of *Pae* and *Tma* SurE reveals conformationally invariant regions.** *(a)* A stereoview is shown for $C_{\alpha}$ traces of *Pae* SurEα (red, thick sticks) superimposed on the *Tma* SurE structure (blue). A red ball denotes every twentieth residue of the *Pae* monomer. Major structural differences are marked by colored arrows and are discussed in the text. These differences are quantified in *(b)* and *(c)* *via* error-scaled difference distance matrices. An ensemble of eight crystallographically-independent *Tma* SurE monomers is analyzed in *(b)*, and a pairwise comparison between matching fragments of one *Tma* and *Pae* SurEα monomer is shown in the lower diagonal matrix of *(c)*. The upper diagonal of *(c)* provides a normal matrix of RMSDs between these two models. Color intensity scales with either the statistical

significance of the model differences (*(b)* and lower diagonal of *(c)*) or the actual values of differences in distances between the atomic coordinates (*(c)*, upper diagonal).

**Figure 4.5: Crystalline *Pae* SurEα is predominantly not domain swapped.** Electron density is shown for the putative hinge loop region of two refined models: *(a)* and *(c)* assuming that the C-terminal α-helix is non-domain swapped (non-DS), or *(b)* and *(d)* assuming that it is domain swapped (DS). $F_o - F_c$ maps are colored red (–3.2σ) and blue (+3.2σ), and $2F_o - F_c$ density is colored green (+1.4σ). The path of the backbone (from N- to C-terminus) is indicated by arrows in *(a)* and *(b)*, and side chains are omitted in *(c)* and *(d)* for clarity. Monomers A and B are distinguished by yellow (A) or gray (B) coloring of carbon atoms, and by subscript letters for the labeled hinge loop residues Ala246 and His247. The positive (blue) and negative (red) $F_o - F_c$ densities for the hinge loop show that neither the non-DS *(a)* nor DS *(b)* model is perfectly accurate, although the non-DS model is better (there is less negative $F_o - F_c$ density for the mainchain atoms of Ala246 and His247 in panel *(a)* compared to *(b)*). The $F_o - F_c$ simulated annealing omit maps in panels *(c)* and *(d)* (+3.0σ) also show that the model of a non-DS dimer *(c)* fits this unbiased density better than the DS model *(d)*. These electron densities, along with details discussed in the text, suggest that crystalline *Pae* SurEα is an inhomogenous mixture of DS and non-DS states, with the non-DS form predominating.

**Figure 4.6: *Pae* SurEα has a much less extensive dimer interface than *Tma* SurE.** Orthogonal views of ribbon diagrams for the *Pae (a)* and *Tma (b)* SurE dimers are shown, with the monomeric subunits colored red and blue (*Pae*) or green and magenta (*Tma*). The total buried surface area in the *Tma* dimer interface is 6892 ± 43 Å$^2$, while in the *Pae* interface this value is 5835 Å$^2$. Comparison of panels *(a)* and *(b)* shows that this large difference is primarily due to swapping of the C-terminal α-helix in *Tma*, but not in *Pae* SurEα.

**Figure 4.7: The putative SurE active site is highly acidic and strongly conserved.**
The molecular surfaces of *Pae* SurEα *(a)* and *Tma* SurE *(b)* are displayed, colored by the calculated electrostatic potential (–10.7 kT (red) to +8.6 kT (blue) for *Pae*, and –12.6 kT to +9.7 kT for *Tma*). Both dimers are in roughly the same orientation, with the C-terminal β-hairpins pointing towards the left (indicated by green arrows). Two orientations of a space-filling model of the SurEα dimer are shown in *(c)*, viewed down the 2-fold NCS axis (this orientation is rotated roughly 90° with respect to panels *(a)* and *(b)*). Conserved residues are colored magenta, with the intensity of coloring reflecting the degree of conservation (presumed active site residues are labeled in red). The putative SurE active site is the highly acidic, concave surface seen conserved in both models.

**Figure 4.8: Hypothetical model of GMP bound to the conserved SurE active site.**
The substantial structural conservation of the putative *Tma* and *Pae* SurE active sites is shown in the stereoviews of *(a)* and more closely in *(b)*. Side chains for the two *Tma* models are shown in blue and cyan. *Pae* SurEα is colored pink and its active site residues are drawn as thicker sticks. Except for the Ser39 loop, the active site structures are nearly identical. Docking of a known *Tma* SurE substrate (GMP) results in a reasonable model in which the phosphate moiety is bound in an identical manner as the inhibitory vanadate (shown in green) found by Lee et al. (2001). The guanine ring stacks above the highly conserved Tyr192 side chain. Additional protein-GMP contacts and solvent molecules are not shown for clarity.

**Figure 4.9: Phylogenetic distribution and genomic organization of SurEs.** An unrooted phylogenetic tree is displayed in *(a)*, as calculated from a multiple sequence alignment of all 43 detectable SurE homologs. Eukaryotic SurEs are shown as zig-zagged lines, and archaeal species are dashed lines. The remaining SurEs are eubacterial. Also, paralogous SurE pairs (α, β) are italicized. Note the great dispersion in the SurE family amongst the eubacteria, eukaryotes, and archaea, and that paralogous *surE* genes do not cluster into clades. The genomic organization of *Pae surE* genes in shown in *(b)*. The circular, 2.2 Mbp genome of *Pae* is marked with the relative locations of its two *surE* genes: *surEα* at 1.73 Mbp and *surEβ* at 2.11 Mbp. The two nearest ORFs to *surE* are a purine NTPase homolog (~1500 bp upstream of *surEα*), and a CTP-synthase homolog (~1200 bp upstream of *surEβ*). These two ORFs are encoded in the same frame as their respective *surE* neighbors.

**Figure 4.10: Representative examples of reactions catalyzed by homologs of SurE gene neighbors.** The reactions catalyzed by homologs of *surE* gene neighbors from *Pae* (NTPase), *M. thermautotrophicum* (PTP), and *A. pernix* (DHPS) are shown in panels *(a)*, *(b)*, and *(c)*, respectively (see text for details). Although these reactions are not closely related, each enzyme is clearly involved in some form of phosphate hydrolysis (gray shaded regions). These chemical activities are consistent with a possible phosphatase role for archaeal SurEs, similar if not identical to the acid phosphatase activity recently reported by Lee *et al*. (2001) and Zhang *et al*. (2001) for the prokaryotic *Tma* SurE.

**Table 4.1: Crystallographic statistics for *Pae* SurEα.**

### Data collection

| Data set | *Inflection* | *Peak* | *High-energy remote* |
|---|---|---|---|
| Wavelength (Å) | 0.97870 | 0.97860 | 0.96485 |
| Resolution range (Å) | 100.0 – 2.40 | 100.0 – 2.00 | 100.0 – 2.85 |
| # reflections (total / unique) | 242,223 / 47,599 | 297,069 / 42,129 | 206,830 / 28,846 |
| Completeness (%)* | 100.0 [100.0] | 99.9 [100.0] | 99.9 [100.0] |
| I / σ(I) | 15.9 [2.3] | 17.4 [2.5] | 14.0 [3.3] |
| $R_{merge}$ (%)† | 9.8 [77.8] | 9.9 [78.1] | 14.7 [94.1] |
| Anomalous signal ($<\chi^2>$)‡ | – | 4.3 [1.9] | – |

### MAD phasing

| | | | | |
|---|---|---|---|---|
| # Se per a.u. (12 expected)§ | | – | 8 | – |
| Resolution range (A) | | 38 – 2.85 | 38 – 2.85 | 38 – 2.85 |
| Phasing power: | acentric | | 0.52 | 0.37 |
| | centric | | 0.44 | 0.33 |
| $R_{cullis}$:¶ | acentric | | 0.94 / 0.65 | 0.96 / 0.95 |
| | centric | | 0.90 | 0.93 |
| Figure of merit‖ | | 0.43 / 0.59 | – | – |

### Model refinement

| | | | | |
|---|---|---|---|---|
| Resolution range (Å) | | 20.0 – 2.0 | $R_{cryst}$ / $R_{free}$ (%)¥ | 18.5 / 22.3 |
| # reflections (working set / test set) | | 38,377 / 2,033 | RMSD bonds (Å) | 0.014 |
| # protein residues (A / B)** | | 276 / 278 | RMSD angles (°) | 1.79 |
| # solvent molecules | water: | 287 | $<B>$ (protein, Å$^2$) | 35.73 |
| | glycerol: | 7 | $<B>$ (water, Å$^2$) | 44.88 |
| | acetate: | 2 | PDB submission code | 1L5X |

\*  Statistics for the highest resolution shell are given in [square brackets].

†  $R_{merge}(I) = \Sigma_{hkl} \left( \left( \Sigma_i |I_{hkl,i} - <I_{hkl}>| \right) / \Sigma_i I_{hkl,i} \right)$

‡  Anomalous signal as measured by the normalized $\chi^2$ for merging Bijvoet pairs $I^+$, $I^-$. That is, $\chi^2 = \Sigma_{I+, I-} \left( \left( I - <I> \right)^2 / \sigma^2(n / n\text{-}1) \right)$. Values > 2 suggest a usefully strong anomalous signal.

§  The number of Se sites per a.u. calculated by SHELXD (12 sites expected per a.u.)

¶  $R_{cullis} = \left( \Sigma_{hkl} \left| |F_{PH} \pm F_P| - F_{H,calc} \right| \right) / \Sigma_{hkl} |F_{PH} \pm F_P|$. Statistics for acentric reflections are given as isomorphous / anomalous.

‖  Values are given before / after density modification and phase extension to 2.0 Å.

¥  $R_{cryst} = \Sigma_{hkl} \left| |F_{obs}| - |F_{calc}| \right| / \Sigma_{hkl} |F_{obs}|$. $R_{free}$ was computed identically, except that 5% of the reflections were omitted as a test set.

\*\*  number of SurEα residues built in monomers A and B, out of 280 residues per monomer of recombinant protein (the His-tag and linker add 14 residues to the wild type sequence)

**Table 4.2: Buried surface area statistics for *Pae* and *Tma* SurEs.**

| | Molecular species | Total surface area ($\text{Å}^2$) |
|---|---|---|
| *Pae* SurEα | Pae monomer sa | 13215.4, 13199.3 |
| | Pae dimer sa | 20579.6 |
| | Pae tetramer sa | 37451.6 |
| | **Pae dimer interface** | **5835.1** |
| | **Pae tetramer interface** | **3716.0** |
| *Tma* SurE (Lee et al.) | 1J9J monomer sa | 13515.4, 13533.5 |
| | 1J9J dimer sa | 20113.9 |
| | 1J9J tetramer sa | 38614.1 |
| | **1J9J dimer interface** | **6935.0** |
| | **1J9J tetramer interface** | **2422.0** |
| *Tma* SurE (Zhang et al.) | 1ILV monomer sa | 13446.1, 13429.4 |
| | 1ILV dimer sa | 20025.6 |
| | 1ILV tetramer sa | 37896.9 |
| | **1ILV dimer interface** | **6849.9** |
| | **1ILV tetramer interface** | **2006.6** |

Note 1: The "sa" entries indicate surface areas of that molecular species and the "interface" entries indicate the total buried surface area in that type of interface (*e.g.*, counting both subunits of a dimer).

Note 2: Only two values are given for monomer surface areas and only one value for the dimer surface area (and dimer interface), even though in a tetramer there should be four independent monomers and two independent values for a particular type of dimer interface. This is because only one SurEα dimer constitutes the crystallographic asymmetric unit, and, therefore, any other calculated numbers would almost certainly be artificially identical (due to crystalline symmetry).

Note 3: The difference between buried surface areas in the two *Tma* structure determinations (6935.0 vs 6849.9 $\text{Å}^2$) is due to minor variations in crystal packing, and probably is not biochemically significant.

**Table 4.3: All 43 known SurE homologs.** The next page provides a comprehensive list of all 43 known SurE homologs along with the species abbreviations, GenPept IDs, and phylogenetic kingdom (P = prokaryote, E = eukaryote, A = archaeon). Also, asterisks denote extremophiles. Genomes with paralogous SurEα/β pairs are boldfaced.

| Species abbreviation | GenPept ID | Kingdom | Species |
|---|---|---|---|
| Ther_mari_SurE | gi15644410 | P* | *Thermotoga maritima* |
| Aqui_aeol_SurE | gi15606188 | P* | *Aquifex aeolicus* |
| Yers_pest_SurE | gi16123508 | P | *Yersinia pestis* |
| Xyle_fast_SurE | gi15837460 | P | *Xylella fastidiosa 9a5c* |
| Vibr_chol_SurE | gi15640553 | P | *Vibrio cholerae* |
| Trep_pall_SurE | gi15639410 | P | *Treponema pallidum* |
| **Syne_sp._SurEα** | **gi16332288** | **P** | ***Synechocystis sp. PCC 6803*** |
| **Syne_sp._SurEβ** | **gi16330072** | **P** | ***Synechocystis sp. PCC 6803*** |
| Sino_meli_SurE | gi1754720 | P | *Sinorhizobium meliloti* |
| Salm_ente_SurE | gi16761699 | P | *Salmonella enterica* |
| Rubr_gela_SurE | gi11280179 | P | *Rubrivivax gelatinosus* |
| Rals_sola_SurE | gi17545923 | P | *Ralstonia solanacearum* |
| Pseu_aeru_SurE | gi15598821 | P | *Pseudomonas aeruginosa* |
| Past_mult_SurE | gi15603477 | P | *Pasteurella multocida* |
| **Nost_sp._SurEα** | **gi17232338** | **P** | ***Nostoc sp. PCC 7120*** |
| **Nost_sp._SurEβ** | **gi17230631** | **P** | ***Nostoc sp. PCC 7120*** |
| Neis_meni_SurE | gi15794586 | P | *Neisseria meningitidis Z2491* |
| Meso_loti_SurE | gi13471179 | P | *Mesorhizobium loti* |
| Legi_pneu_SurE | gi5771428 | P | *Legionella pneumophila* |
| He_pylJ99_SurE | gi15611932 | P | *Helicobacter pylori J99* |
| He_pyl26695_SurE | gi15645546 | P | *Helicobacter pylori 26695* |
| Haem_infl_SurE | gi16272643 | P | *Haemophilus influenzae Rd* |
| Esch_coli_SurE | gi15803261 | P | *Escherichia coli O157* |
| Dein_radi_SurE | gi15807387 | P | *Deinococcus radiodurans* |
| Coxi_burn_SurE | gi8141682 | P | *Coxiella burnetii* |
| Chla_pneu_SurE | gi15618182 | P | *Chlamydophila pneumoniae* |
| Chla_trac_SurE | gi15604938 | P | *Chlamydia trachomatis* |
| Chla_muri_SurE | gi14195226 | P | *Chlamydia muridarum* |
| Caul_cres_SurE | gi16126241 | P | *Caulobacter crescentus* |
| Camp_jeju_SurE | gi15791661 | P | *Campylobacter jejuni* |
| Bruc_meli_SurE | gi17987364 | P | *Brucella melitensis* |
| Agro_tume_SurE | gi15889009 | P | *Agrobacterium tumefaciens* |
| Yarr_lipo_SurE | gi400781 | E | *Yarrowia lipolytica* |
| Sacc_cere_SurE | gi6319570 | E | *Saccharomyces cerevisiae* |
| **Arab_thal_SurEα** | **gi7485145** | **E** | ***Arabidopsis thaliana*** |
| **Arab_thal_SurEβ** | **gi15218620** | **E** | ***Arabidopsis thaliana*** |
| **Pyro_aero_SurEα** | **gi18313680** | **A*** | ***Pyrobaculum aerophilum*** |
| **Pyro_aero_SurEβ** | **gi18314130** | **A*** | ***Pyrobaculum aerophilum*** |
| Meth_ther_SurE | gi15679432 | A* | *Methanobacterium thermautotrophicum* |
| Meth_jann_SurE | gi15668739 | A* | *Methanococcus jannaschii* |
| Halo_sp._SurE | gi15790299 | A* | *Halobacterium sp. NRC* |
| Arch_fulg_SurE | gi11498547 | A* | *Archaeoglobus fulgidus* |
| Aero_pern_SurE | gi14600980 | A* | *Aeropyrum pernix* |

**Figure 4.1: Structure validation and sample electron density for the refined *Pae* SurEα model.**

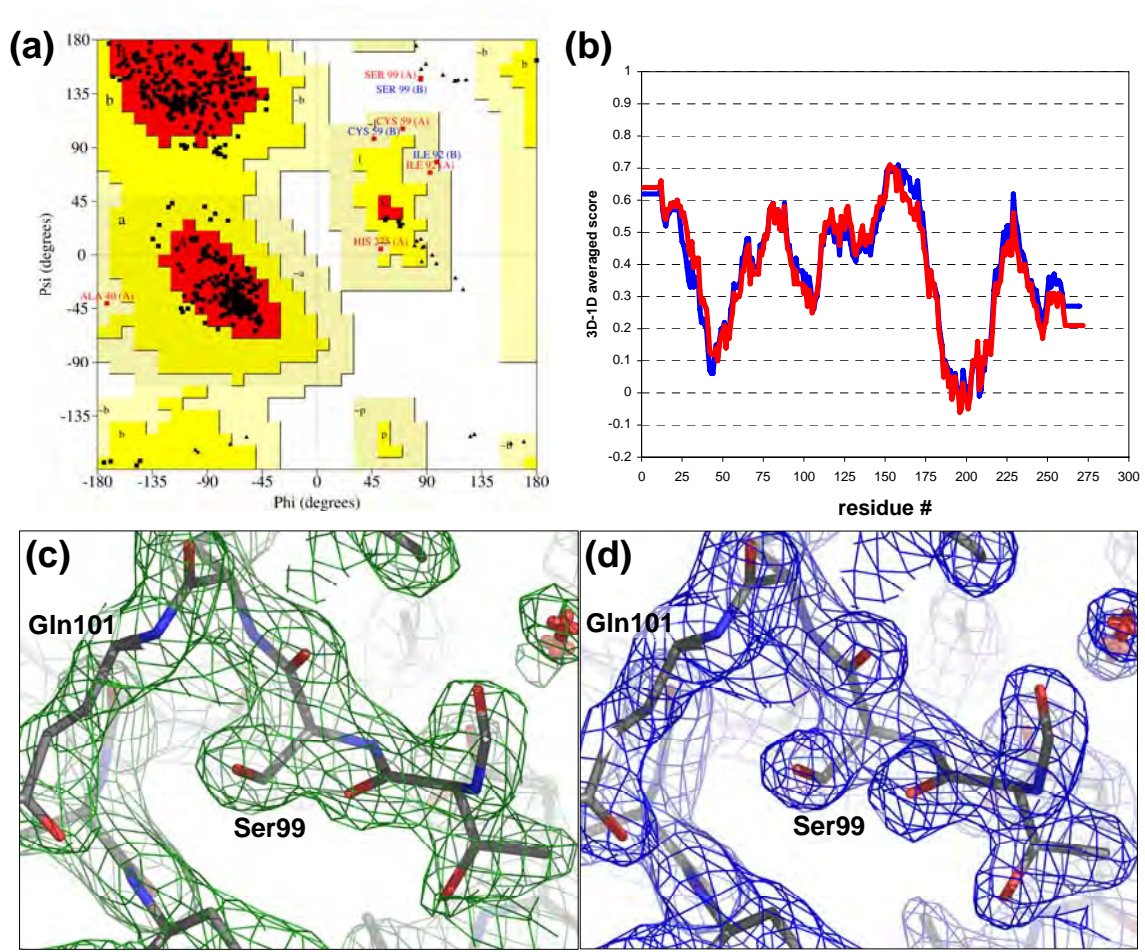

**Figure 4.2: Sequence analysis of all archaeal SurEs.**

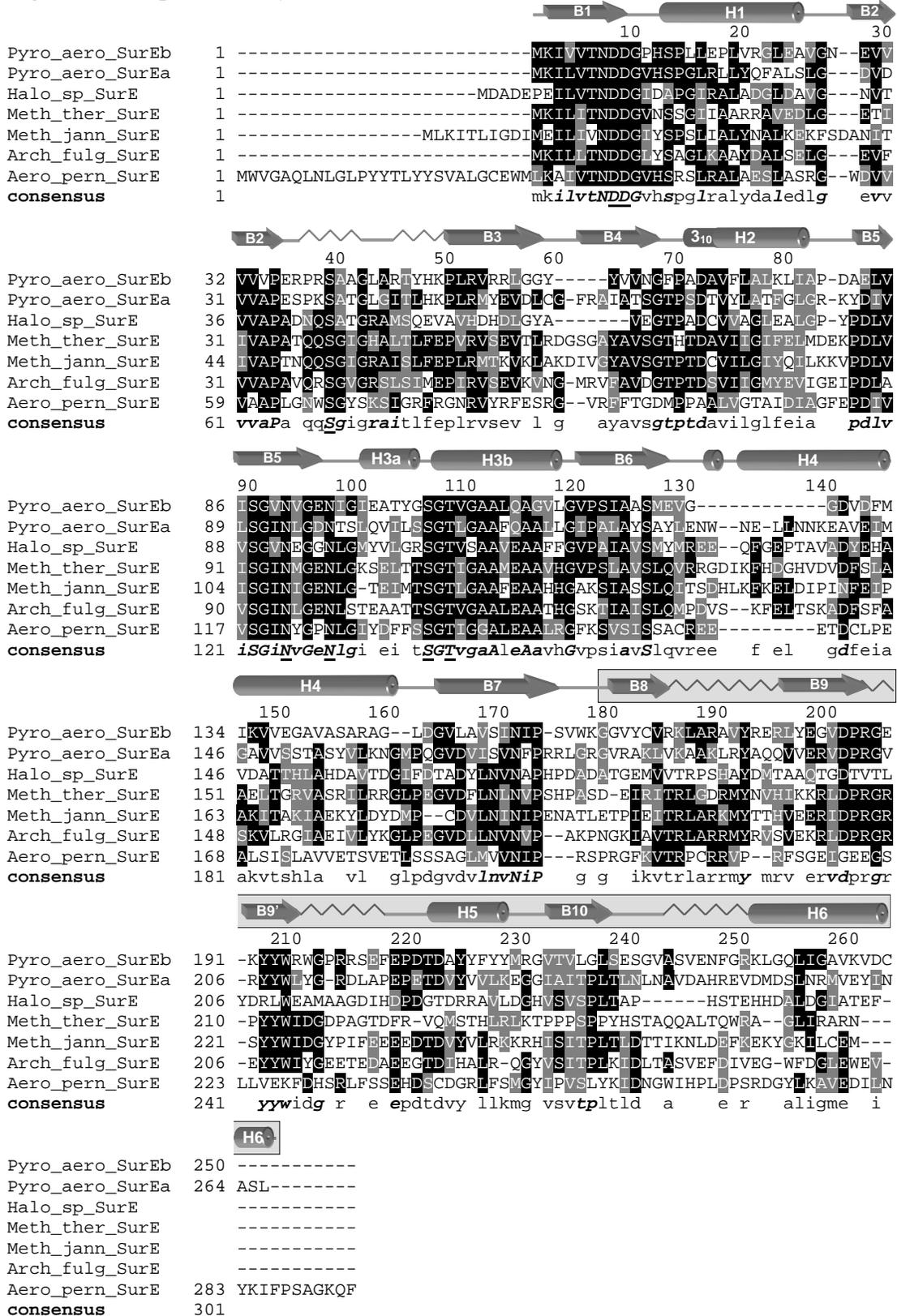

**Figure 4.3:** *Pae* SurEα is a Rossmann-like fold with an extended C-terminal domain.

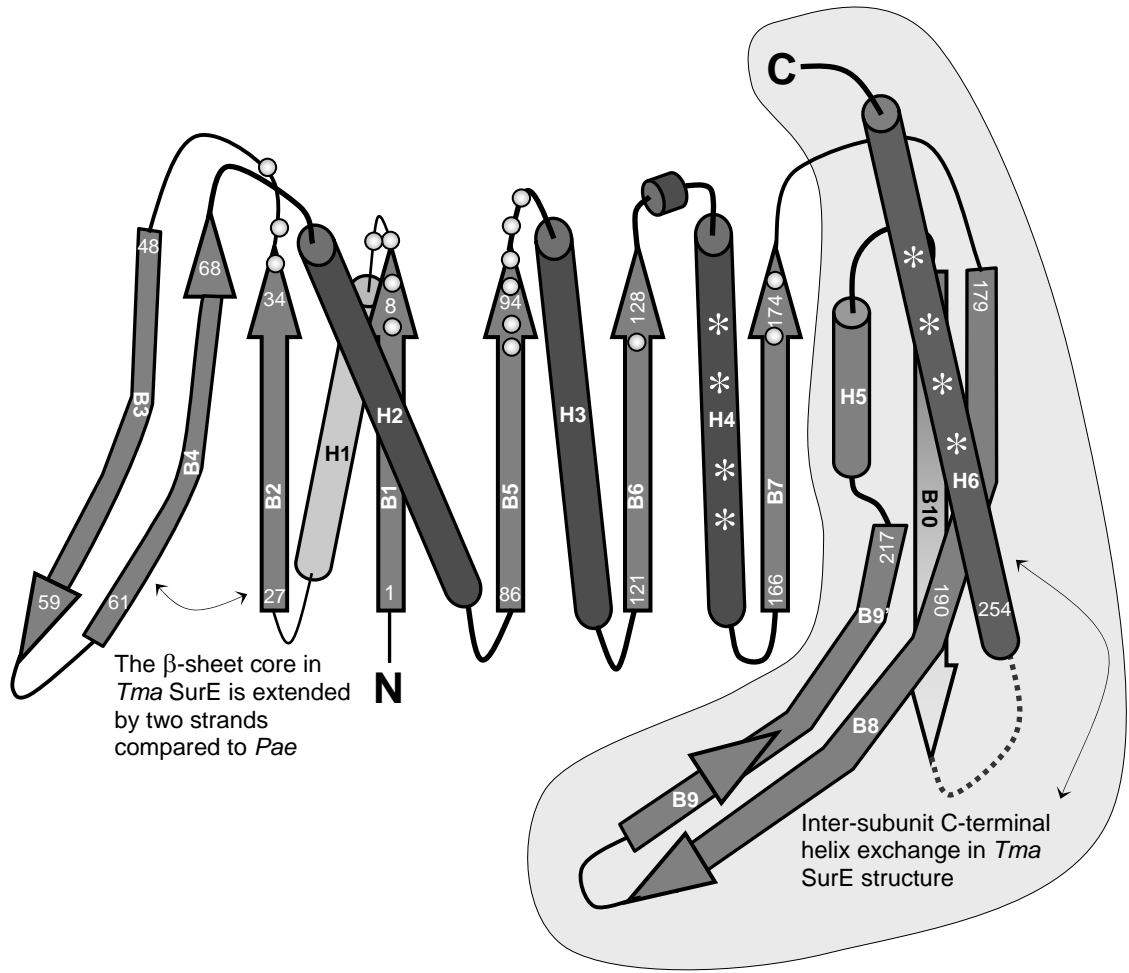

The β-sheet core in *Tma* SurE is extended by two strands compared to *Pae*

Inter-subunit C-terminal helix exchange in *Tma* SurE structure

**Figure 4.4: Structural comparison of *Pae* and *Tma* SurE reveals conformationally invariant regions.**

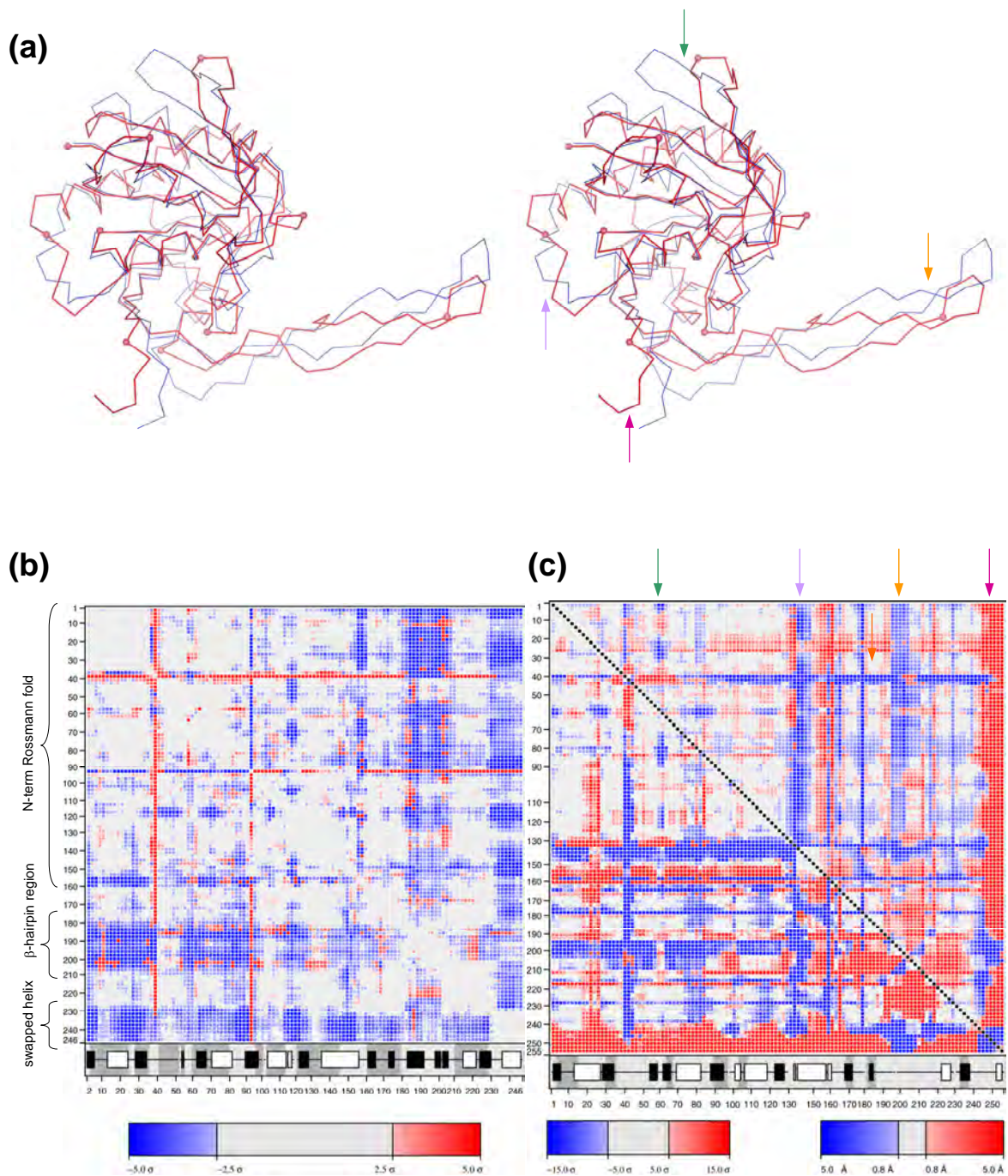

**Figure 4.5: Crystalline *Pae* SurEα is predominantly non-domain swapped.**

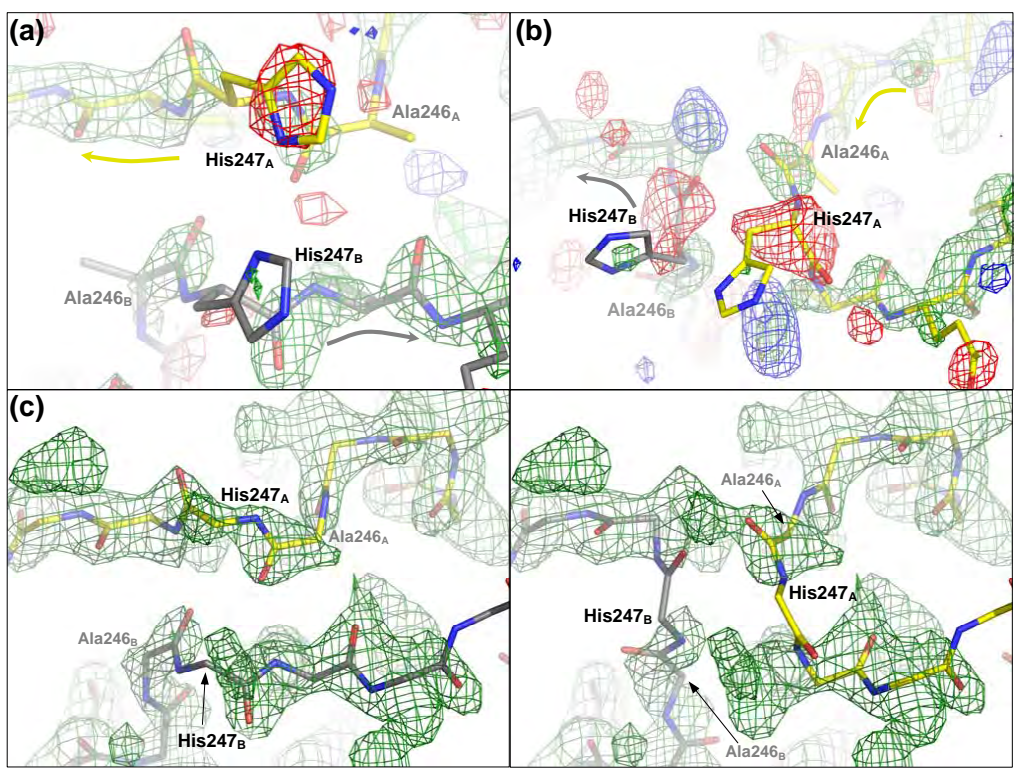

**Figure 4.6:** *Pae* SurEα has a much less extensive dimer interface than *Tma* SurE.

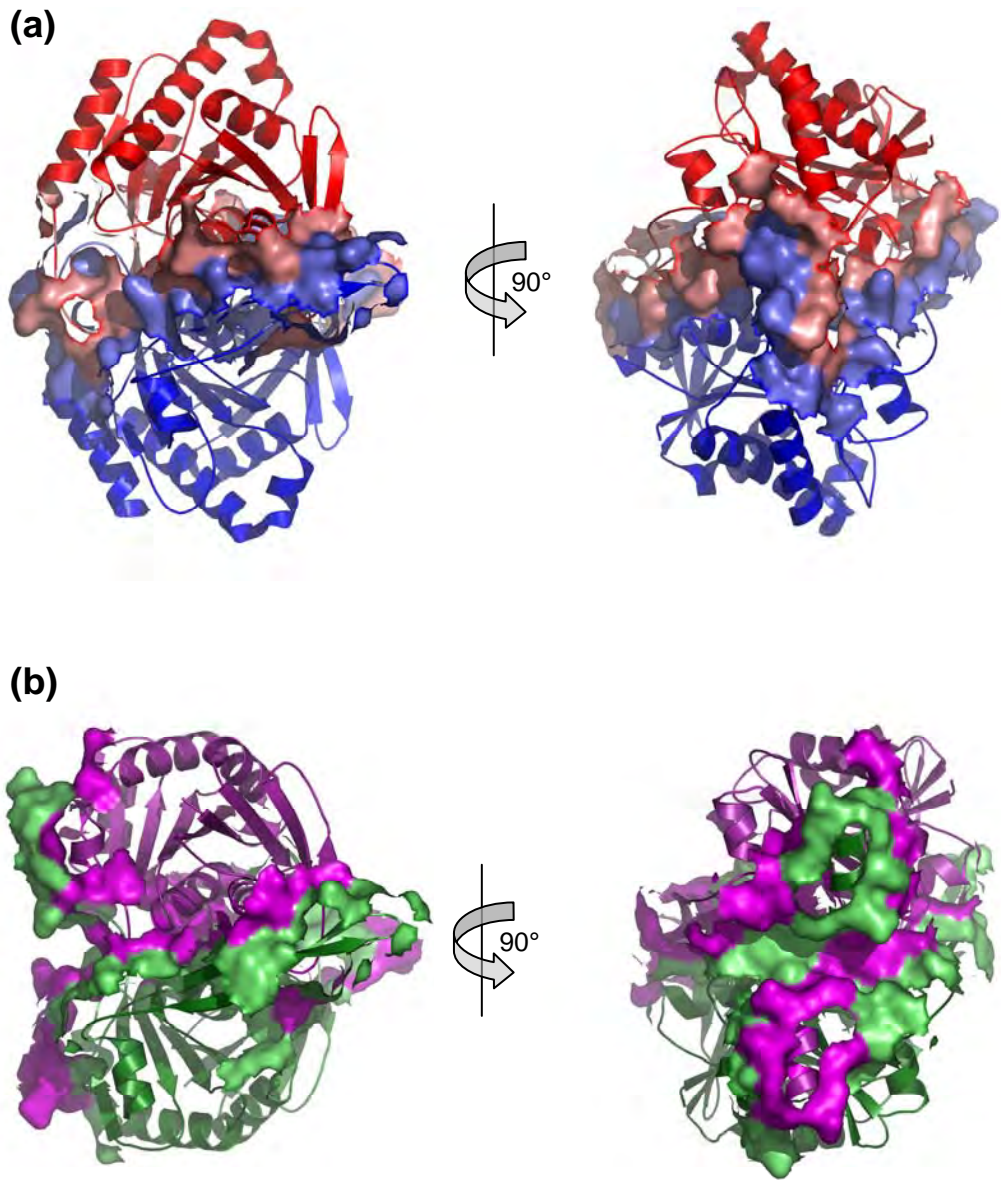

**Figure 4.7: The putative SurE active site is highly acidic and strongly conserved.**

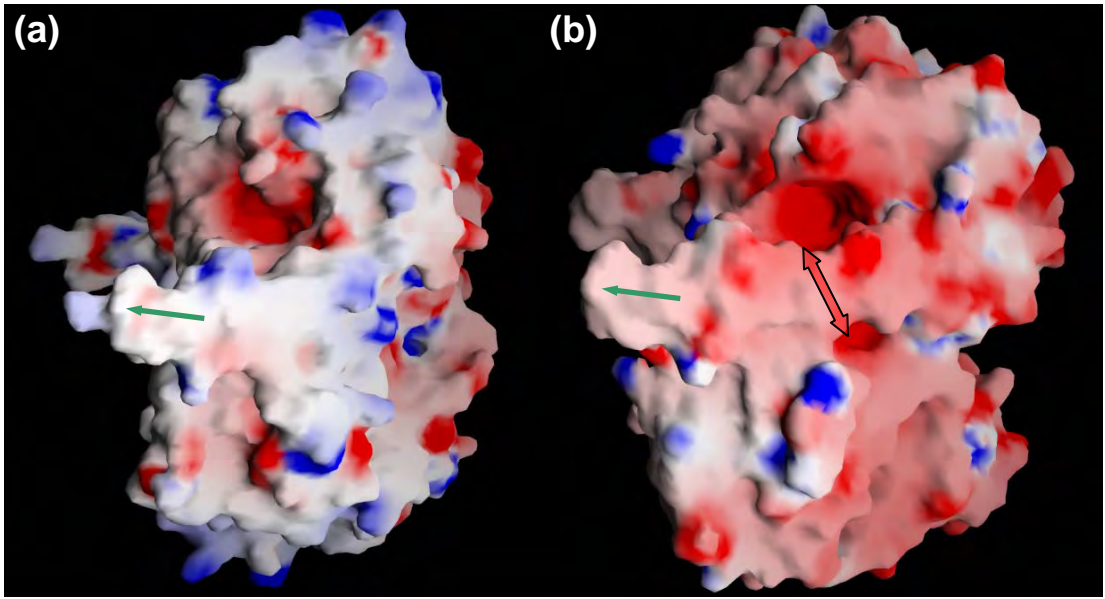

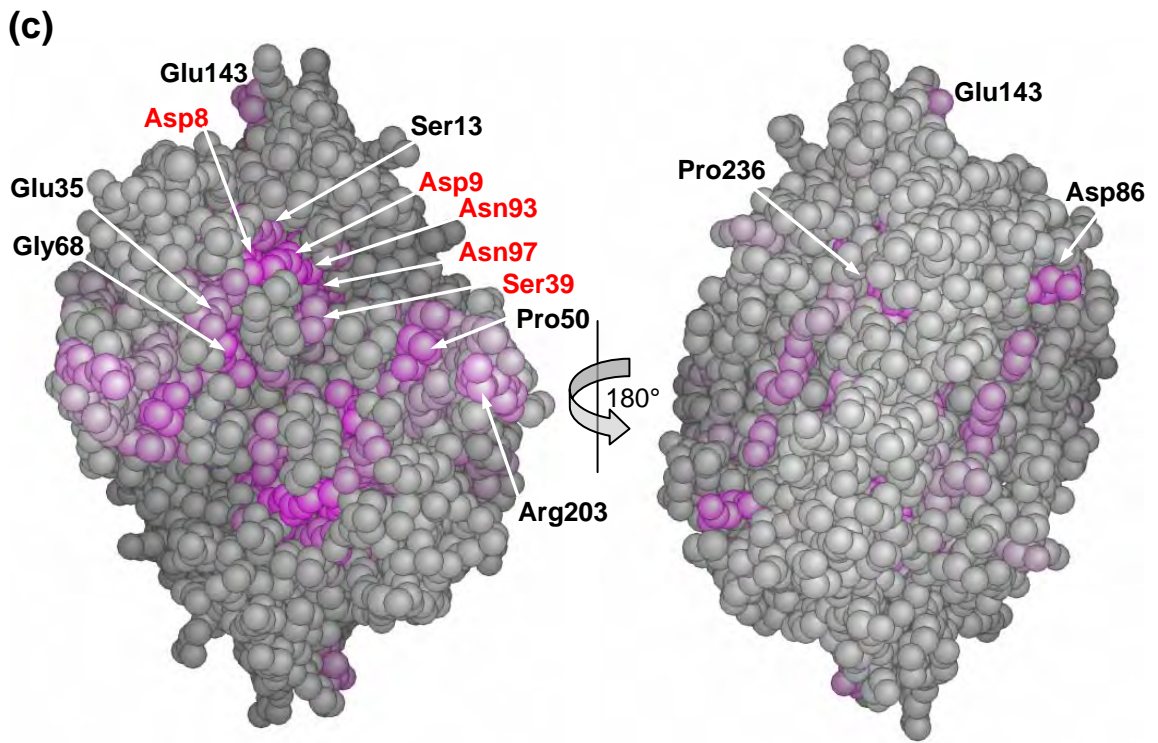

**Figure 4.8: Hypothetical model of GMP bound to the conserved SurE active site.**

**(a)**

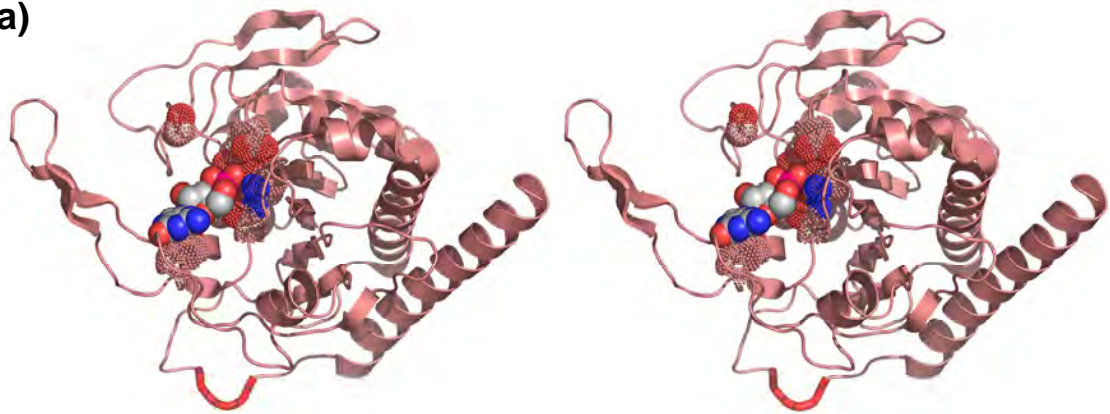

**(b)**

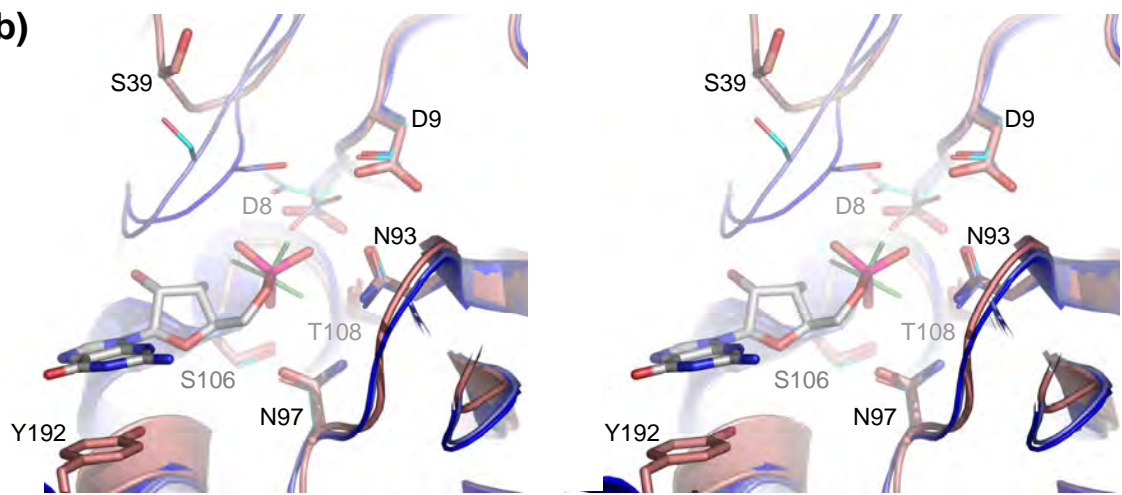

**Figure 4.9: Phylogenetic distribution and genomic organization of *surE* genes.**

**(a)**

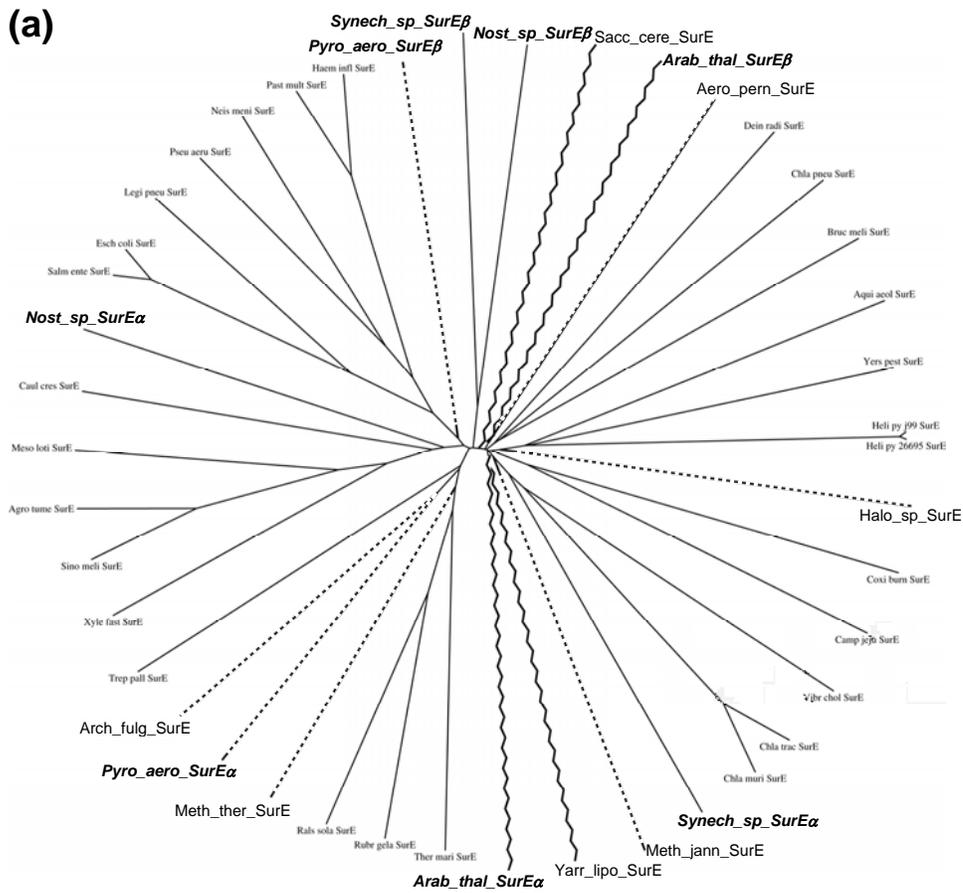

**(b)**

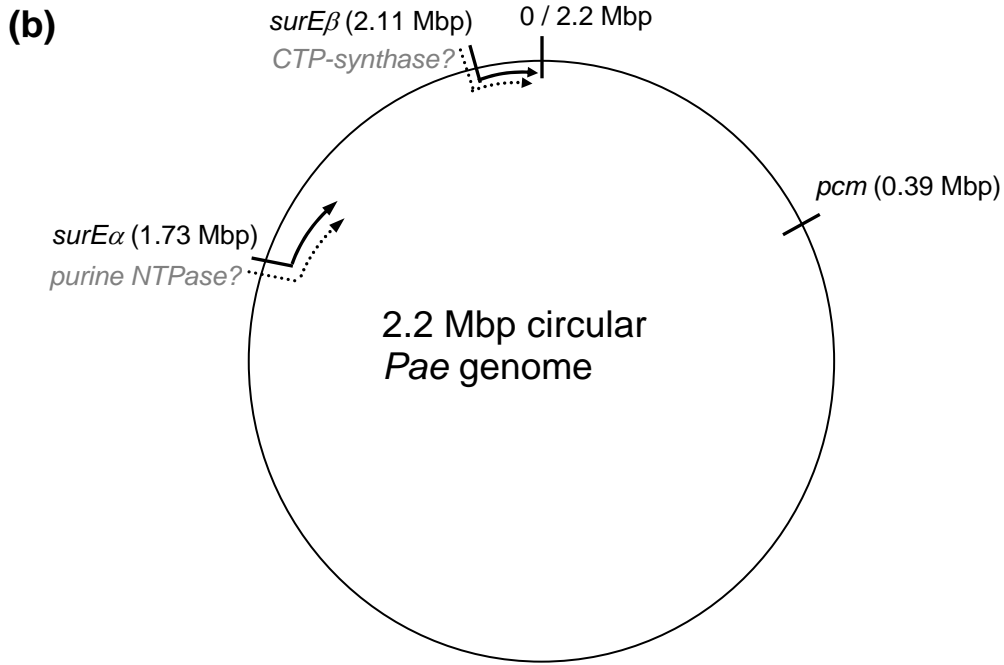

**Figure 4.10: Representative reactions catalyzed by homologs of SurE gene neighbors.**

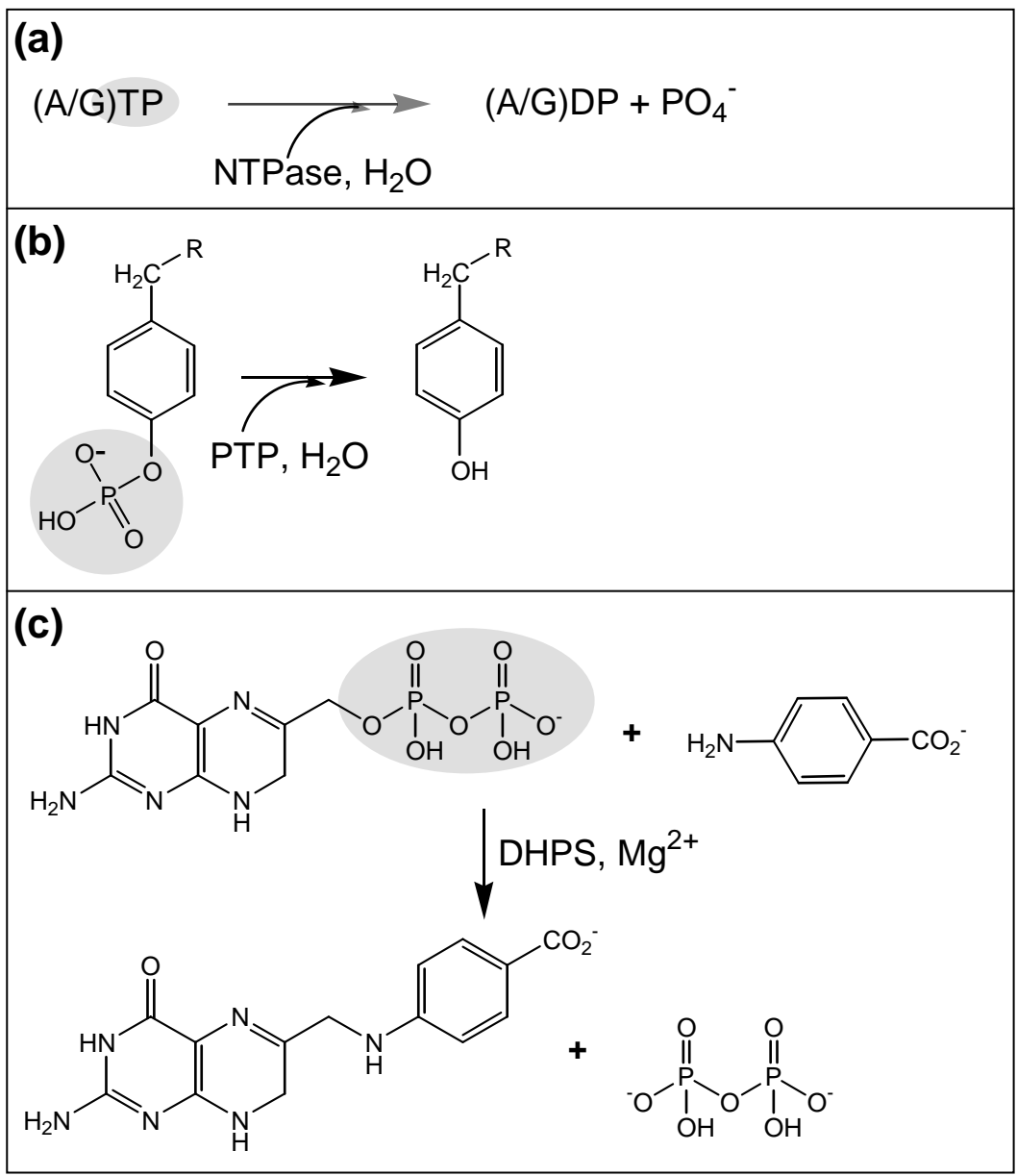

**Chapter 5:**

**The crystal structure of a Nudix protein from *Pyrobaculum aerophilum* reveals a dimer with two intersubunit β-sheets\***

\* This chapter is an adaptation of the following published article:

## Abstract

Nudix proteins, formerly called MutT homolog proteins, are a large family of proteins that play an important role in reducing the accumulation of potentially toxic compounds inside the cell. They hydrolyze a wide variety of substrates that are mainly composed of a nucleoside diphosphate linked to some other moiety *X* and thus are called Nudix hydrolases. Here, the crystal structure of a Nudix hydrolase from the hyperthermophilic archaeon *Pyrobaculum aerophilum* is reported. The structure was determined by the single-wavelength anomalous scattering method with data collected at the peak anomalous wavelength of an iridium-derivatized crystal. It reveals an extensive dimer interface, with each subunit contributing two strands to the β-sheet of the other subunit. Individual subunits consist of a mixed, highly twisted and curved β-sheet of 11 β-strands and two α-helices, forming an α-β-α sandwich. The conserved Nudix box signature motif, which contains the essential catalytic residues, is located at the first α-helix and the β-strand and loop preceding it. The unusually short connections between secondary-structural elements, together with the dimeric form of the structure, are likely to contribute to the thermostability of the *P. aerophilum* Nudix protein.

## Introduction

Nudix proteins form a large family of proteins that are found in all organisms. They hydrolyze a wide variety of substrates that contain a nucleoside diphosphate linked to some other moiety, *X*. Therefore, these enzymes are called Nudix hydrolases (Bessman *et al*., 1996). Their functions are mainly to reduce the level of the potentially toxic compounds and the accumulation of metabolic intermediates (Safrany *et al*., 1998; O'Handley *et al*., 2001). The best-studied Nudix enzyme is the MutT protein from *E. coli* (Shimokawa *et al*., 2000; Abeygunawardana *et al*., 1995; Taddei *et al*., 1997; Wagner *et al*., 1997; Bhatnagar *et al*., 1991; Frick *et al*., 1995; Porter *et al*., 1996). It reduces the rate of AT to CG transversion several thousand-fold by hydrolyzing the mutagenic nucleotide 8-oxo-dGTP to give 8-oxo-dGMP, thus preventing it from being incorporated into DNA (Bessman *et al*., 1996). Its catalytic activity is reported to depend on the ligation of an essential metal ion. Structural and mutagenesis studies have identified the metal ligands as part of a signature motif conserved among the Nudix family, $GX_5EX_7REUXEEXGU$, where *U* is a hydrophobic residue and *X* is any amino acid (Harris *et al*., 2000; Shimokawa *et al*., 2000; Lin *et al*., 1996; Lin *et al*., 1997). Furthermore, the structure of this motif appears to be conserved among three members of the Nudix family for which structures have been determined: *E. coli* MutT, diadenosine tetraphosphate hydrolase ($A_{P4}A$ hydrolase) from *Lupinus angustifolius* (Sawrbrick *et al*., 2000) and an ADP-ribose pyrophosphatase (ADPRase) from *E. coli* (Gabelli *et al*., 2001). All three structures contain an α-β-α sandwich in which the conserved residues in the Nudix motif are located in helix α1 and in the loop preceding it.

Beyond the conserved catalytic core structure, Nudix proteins demonstrate variations in peripheral structure and oligomerization state, perhaps contributing to the wide variety of substrate specificities observed in this family. For example the $A_{P4}A$ hydrolase has a structure similar to that of the *E. coli* MutT but with one extra β-strand and two more α-helices. On a larger scale, the structures of MutT and $A_{P4}A$ hydrolase are monomeric, whereas ADPRase is a dimer with its N-terminal domain swapping between the subunits. Furthermore, in *E. coli* MutT, the entire adenosine moiety of the substrate analog AMPCPP binds in the active site cleft behind helix α1 (Lin *et al*., 1997); however, in ADPRase, the terminal ribose of the substrate binds in the cleft, while the adenine moiety binds to the N-terminal domain.

Here, we report the crystal structure of a Nudix protein from the hyperthermophilic crenarchaeote *Pyrobaculum aerophilum* (PA) determined with single-wavelength anomalous scattering data. The PA Nudix protein was found to be a dimer, with each subunit having a similar fold as the *E. coli* MutT. Unlike the structure of ADPRase, the PA Nudix protein dimerizes through two intersubunit β sheets that lie in the center of the polypeptide sequence. The active site cleft has a distinctly different structure from those of other Nudix hydrolases with known structures. In this paper, the structure is analyzed in terms of the putative catalytic function and thermostability.

## Materials and Methods

### *Cloning, expression, and purification of the PA Nudix*

The open reading frame (ORF) that encodes the putative Nudix protein reported here was found *via* a BLAST search of the PA genome with the sequence of the *E. coli* MutT protein.  A phosmid clone containing the PA *mutT* ORF was kindly provided by the laboratory of Dr. Jeffrey Miller (UCLA).  The following primers were utilized for polymerase chain reaction (PCR) amplification of the gene with Deep Vent$_R$ DNA polymerase (New England Biolabs):  ACTATGATCGTTACCAGCGGCGTTTTA (sense) and AAGCTTTGAAATTTTTCCCAGTCTATATAG (antisense).  Amplified DNA was sub-cloned into the pCR-Blunt vector (Invitrogen) to give an intermediary vector that was then digested with the restriction enzymes *NdeI* and *HindIII*.  The fragment containing the *mutT* gene was purified by gel-extraction and ligated into a bacterial expression vector (pET22b(+), Novagen) that had been linearized by double digestion with the same enzymes.  The resulting recombinant plasmid (designated pET22b(+)-mutT) adds a C-terminal 6xHis tag to the expressed protein.  The plasmid was transformed into chemically competent NovaBlue *E. coli* cells (Novagen) for screening, and the DNA sequence of the *mutT* gene was confirmed by sequencing (Davis Sequencing, Inc.)  A point mutant of PA Nudix (M16L), in which a Leu was substituted for Met16, was prepared by site-directed mutagenesis (*via* overlap-extension PCR) in order to avoid heterogeneity resulting from a second translation initiation site at this position.

Recombinant PA Nudix was over-expressed by transformation of the pET22b(+)-mutT plasmid into chemically competent BL21(DE3) *E. coli* cells. Single colonies were used to inoculate Luria-Bertani broth supplemented with 100 μg/ml ampicillin, and the cultures were grown at 37°C to an OD (at 600 nm) of about 0.8. Isopropyl-β-D-thiogalactoside (IPTG) at 1 mM was then added to induce over-expression, and cells were further grown for 3-4 hours. Cells were harvested by centrifugation for 10 min at 8000*g*, and stored at -20°C (253 K).

Thawed cell pellets were resuspended at room temperature in a high-ionic strength buffer containing 20 mM HEPES pH 7.8, 0.5 M NaCl, 0.5 mM PMSF, with or without 0.5% v/v Triton X-100. Cells were lysed by lysozyme treatment (~0.5 mg/ml for 30 min at room temperature) followed by 2-3 passes through a French press operating at ~ 1100 psi, or by sonication on ice for 5x1 min with 1 min intervals. Lysed cells were maintained on ice, and cell debris was cleared by centrifugation at ~ 37,000 g for 30 min. The supernatant of the cell lysate was either directly loaded onto a $Ni^{2+}$-charged HiTrap chelating column (Pharmacia), or pretreated by heating to ~ 75-80 °C for 10 min and clearing the denatured, insoluble *E. coli* proteins by centrifugation at 37,000 g for 35 min (> 80% purity was achieved with this heat treatment). The final, full-length protein was > 99% pure as estimated by several independent techniques (SDS-PAGE, MALDI-TOF and electrospray mass spectrometries).

### Crystallization and data collection

The purified protein was exchanged by dialysis, or by 3 cycles of dilution and concentration, into a dilute buffer (10 mM TRIS, 5 mM EDTA, pH 8.0, or 5 mM HEPES, 10 mM NaCl, pH 7.5), and then concentrated to above 10.0 mg/ml at 4 °C in a Centriprep ultrafiltration device. Crystallization experiments were carried out using the hanging drop vapor diffusion method. In each drop, 3 to 5 μl of protein solution was mixed with an equal volume of well solution. Hampton crystal screening kits I and II were used for initial screening of crystallization conditions. Crystals were obtained under several conditions. The best crystals were obtained from drops set up with well solutions containing 5-8% PEG 4000 or PEGMME (polyethylene glycol monomethyl ether) 2000, 5% isopropanol, 50 mM $(NH_4)_2SO_4$, and 100 mM NaOAc pH 4.8 (Native-1 crystals), or well solutions consisting of 100 mM MES pH 6.2 and 15% 2-methyl-2,4-pentanediol (MPD) (Native-2 crystals), at room temperature. Crystals were soaked for 5 minutes in the well solution but with 30% of glycerol, or 50% of MPD for Native-2 crystals, and flashed frozen in a stream of $N_2$ gas of 100 K. An iridium derivative was obtained by soaking a crystal, which was obtained under similar conditions as Native-1 crystals, in the well solution containing 20 mM of $IrCl_3$ for 8 hours. The crystal was then similarly flash-frozen. Diffraction data from Ir-derivatized and Native-1 crystals were collected at beamline X8C, NSLS, Brookhaven National Lab, and those from the Native-2 crystal were collected on an in-house RIGAKU FRD generator with an RAXIS IV++ image plate detector. For the iridium derivative crystal, a single set of data was collected at the peak anomalous wavelength of iridium. The diffraction data were indexed, integrated and scaled with the programs DENZO and SCALEPACK (Otwinowski and Minor, 1996). Data processing statistics of the iridium derivative data and two native data sets are listed in Table 5.1. The iridium and Native-1 crystals are not isomorphous, despite their close similarity in cell dimensions. Thus, the native data were not used for initial phasing. After solving the structures, it became clear that the crystals have different packing interfaces along the crystal Z-axis. The two crystals pack in the same way in the XY plane, but the packing between layers is shifted along the Y-axis by about 18.6 Å.

### *Phasing and structure determination*

Initial phases were determined for the data collected at the peak anomalous wavelength from the iridium derivative crystal. Heavy atom positions were identified with programs XtalView (McRee, 1992) and SHELXD (Uson and Sheldrick, 1999), and then refined with the program MLPHARE (Otwinowski, 1991; CCP4, 1994). Density modification, *i.e.* solvent flattening, histogram matching, multi-resolution modification, and two-fold NCS averaging were carried out with the program DM (Cowtan, 1994).

The automatic refinement procedure, ARP/warp (Perrakis *et al.*, 1999), was then used for automatic model building. The initial model and the electron density maps were displayed with the graphics program O (Jones *et al.*, 1991), and the model was manually rebuilt to fix the side chains and to add residues that have clear electron density. The model was then refined using CNS (Brunger *et al.*, 1998) with NCS restraints. Sigma weighted $2F_o – F_c$ and $F_o – F_c$ electron density maps were calculated from the refined model, and the model was manually fixed using the program O. This cycle was repeated until the refinement converged. This refined model was used as a model for molecular replacement of the Native-1 data with the program AMORE (Navaza, 1994), or for the Native-2 data with the program EPMR (Kissinger *et al.*, 1999), and refinement was similarly carried out with the programs CNS and O.

The final refinement statistics are listed in Table 5.2. Over 91% of residues in the native-1 structure and over 93% in the Ir derivative structure fall into the most favored regions in the Ramachandran plot. Some glycerol, or MPD molecules from the cryogenic solution were found in the refined structures (see Table 5.2). Positive $F_o - F_c$ density was found near a cluster of four His side chains, His19, His31, His89 and His91, in the active site cleft. A water molecule modelled into this site was not sufficient to account for this density. Since the protein was purified from a Ni-chelating column, a $Ni^{2+}$ ion was put into the density. The hydroxyl group of Tyr84 side chain is also at a good position to function as a ligand.

In the Ir derivative structure, all the iridium atoms were found to bind in surface cavities on the edge of bulk solvent channels. Three of them, Ir1, Ir2 and Ir3, were found at the crystal-packing interface, while Ir4 was found near the Nudix box residues of subunit A. The strongest iridium peak, Ir1, was found at a packing interface of three molecules, and surrounded by five water molecules, which take five positions of an octahedral coordination. The interactions between the protein atoms and Ir1 and Ir4 seem to be mediated by water. Ir2 and Ir3 are related approximately by the two fold NCS, each bound near a sulfate ion, as well as side chains of Asp65, Asp66 and Asn67 of a symmetry-related molecule.

## Results

### *Protein expression and purification*

The PA Nudix gene was cloned and over-expressed in *E. coli* with a His tag at the C-terminus. It was found that the wild type protein was expressed as a mixture of the full length and a truncated form that lacks the first 15 N-terminal residues. Apparently, the Met16 codon acts as a second start site for translation of this truncated protein. The DNA sequence ~ 10 bp upstream of the Met16 codon corresponds to a bacterial ribosome binding site, thus exacerbating the problem. Similar problems have been reported with other proteins (Matsumiya *et al*., 2001). In order to circumvent the alternative start site, a point mutant (M16L) was made, in which a Leu was substituted for Met16. This mutant was expressed as a single full-length polypeptide and was purified to better than 99% pure. It differs from the wild type sequence only in the M16L mutation and 14 amino acids appended to the C-terminus (the 6xHis tag and an 8-aa linker). This M16L mutant was used for all the studies reported here.

### *Phasing with single-wavelength anomalous scattering data*

The crystal structure was determined using single wavelength anomalous scattering data collected at a wavelength near the peak of fluorescence at the LIII edge of iridium. An anomalous difference Patterson map revealed a single well-occupied Ir site and a few much weaker potential sites. SHELXD (Uson and Sheldrick, 1999) was used to evaluate the heavy atom sites and how well they correlate with the Patterson map. The two top sites were input to MLPHARE (Otwinowski, 1991; CCP4, 1994) for further refinement and phase calculation. The anomalous difference Fourier map calculated from these phases gave four unique peaks above the 5.5 sigma contour level. The positions of the two new peaks also matched the third and fifth sites predicted by SHELXD. These four sites were then input to MLPHARE for refinement and phase calculation. The four iridium atoms (designated as Ir1, Ir2, Ir3 and Ir4) have occupancies of 0.74, 0.50, 0.36 and 0.39, and temperature factors of 57, 59, 64 and 67, respectively after refinement with MLPHARE. The atom Ir1, which has the highest occupancy and lowest B-factor, gave the strong peak in the Patterson map.

Density modification with solvent flattening, histogram matching and multi-resolution modification by DM (Cowtan, 1994) was carried out to improve the phases and to identify the correct absolute configuration. The Fourier map after DM showed enough secondary structure elements to locate the two-fold non-crystallographic symmetry (NCS) axis. Two-fold NCS averaging with DM significantly improved the quality of the electron density map (Figure 5.1), and the entire map was continuous and could be easily traced. Automatic model building by ARP/wARP (Perrakis *et al*., 1999) at a resolution of 1.85 Å built more than 80% of the model. The map calculated at this resolution with phases from the auto-built model showed clear side chains for more than 80% of the residues, and a complete model was built in a single round of manual rebuilding.

### *Overall structure*

The refined crystal structure of the PA Nudix protein shows that it forms a dimer. Each subunit also contributes two β-strands to the β-sheet of the other subunit to form an extensive dimer interface. The dimer is held together mainly by the two β-sheets to which both subunits contribute strands. Hydrogen bonding interactions occur between strand β5b of one subunit and β6 of the other subunit (Figure 5.2 and 3). There are eight main chain hydrogen bonds between two subunits in the dimer. In addition, a patch of residues with hydrophobic side chains, Ile2, Ile77, Leu94, Pro38, Pro74, Tyr61, Tyr96, Phe92 and Met73, located at the center of the dimer interface, also contribute to the binding energy of the dimer. The molecular surface of each monomer, calculated with the program GRASP (Nicholls *et al.*, 1991), is about 7500 $\mathring{A}^2$, and the total buried surface in the dimer interface is about 2000 $\mathring{A}^2$. Dimer (and possibly tetramer) formation in solution was confirmed by analytical ultracentrifugation studies (data not shown).

Each subunit (referred to as A and B) is composed of a mixed β-sheet and two α-helices (Figure 5.2). The β-sheet is highly curved and twisted, and wraps around the helices. The topology of the protein is shown in Figure 5.3. Most of the strands are connected by tight turns or short loops, with the longest loop consisting of only six residues. The β-sheet can be divided into a central sheet and a sub-sheet, which are connected through strand β2. The N-terminal part of strand β2 forms hydrogen bonds with the C-terminal part of β1, but then curves and makes a sub-sheet with strands β3, β8 and β9. Strands β1, β4 to β7, together with β5a′ and β5b′ from the other subunit, form the central β sheet.

The two dimer subunits are similar to each other in both Native-1 and Native-2 crystal forms. However, in the Native-1 form subunit A has a long C-terminal helix that extends away from the molecular surface, whereas in subunit B the C-terminal residues turn back and form a more compact structure, making the helix shorter (Figure 5.4A).

Interestingly, the residues in the linker to the His6x tag are ordered in $2F_o - F_c$ maps for both the Native-1 ($P2_12_12_1$) and Native-2 ($P2_1$) Nudix crystals, but adopt different conformations that give rise to these two different crystal forms in an easily understandable way. As mentioned, in the Native-1 form the subunit A linker residues extend the C-terminal helix (α2 in Figures 2 and 3), thereby making crystal contact interactions with adjacent, crystallographically-related A+B dimers in the $P2_12_12_1$ lattice. However, in the Native-2 crystals the C-terminal helices of *both* subunits (A+B) are fully extended, thereby allowing four Nudix monomers (A+B+A'+B') to pack into the asymmetric unit of the lower-symmetry $P2_1$ lattice as a tetramer with 222 symmetry (Figure 5.4B). The A+B and A'+B' dimers associate *via* their extended C-terminal helices and form a tetramer which is composed of two Nudix dimers that are nearly identical to the ones seen in the Native-1 form (except for the C-terminal helices of subunits B, B', Figure 5.4B). Since the scant interactions between the helices of these two dimers mainly involve residues from the hydrophobic linker to the His6x tag, this Nudix tetramer is not likely to be biologically significant. Nonetheless, it is a peculiar example of a His tag resulting in an artefactual oligomerization state for a crystallized protein.

### The conserved residues and the active site cleft

The Nudix box sequence signature motif, $G^{30}X_5E^{36}X_7R^{44}E^{45}UXE^{48}E^{49}XG^{51}U$ (the superscripts refer to the residue number in the PA Nudix sequence) spans the strand β4, helix α1 and the N-terminal part of strand β5 (Figure 5.5 and 6). The conserved residues are at the same relative positions in three-dimensional space as those in the *E. coli* MutT structure (Figure 5.7). These residues probably play similar roles throughout the Nudix enzyme family. All side chains of the conserved residues are on the same side of helix α1. The semi conserved hydrophobic residues, Phe46 and Ile52, interact with the residues on the central β sheet, and are probably important for anchoring the catalytic helix to the central β sheet. The conserved glutamate side chains are involved in hydrogen bonds with nearby side chain and main chain atoms (Figure 5.6), and are in conformations that are not capable of coordinating a metal ion. It is possible that when in solution or at high temperatures the side chains can rearrange to bind the metal ion. However, the carbonyl oxygen of conserved Gly30 is involved in a hydrogen bond to the strand β1, and is not available for coordinating the metal ion. The carbonyl oxygen of the corresponding glycine in *E. coli* MutT was found to be a ligand to the enzyme bound metal ion (Lin *et al*., 1997). However, in both the $A_{P4}A$ hydrolase and the ADPRase, the corresponding glycine was also found to have a conformation incapable of coordinating the metal ion. It is likely that it is the small size of this conserved glycine residue that is important for the catalytic function of the Nudix enzymes. The side chain of conserved Arg44 in the PA Nudix is well ordered, and its guanidinium group interacts with the side chains of Glu45, Glu36 and Glu48 (Figure 5.6). The putative function of this conserved arginine in the *E. coli* MutT is to orient the side chain of Glu45, which is proposed to function as a general base (Harris *et al*., 2000; Lin *et al*., 1997).

There is a deep narrow cleft connecting the catalytic helix α1 at the front to the N-terminus of helix α2 at the back (Figure 5.2B). Strands β1, β4 and β7 form the floor of the cleft. Strand β6 and part of β7, together with β5a′ and β5b′ from the other subunit, form one wall of the cleft, while β2-loop-β3 and β9 to the N-terminal end of α2 form the other wall.  In the NMR solution structure of the *E. coli* MutT, the nucleoside group of the substrate analog AMPCPP was found to bind in the corresponding cleft (Figure 5.7).  It is possible that in the PA Nudix protein, this cleft also functions in the recognition of the substrate.  The loop between β2 and β3 and the turn between β6 and β7 tilt toward the center of the cleft, partially covering the top of the cleft.  The positions of these residues differ most greatly between the two subunits in the dimer (Figure 5.8), thus indicating their potential flexibility for accommodating the binding of substrates.

***Sulfate-binding sites***

For the Ir derivative and Native-1 crystals, which were obtained in the presence of 50 mM $(NH_4)_2SO_4$, a sulfate ion was found at the N-terminus of the last helix (α2) of each subunit, sitting at one end of the active site cleft (Figures 2B and 7).  The oxygen atoms of the sulfate ion form hydrogen bonds with the main chain nitrogen atoms of Asn124, Val125, Arg126 and Lys127.  In molecule A, both side chains of Arg126 and Lys127 are well ordered, and form hydrogen bonds with the oxygen atoms of the sulfate.  However, in subunit B these two residues have partially disordered side chains.  The sulfate bound to subunit B has a higher B factor (Table 5.2), possibly indicating a lower occupancy.  Asn124 has main chain dihedral angles in the left-handed helix region of the Ramachandran plot ($\phi \sim 70°$ and $\psi \sim 0°$).  This is necessary for the optimal interaction of its backbone amide group with the sulfate ion.  In one crystal form ($P2_1$ space group with two molecules in one asymmetric unit, data not shown), molecule B has a very low occupancy of sulfate, and the electron density maps indicated that there are two conformations for Asn124, with one conformation having regular main chain dihedral angles ($\phi \sim -75°$ and $\psi \sim -25°$).  We note that in *E. coli* MutT, the amino terminal Asn residue of helix 2 (N119) interacts with the 6-oxo group of dGTP (Lin *et al*., 1997).

## Discussion

### Comparison with other Nudix hydrolases of known three-dimensional structure

The PA Nudix protein adopts the same fold as the *E. coli* MutT. However, the two structures are quite different. Figure 5.7 shows the superposition of the Cα traces of *E. coli* MutT onto that of the subunit A of PA Nudix. Only β1, α1, and part of β7 can be aligned. The rmsd over the 30 Cα atoms aligned is 1.48 Å. When structures are aligned based on these secondary structure elements, most of the long β-strands are close in space in the two structures. However, the short β-strands and α2 are shifted relative to each other. The β-strands of the PA Nudix enzyme are highly curved compared to those in the MutT and the $A_{P4}A$ hydrolase. As a result, the active site cleft is narrower and deeper in the PA Nudix structure. This indicates that the PA Nudix hydrolase is likely to have a different substrate from those of MutT or $A_{P4}A$ hydrolase. In the crystal structure of ADPRase complexed with ADP-ribose (Gabelli *et al.*, 2001), the terminal ribose group binds in the active site cleft, while the adenine moiety interacts with the N-terminal domain which is involved in dimer formation through domain swapping. The PA Nudix does not have this N-terminal domain, and thus is unlikely to have the same ADP-sugar hydrolase activity.

In the PA Nudix hydrolase structure, the N-terminal end of the long helix α2 is located at the other end of the active site cleft opposite to the catalytic helix, and a sulfate ion was found binding at the N-terminal end of α2 (Figure 5.2B and 7). The distance between the Nudix box residues to the sulfate binding site is about 20 Å. This sulfate-binding site could potentially be a binding site for a phosphate group on the substrate, although the distance is large. It is possible that this Nudix enzyme hydrolyzes substrates that have a phosphate group away from the scissile bond that can bind to the N-terminal end of the helix α2. An attempt to identify potential substrates for the PA Nudix protein showed that it is inactive against 14 typical substrates for known Nudix enzymes (ADP-ribose, ADP-mannose, ADP-glucose, GDP-mannose, GDP-glucose, UDP-mannose, UDP-glucose, Ap$_2$A, Ap$_3$A, Ap$_4$A, NADH, deamino-NADH, NAD and FAD) (Yang, Wang and Mura, unpublished results). These results suggest that PA Nudix is potentially a novel Nudix enzyme. Alternatively, PA Nudix may be inactive because of the M16L mutation, or because the active form lacks residues 1-15.

### Thermostability

*P. aerophilum* is a hyperthermophile that grows optimally at 100°C (373 K, Volkl *et al*., 1996). The dimeric form of the PA Nudix might contribute to the thermostability of this enzyme. Intersubunit interactions have been proposed to be a major stabilization mechanism for hyperthermophilic proteins (Vieille and Zeikus, 2001). In addition, ion pairs and hydrophobic interactions between subunits make the dimer more resistant to dissociation. Two ion pairs were found between the subunits, Glu79 of one subunit to Arg71 of the other. Another factor that may contribute to the thermostability of this protein is that PA Nudix has very few residues in loops. Thompson and Eisenberg (1999) compared the sequences of about 20 complete genomes and found that thermophilic proteins generally have loop deletions relative to their mesophilic homologs. Sequence alignment between *E. coli* MutT and PA Nudix showed a deletion at loop1 of PA Nudix sequence (Figure 5.5). In addition, the loops are shortened by incorporating more residues in the secondary structure elements. About 75% of residues are in β-strands and α-helices, compared to only about 49% in the *E. coli* MutT protein (Figure 5.5). More β-strands are present in the PA Nudix structure and the β- strands are connected mostly by tight β-turns.

## Figure legends

**Figure 5.1: A section of the initial electron density map superimposed on the final coordinates.** The map was calculated at 2.5 Å with phases after 2-fold NCS averaging, and contoured at 1.3 σ. The figure was generated with the program O and rendered with POV-RAY.

**Figure 5.2: Ribbon diagrams of the PA Nudix structure.** Panel A shows the dimer with the 2-fold NCS axis perpendicular to the paper plane. Each subunit is composed of a long, highly twisted central β sheet and a small sub-sheet connected through strand β2, sandwiched between two α helices. Each subunit contributes two β strands (β5a and β5b) to the β sheet of other subunit to form the dimer interface. Subunit A (on the left) has a long C-terminal helix α2 with the 8 amino acid linker at the C-terminal end extending the helix, while subunit B has a shorter helix α2 and its C-terminus and the linker fold back to form a more compact structure. Panel B is a near 180° rotation of panel A along a horizontal axis. A cleft formed by the β strands connects helix α1 in the front to the N-terminal end of helix α2 in the back. The conserved arginine and glutamate side chains, as well as two sulfate ions found at the N-terminal end of α2 of each subunit are shown as sticks. The figure was generated with the program RIBBONS (Carson, 1997).

**Figure 5.3: The topology diagram of the PA Nudix structure.** The strand β1 is connected by a type II β turn to β2, which forms hydrogen bonds with the C-terminal part of β1 and then makes a curve to form hydrogen bonds to β3, making a sub-sheet with strands β8 and β9. A proline makes a kink at the end of β3, allowing β4 to have hydrogen bonds to the N-terminal half of β1. A small loop connects β4 to helix α1. Strands β5 and β6 together form an almost continuous strand antiparallel to the longest strand β7. Strands β5a and β5b go to the other subunit and form part of its central β sheet, while strands β5a′ and β5b′ are from the other subunit. The last helix α2 is connected to β9 by a type I β turn. There are more β strands in this protein structure than in the *E. coli* MutT. The shaded β strands are the strands that are present in the *E. coli* MutT structure.

**Figure 5.4: A comparison of the native Nudix crystal structures in the tetrameric *P*2₁2₁2₁ (Native-1) and dimeric *P*2₁ (Native-2) crystal forms.** In panel A the Cα traces of the two dimers in the asymmetric unit of the Native-2 form (blue and lavender) are superimposed in two orthogonal views, along with the single dimer from the asymmetric unit of the Native-1 form (red). These structures superimpose very well (~0.5 Å rmsd for all three unique pairs of dimers), and it can be seen that the only significant differences lie in the C-terminal helices (which are extended in the Native-2 structure). The extended helices allow two Nudix dimers to crystallize as a tetramer in the *P*2₁ form – shown as Cα traces in panel B. The contents of one unit cell are drawn as viewed down the crystallographic 2₁ axis, with each of the two tetramers colored various hues of red or blue. Note the inter-dimer helix-helix contacts.

**Figure 5.5: Sequence alignment between the *E. coli* MutT and the PA Nudix.** The sequences were aligned with CLUSTAL-W (Thompson *et al.*, 1994) and the figure was generated by ALSCRIPT (Barton, 1993). The sequence of the PA Nudix protein also strands β8 and β9. A proline makes a kink at the end of β3, allowing β4 to have hydrogen bonds to the N-terminal half of β1. A small loop connects β4 to helix α1. Strands β5 and β6 together form an almost continuous strand antiparallel to the longest strand β7. Strands β5a and β5b go to the other subunit and form part of its central β sheet, while strands β5a′ and β5b′ are from the other subunit. The last helix α2 is connected to β9 by a type I β turn. There are more β strands in this protein structure than in the *E. coli* MutT. The shaded β strands are the strands that are present in the *E. coli* MutT structure.

**Figure 5.4: A comparison of the native Nudix crystal structures in the tetrameric $P2_12_12_1$ (Native-1) and dimeric $P2_1$ (Native-2) crystal forms.** In panel A the Cα traces of the two dimers in the asymmetric unit of the Native-2 form (blue and lavender) are superimposed in two orthogonal views, along with the single dimer from the asymmetric unit of the Native-1 form (red). These structures superimpose very well (~0.5 Å rmsd for all three unique pairs of dimers), and it can be seen that the only significant differences lie in the C-terminal helices (which are extended in the Native-2 structure). The extended helices allow two Nudix dimers to crystallize as a tetramer in the $P2_1$ form – shown as Cα traces in panel B. The contents of one unit cell are drawn as viewed down the crystallographic $2_1$ axis, with each of the two tetramers colored various hues of red or blue. Note the inter-dimer helix-helix contacts.

**Figure 5.5: Sequence alignment between the *E. coli* MutT and the PA Nudix.** The sequences were aligned with CLUSTAL-W (Thompson *et al.*, 1994) and the figure was generated by ALSCRIPT (Barton, 1993). The sequence of the PA Nudix protein also includes an 8 amino acid linker at the C-terminal end. The identical residues were highlighted in black. Helices and β strands of the two structures are marked with cylinders and arrows, respectively. The two sequences align well only around the Nudix box signature motif (boxed residues). The PA enzyme has more β strands and much shorter loops.

**Figure 5.6: Conserved Nudix box residues.** The structure around the active site is shown as a ribbon representation with residues in strand β4 and conserved side chains of the catalytic helix shown as sticks. The semi conserved hydrophobic side chains, Phe46 and Ile52, are also shown. All the conserved side chains are on one side of helix α1, while the side chains of Phe46 and Ile52 are buried and face the central β sheet. Hydrogen bonds from the conserved side chains are shown in red dashed lines.

**Figure 5.7: Structure superposition between the subunit A of PA Nudix and the *E. coli* MutT.** The *E. coli* MutT structure (PDB code 1tum, Lin *et al*., 1997) is shown in red and its bound AMPCPP shown in ball-and-stick. The two structures were aligned based on β1, α1 and part of β7 of the PA structure. They have similar folds and the conserved residues in the Nudix motif (shown as black dots) are at similar positions. However, away from helix α1 the two structures are quite different. The active site cleft is much narrower and deeper in the PA Nudix structure. The adenosine group of the AMPCPP in the *E. coli* MutT structure binds in the active site cleft. A sulfate ion was found at the N-terminus of α2 of the PA protein (shown as ball-and-stick), indicating a possible binding site for a phosphate group. The figure was generated with MOLSCRIPT (Kraulis, 1991).

**Figure 5.8: Rmsd in Å of the main chain atomic positions for the superimposed subunits of the Native-1 dimer.** The solid line is from the structure of the iridium derivative crystal and the dashed line is from the native crystal. Higher rmsd was observed for residues 18 – 25, 79 – 90 and 64 – 69, which form the walls of the active site cleft, indicating their flexibility. The peak around residue 32 is due to partially disorder of the loop (higher B factors), while peaks around residues 110 – 120 and after 135 are due to differences in crystal contacts of the two subunits. A plot of the rmsd atomic positions between the dimers of the two different crystals (not shown) also gives the same peaks at residues 18 – 25, 79 – 90 and 64 – 69, but with a smaller scale.

**Table 5.1. Statistics of data collection for PA Nudix.**

| | Ir derivative[a] | Native-1[a] | Native-2[b] |
|---|---|---|---|
| Space group | $P2_12_12_1$ | $P2_12_12_1$ | $P2_1$ |
| Cell parameters (Å) | a = 52.42<br>b = 71.66<br>c = 85.76 | a = 52.71<br>b = 72.61<br>c = 85.18 | a = 52.71<br>b = 72.61<br>c = 85.18<br>β = 99.96° |
| Resolution (Å) | 100 – 1.85 | 100 – 1.8 | 90 – 2.4 |
| Unique reflections<br>(Observed reflections) | 25455<br>(348337) | 30654<br>(398012) | 23707<br>(84048) |
| Completeness (last bin)[c] (%) | 89.3 (52.0) | 98.8 (96.4) | 98.2 (97.0) |
| Anomalous completeness (last bin)[c] (%)[d] | 87.6 (69.8) | | |
| I/σ (last bin)[c] | 18.7 (3.2) | 18.4 (3.2) | 14.1 (3.0) |
| $R_{sym}$[e] (last bin)[c] | 0.076 (0.470) | 0.080 (0.443) | 0.093 (0.465) |
| # subunit per asymmetric unit | 2 | 2 | 4 |
| X-ray wavelength (Å) | 1.1053 | 1.10 | 1.5418 |

**a.** Data for the Ir derivative and Native-1 were collected at the synchrotron. The cell dimensions for the two data sets are similar. However, the two crystals have different packing, and therefore are not isomorphous.

**b.** Native-2 crystals were obtained from drops equilibrated against 100 mM MES pH 6.2 and 15% MPD. Data were collected on an RAXIS IV++ detector.

**c.** The highest resolution bin is from 1.92 to 1.85 Å for the Ir derivative data and from 1.86 to 1.80 Å for the native data.

**d.** Anomalous data were processed to 2.0 Å, with the last bin from 2.07 Å to 2.0 Å.

**e.** $R_{sym} = \Sigma|(I_{hkl} - <I_{hkl}>)|/\Sigma I_{hkl}$, where $<I_{hkl}>$ is the average of $I_{hkl}$ over all symmetry equivalents.

**Table 5.2. Statistics for atomic refinement of PA Nudix.**

| | Ir derivative | Native-1 | Native-2 |
|---|---|---|---|
| Resolution range for refinement (Å) | 20 – 1.85 | 20 – 1.8 | 20 – 2.4 |
| # protein atoms | 2374 | 2416 | 4732 |
| $<B>$-factor protein atoms (Å$^2$)$^a$ | 26.1 | 19.65 | 31.9 |
| # other molecules ($<B>$-factors) | 2 SO$_4$ (25.2, 36.6); 4 Ir (34.7); 2 Ni (74.4); 3 glycerol (37.1); 5 acetate (46.2); 149 H$_2$O (34.4) | 2 SO$_4$ (21.1, 42.6); 2 Ni (66.1); 6 glycerol (38.8); 2 acetate (30.9); 244 H$_2$O (29.6) | 6 MPD (64.8); 173 H$_2$O (30.9) |
| Rmsd bond length (Å) | 0.017 | 0.019 | 0.013 |
| Rmsd bond angle (Å) | 1.8 | 1.8 | 1.8 |
| $R_{\text{free}}$[b] | 0.218 | 0.219 | 0.274 |
| $R_{\text{work}}$[b] | 0.182 | 0.183 | 0.190 |

**a.** The Ir derivative crystal has a higher mosaicity (about 1.2°) than the native crystal (about 0.7°).

**b.** R-factors were calculated using data in the resolution range for refinement without a σ cutoff. $R_{\text{free}}$ was calculated with a set of data (8%) never used in the refinement. $R_{\text{work}}$ was calculated against the data used in the refinement.

**Figure 5.1: A section of the initial electron density map superimposed on the final coordinates.**

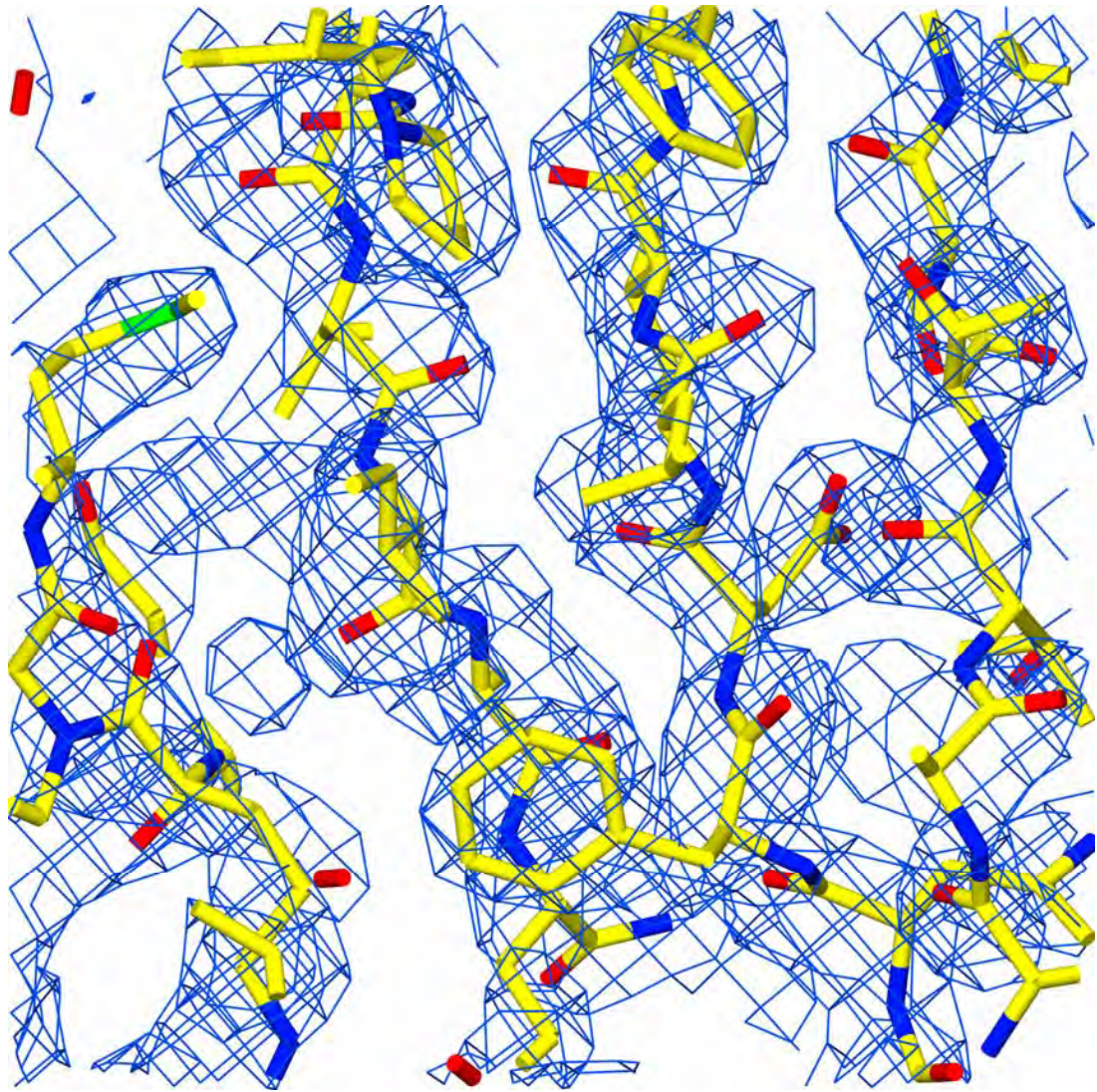

**Figure 5.2: Ribbon diagrams of the PA Nudix structure.**

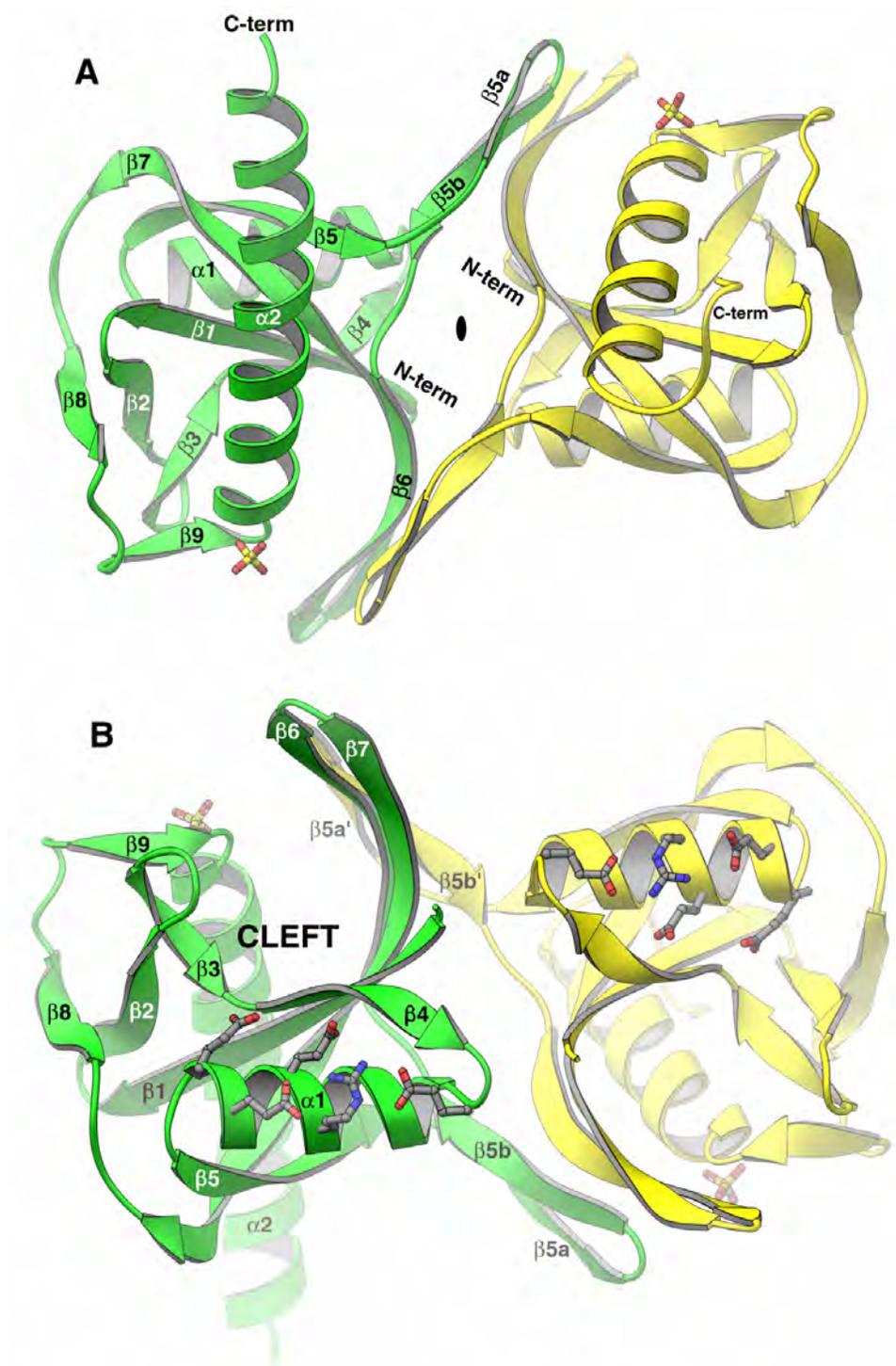

**Figure 5.3: The topology of the PA Nudix structure.**

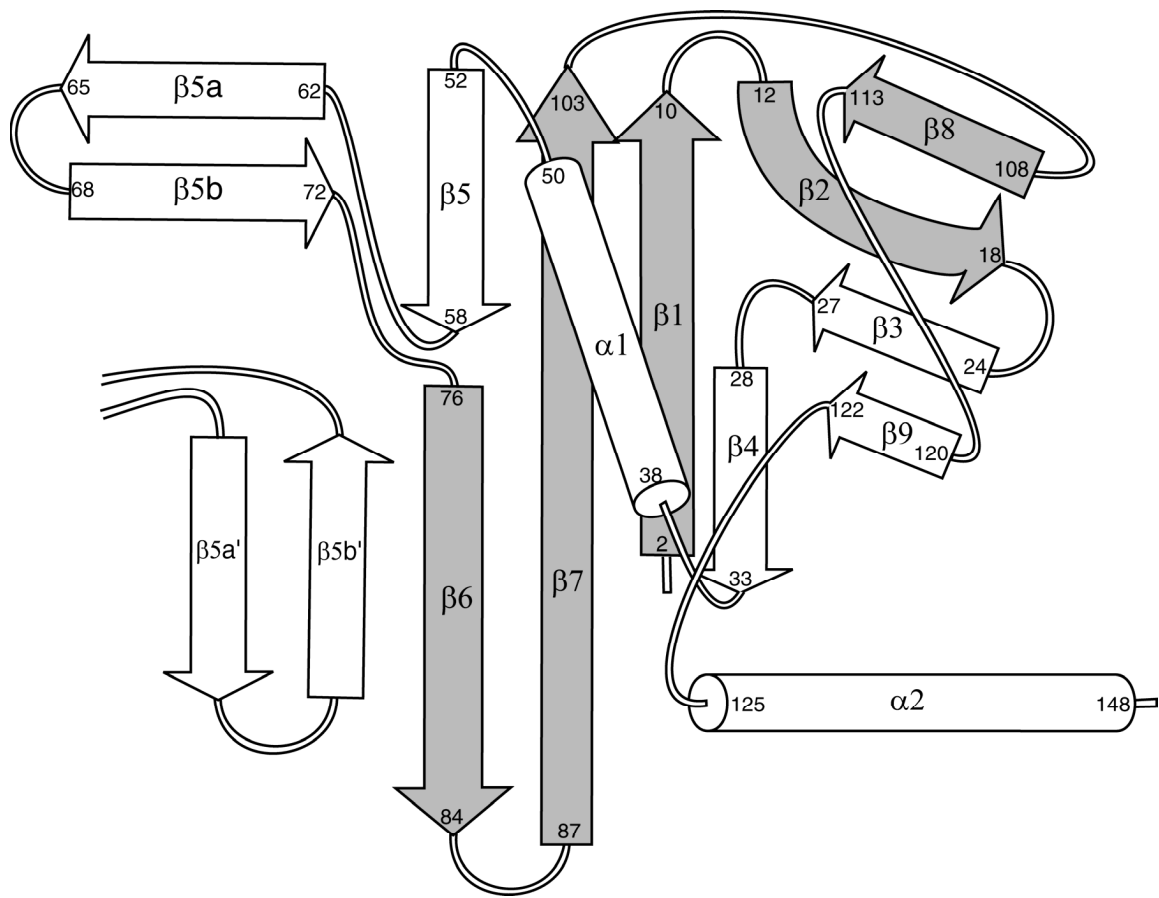

**Figure 5.4: A comparison of the Nudix crystal structures in the *P*2$_1$2$_1$2$_1$ (Native-1) and *P*2$_1$ (Native-2) crystal forms.**

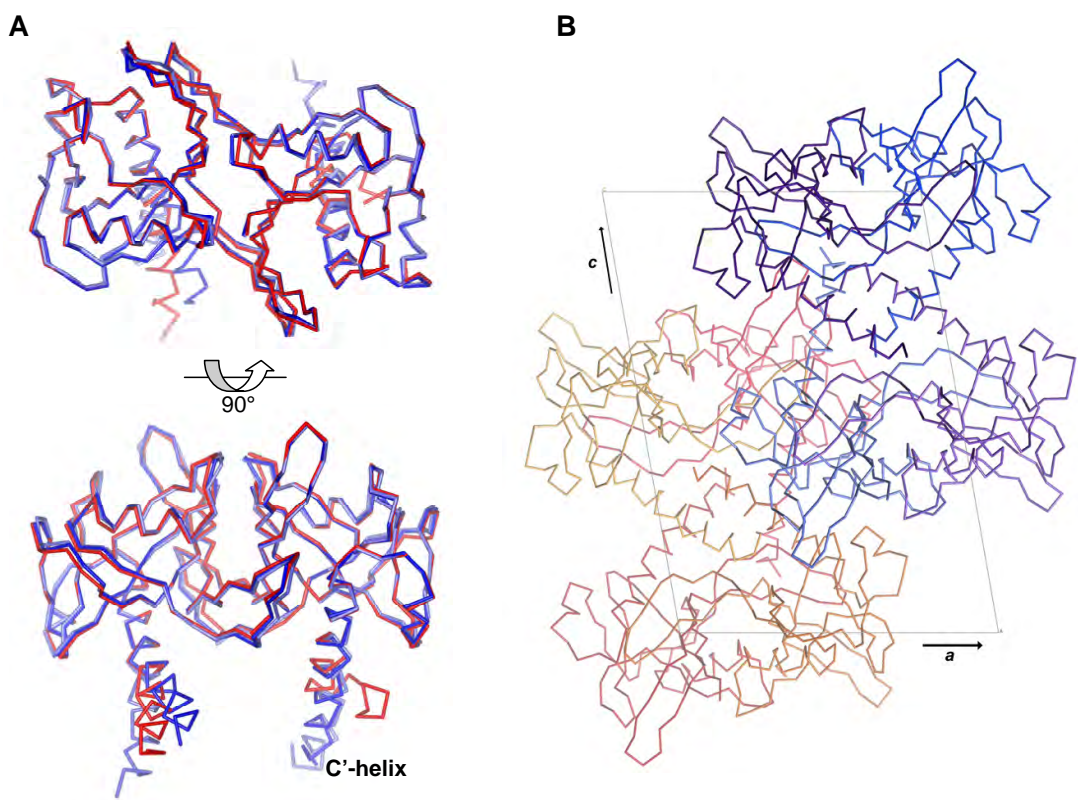

**Figure 5.5: Sequence alignment between the *E. coli* MutT and PA Nudix.**

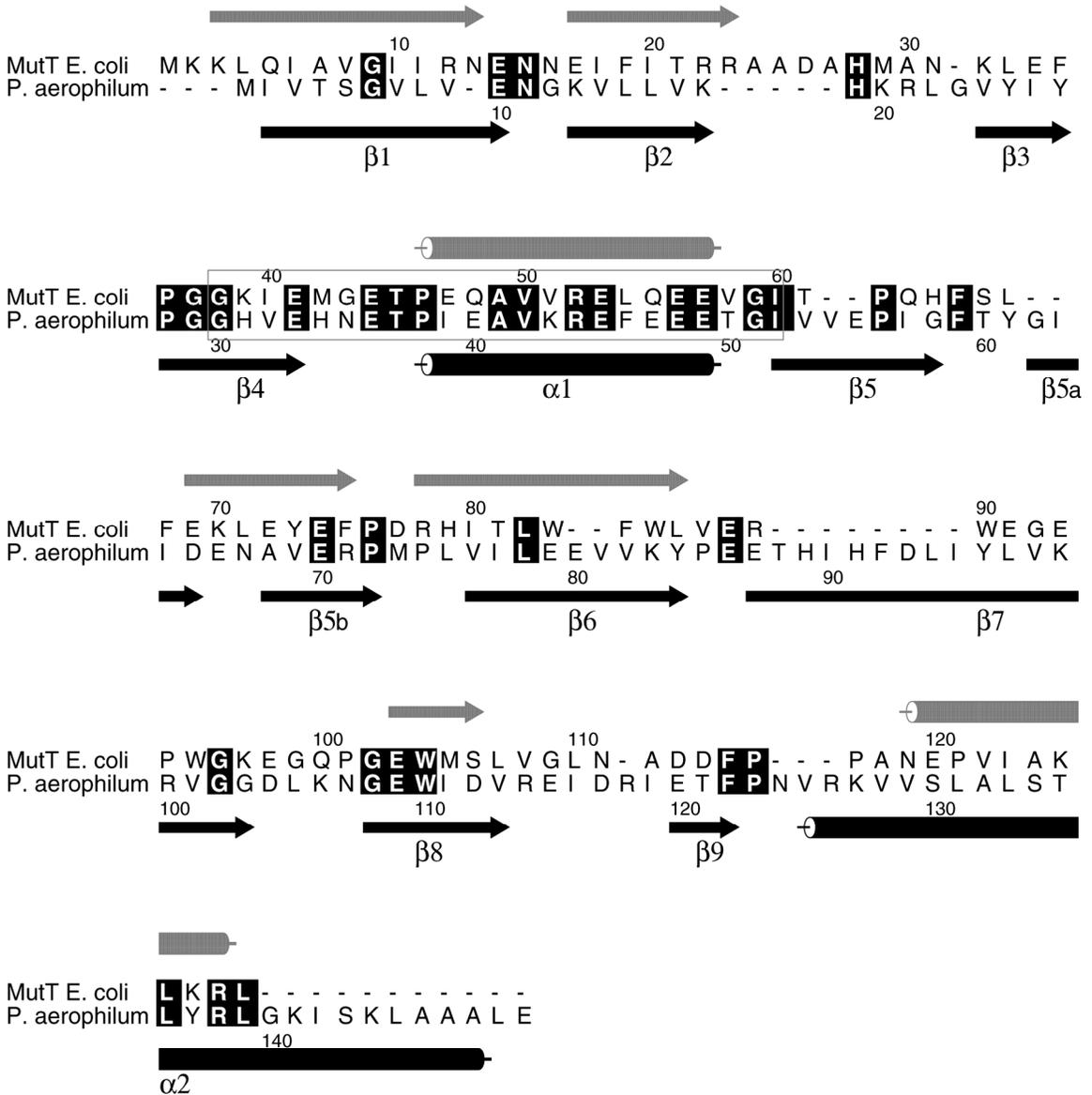

**Figure 5.6: Conserved Nudix box residues.**

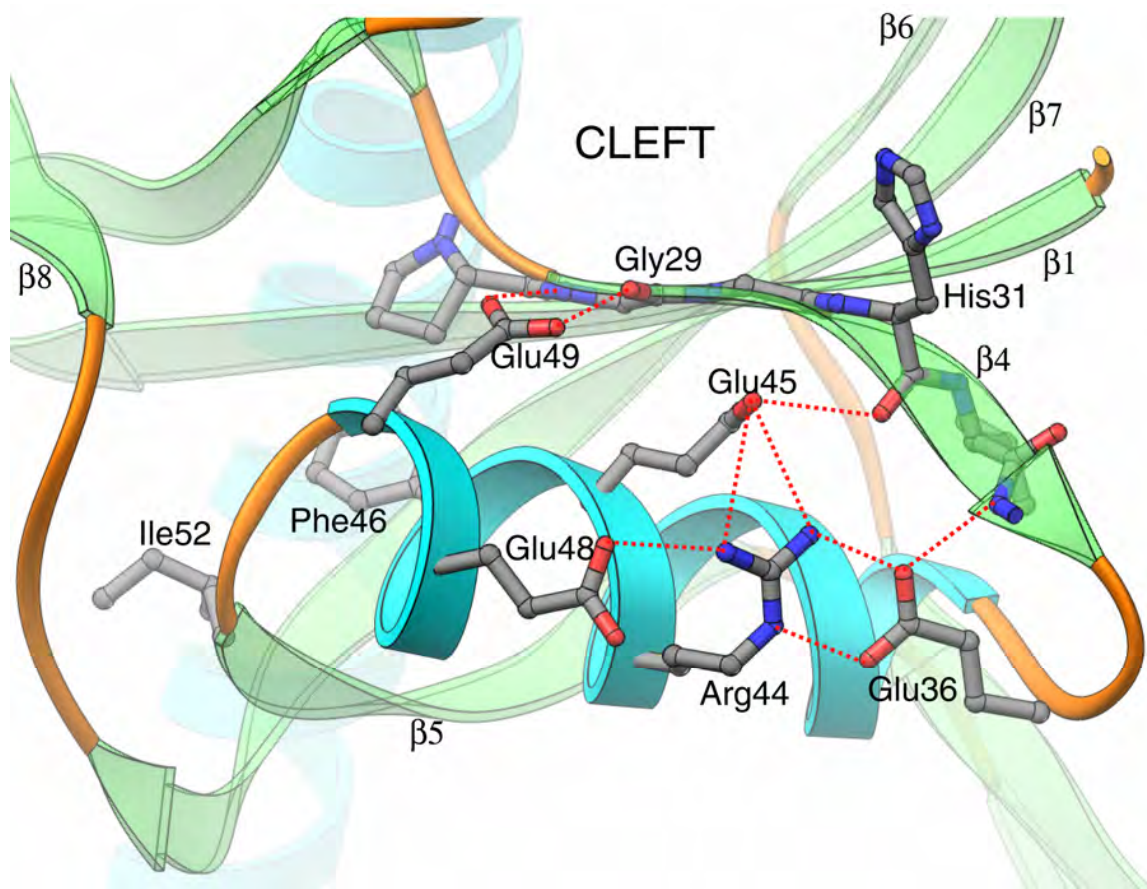

**Figure 5.7: Structure superposition between the subunit _A_ of PA Nudix and the _E. coli_ MutT.**

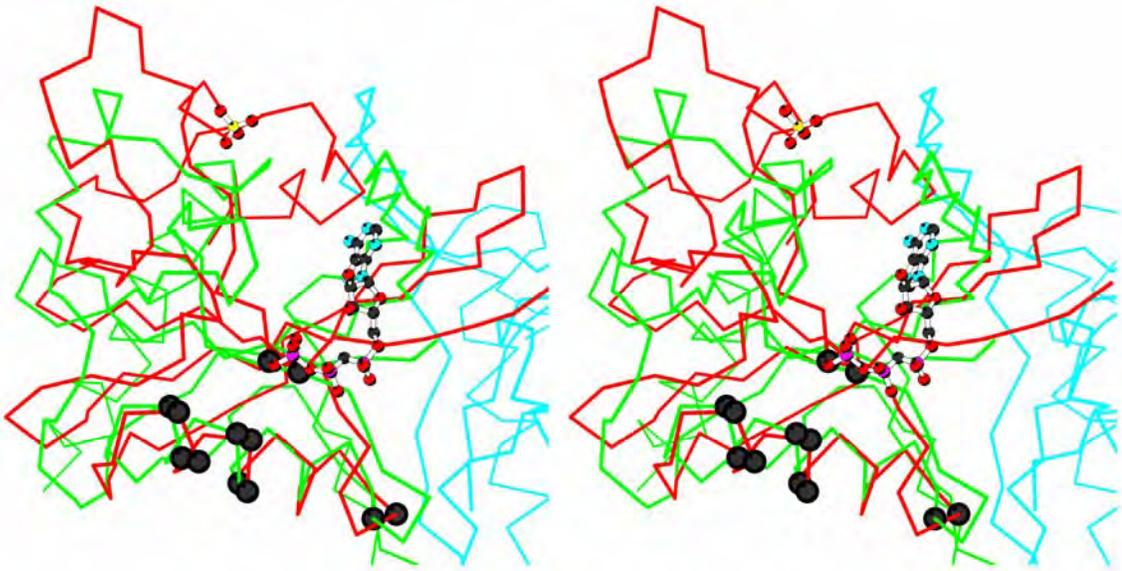

**Figure 5.8: RMSD in Å of the main-chain atomic positions for the superimposed subunits of the Native-1 dimer.**

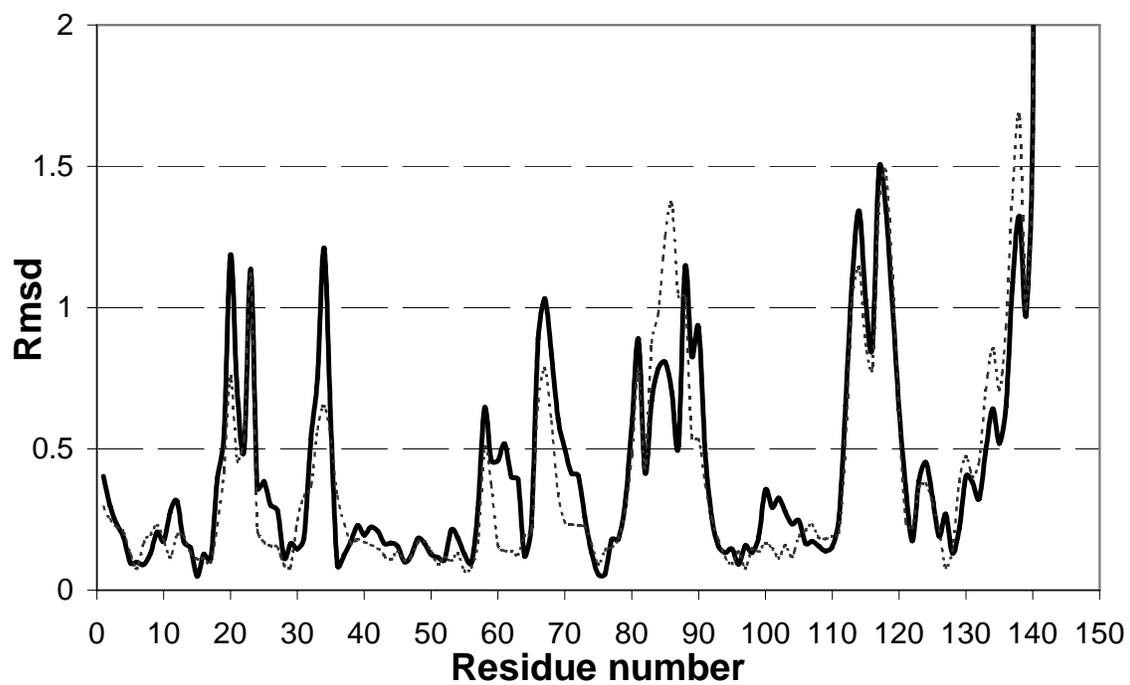

## Appendix to the Dissertation:

## Various UNIX and Perl scripts used in this research

A brief description of each script:

***make14mer.com*** – generates a 14-mer from a 7-mer input PDB file

***racc.pl*** –  rare codon calculator, to predict if rare Arg, Leu, Ile, or Pro codons may be a problem in protein over-expression in *E. coli*

***make_homology_model.pl*** –  to create molecular replacement homology models from an input sequence alignment and PDB file

***scripted_glrf.sh***, ***alter.pl***, **and** ***process_bigrun.pl*** –
to calculate cross-rotation functions that are systematically varied over the integration radius and resolution limits

***map_conservation_to_B.pl*** –  to map sequence conservation values to the B-factor fields of a PDB file

***symm.csh*** **and** ***write_pdbset.pl*** –
calculates all non-crystallographic symmetry transformations and expands the monomeric subunit into the multimer (matrix of all pairwise RMSDs between subunits of the multimer is output as well)

```
#!/bin/csh -f

# This c-shell script builds a 14-mer model given the most recent 7-mer structure
# available
#
# The input consists of:
#
# (1) A PDB file containing most recent 7-mer structure ("hept_exp_from_mono6.pdb")
# (2) A PDB file containing the atoms from which the center of gravity and rotational
#     axes will be calculated ("cys7sg.pdb")
# (3) Command line-input values for the desired translation distance (in Angstroms)
#     and rotation angle (in degrees)
# (4) NOTE: The input PDB files must contain the "CRYST" "SCALE" and "ORIGX" lines
#           The chains must be labelled by distinct chainid's, starting from 'A'.
#
# The output consists of:
#
# (1) A PDB file containing the input 7-mer translated so that its desired COG lies at
#     the origin ("translate1.pdb")
# (2) A PDB file of the rotated heptamer ("lsqkab.pdb")
# (3) A PDB file of the composite 14-mer ("make14mer_output.pdb")
# (4) plus some log files if the last clean-up step is commented out

##### INITIALIZE VARIABLES TO NULL #####

set v1; set v2; set v3
set norm_v1; set norm_v2; set norm_v3
set count
set dist
set angle

##### ASK FOR THE FINAL TRANSLATION DISTANCE (IN ANGSTROMS) IF NOT INPUT ON COMMAND LINE
#####
##### ASK FOR THE FINAL DESIRED ROTATION ANGLE (IN DEGREES) IF NOT INPUT ON COMMAND LINE
#####

if ( ${1} == "" ) then
   echo ""; echo "Program needs command line-input TRANSLATION DISTANCE \!\!\!"; echo ""
   kill -9 $$
else set dist = $argv[1]; echo ""; echo "translation distance will be: $dist angstroms"
endif

if ( ${2} == "" ) then
  echo ""; echo "Program needs a command line-input ROTATION ANGLE \!\!\!"; echo ""
   kill -9 $$
else set angle = $argv[2]; echo ""; echo "rotation angle will be: $angle degrees"
endif

##### CALCULATE THE CENTER OF GRAVITY FOR THE SG atoms OF THE CYS RING OF THE 7-MER #####

pdbset xyzin cys7sg.pdb << eof-1 > cog.log
SPACEGROUP C2
COM
eof-1

##### TRANSLATE INPUT 7-MER SO THAT CENTER OF GRAVITY OF CYS SG LIES AT ORIGIN #####

echo "#\!/bin/csh -f" > translate1.com
echo "pdbset XYZIN almost_pruned.pdb XYZOUT translate1.pdb << eof-2 > translate1.log" >>
translate1.com
echo "" >> translate1.com
echo "TRANSFORM 1 0 0 -" >> translate1.com
echo "          0 1 0 -" >> translate1.com
echo "          0 0 1 -" >> translate1.com
```

```
echo -n "              " >> translate1.com
echo -n "-"`(grep "    Center of Mass:" cog.log) | awk '{print $4}' ` >> translate1.com
echo -n "  " >> translate1.com
echo -n "-"`(grep "    Center of Mass:" cog.log) | awk '{print $5}' ` >> translate1.com
echo -n "  " >> translate1.com
echo -n "-"`(grep "    Center of Mass:" cog.log) | awk '{print $6}' ` >> translate1.com
echo -n "  " >> translate1.com
echo "" >> translate1.com
echo "SPACEGROUP C2" >> translate1.com
echo "eof-2" >> translate1.com

chmod 755 translate1.com
translate1.com

##### USE AL_MOLEMAN TO SPLIT THE 7-MER INTO MONOMERS FOR UPCOMING USE IN ALIGN_V2 #####

al_moleman << eof > al_moleman.log

translate1.pdb
split
split_
quit
eof

##### CALCULATE AND NORMALIZE THE VECTOR ABOUT WHICH ROTATION WILL BE PERFORMED #####

set v1 = `grep SG translate1.pdb | head -1 | awk '{print $7}' `
set v2 = `grep SG translate1.pdb | head -1 | awk '{print $8}' `
set v3 = `grep SG translate1.pdb | head -1 | awk '{print $9}' `

echo ""
echo "unnormalized rotation axis = "$v1 $v2 $v3

set count=0

foreach z ($v1 $v2 $v3)
set count=`expr $count + 1`
echo "scale=5" > bc.script
echo "l=sqrt(`echo $v1`^2 + `echo $v2`^2 + `echo $v3`^2)" >> bc.script
echo "i=$z/l" >> bc.script
echo "i" >> bc.script
echo "quit" >> bc.script

if ($count == 1) then
set norm_v1 = `bc bc.script`
endif
if ($count == 2) then
set norm_v2 = `bc bc.script`
endif
if ($count == 3) then
set norm_v3 = `bc bc.script`
endif

end

echo ""
echo "normalized rotation axis = "$norm_v1  $norm_v2  $norm_v3; echo " ";

##### CALCULATE THE VECTOR ALONG WHICH THE TRANSLATION WILL TAKE PLACE
##### (i.e. THE 7-FOLD ROTATION AXIS)
##### USING ALIGN_V2 PROGRAM. NOTE THAT IT'S ALREADY NORMALIZED IN THE ALIGN_V2 OUTPUT
#####

align_v2 << eof > split.log

1
```

```
split_a.pdb

PDB
split_b.pdb

PDB
AUTO
split_a_to_b.rot
split_a_to_b.show
stop
eof

set norm_t1 = `grep direction split.log | awk '{print $4}' `
set norm_t2 = `grep direction split.log | awk '{print $5}' `
set norm_t3 = `grep direction split.log | awk '{print $6}' `

echo
echo -n "normalized vector along which translation will be performed (i.e. 7-fold axis) =
"
echo $norm_t1 $norm_t2 $norm_t3

##### NOW SCALE THE TO-BE-APPLIED TRANSLATION VECTOR BY THE COMMAND LINE-INPUT AMOUNT
#####

bc << eof > temp1
-`echo $dist` * $norm_t1
quit
eof
set use_t1 = `awk '{print $1}' temp1`

bc << eof > temp2
-`echo $dist` * $norm_t2
quit
eof
set use_t2 = `awk '{print $1}' temp2`

bc << eof > temp3
-`echo $dist` * $norm_t3
quit
eof
set use_t3 = `awk '{print $1}' temp3`

echo ""
echo -n "the scaled translation vector that will be applied = "
echo $use_t1 $use_t2 $use_t3; echo " "; echo " ";

##### PERFORM DESIRED ROTATION & TRANSLATION ALONG CALCULATED AXES VIA DIRECTION COSINES
#####

echo "#\!/bin/csh -f" > lsqkab.com
echo "lsqkab XYZIN2 translate1.pdb XYZOUT lsqkab.pdb << eof-z > lsqkab.log" >> lsqkab.com
echo "ROTATE DCS  "$norm_v1"  "$norm_v2"  "$norm_v3"   "${angle}  >> lsqkab.com
echo "TRANSLATE  "$use_t1"  "$use_t2"   "$use_t3  >> lsqkab.com
echo "END" >> lsqkab.com
echo "eof-z" >> lsqkab.com

chmod 755 lsqkab.com
lsqkab.com

########### END OF SCRIPTS, so...#####
########### clean up stuff  ##########

cat translate1.pdb lsqkab.pdb > make14mer_output.pdb

rm split* temp* cog.log XYZOUT crap* al_moleman.log
rm translate1.com translate1.log bc.script lsqkab.com lsqkab.log
```

```perl
#!/joule2/programs/bin/perl

# Calculates the number of rare (for E. coli) Arg, Leu, Ile, and Pro codons in input
# nucleic acid sequence file, and number of consecutive rare Arg, Leu, Ile, and Pro
# codons if > 1

# Ask for the nucleic acid sequence file name (if it was not entered on the command line)
# and open it:

if ($ARGV[0] eq undef)
 {
  print "Enter the name of the nucleic acid sequence file (in GCG format):";
  chomp ($naseq = <STDIN>);
  if ($naseq =~ /\./)
        {$output_file = $naseq . "..racc";
         $output_file =~ s/\..*?\.//g;}
  else {$output_file = $naseq . ".racc";}
  print "\nA file named \"$output_file\" will contain the results that follow\n\n";
  open (OUTFILE, ">$output_file") || die "Can't open \"$output_file\" for writing\n\n";
 }

else {chomp ($naseq = $ARGV[0]);}

open (NASEQ, $naseq) || die "Can't open sequence file: $naseq\n";

# check that the NA sequence file is in GCG format:

chomp ($okornot = <NASEQ>);
unless ($okornot =~ /!!NA_SEQUENCE\b/)
{
 print "\nThe nucleic acid sequence file \"$naseq\" is not in GCG format!\n";
 if ($ARGV[0] eq undef)
    {print OUTFILE "\nNucleic acid sequence file \"$naseq\" is not in GCG format!\n";}
}

# read in and process the NA sequence:

@seq = <NASEQ>;

for ($i = 0; $i <= $#seq; $i++)
{
   if (($seq[$i] =~ /[bdefhijklmnopqrsuvwxyz]/i) |
       !($seq[$i] =~ /\d/) | !($seq[$i] =~ /[a-z]/i))
         {$seq[$i] = "";}
}

$whole_seq = join("", @seq);
$whole_seq =~ s/[^a-z]//ig;
#$whole_seq =~ s/(\d)|(\s)//g;    an almost equivalent way to do it

# calculate properties of interest:

$seq_length = rindex($whole_seq, /\w/);
print "\nFor the following sequence from the input file \"$naseq\":\n";
print "\n$whole_seq\n";
print "\nThe length is: $seq_length nucleotides\n\n";

if ($ARGV[0] eq undef)
{
 print OUTFILE "For following sequence from input file \"$naseq\":\n";
 print OUTFILE "\n$whole_seq\n";
 print OUTFILE "\nThe length is: $seq_length nucleotides\n";
}
```

```perl
#warn user if DNA sequence isn't an ORF (i.e. multiple of 3 in length)

unless ($seq_length % 3 == 0)
{
 warn "\nNote: the sequence file \"$naseq\" is not an ORF\n";
 if ($ARGV[0] eq undef)
    {print OUTFILE "\nNote: the sequence file \"$naseq\" is not an ORF\n";}
}

# initialize rac hash variables:

%rac = ();

$rac{rac_1x} = 0;                       # number of single rare Arg codons
$rac{rac_2x} = 0;                       # number of tandem rare Arg codon repeats
$rac{rac_3x} = 0;                       # number of triple rare Arg codon repeats

@rac{10, 11} = ("(aga)|(agg)|(cga)", 0);    # no. single rare Arg codons
@rac{10, 12} = ("(agg)|(aga)|(cga)", 0);    # no. tandem rare Arg codon rpts
@rac{10, 13} = ("(agg)|(aga)|(cga)", 0);    # no. triple rare Arg codon rpts

@rac{20, 21} = ("cta", 0);              # number of single rare Leu codons
@rac{20, 22} = ("cta", 0);              # number of tandem rare Leu codon repeats
@rac{20, 23} = ("cta", 0);              # number of triple rare Leu codon repeats

@rac{30, 31} = ("ata", 0);              # number of single rare Ile codons
@rac{30, 32} = ("ata", 0);              # number of tandem rare Ile codon repeats
@rac{30, 33} = ("ata", 0);              # number of triple rare Ile codon repeats

@rac{40, 41} = ("ccc", 0);              # number of single rare Pro codons
@rac{40, 42} = ("ccc", 0);              # number of tandem rare Pro codon repeats
@rac{40, 43} = ("ccc", 0);              # number of triple rare Pro codon repeats

# parse the sequence from $whole_seq into codons and count away:

for ($i = 1; $i <= ($seq_length/3); $i++)
{
  $codon[$i-1] = substr($whole_seq, ($i-1)*3, 3);

 if ($codon[$i-1] =~ /(agg)|(aga)|(cga)/i)
 { $rac{rac_1x}++; push(@rac_1x_positions, $i); }

 if ($i >= 2)
 {
 if (($codon[$i-1] =~ /(agg)|(aga)|(cga)/i) & ($codon[$i-2] =~ /(agg)|(aga)|(cga)/i))
  { $rac{rac_2x}++; push(@rac_2x_positions, $i); }
 }

 if ($i >= 3)
 {if (($codon[$i-1] =~ /(agg)|(aga)|(cga)/i) & ($codon[$i-2] =~ /(agg)|(aga)|(cga)/i)
                                 & ($codon[$i-3] =~ /(agg)|(aga)|(cga)/i))
 { $rac{rac_3x}++; push(@rac_3x_positions, $i); }
 }
}

######### now repeat the same for Leu, Ile, and Pro codons:   ##########

for ($a = 10; $a <= 40; $a += 10)
{

for ($i = 1; $i <= ($seq_length/3); $i++)

{ $a1 = $a + 1; $a2 = $a + 2; $a3 = $a + 3;

 if ($codon[$i-1] =~ /($rac{$a})/i)
```

```perl
   { $rac{$a1}++;  push(@posit_1x, $i);
   }

   if ($i >= 2)
   {
   if (($codon[$i-1] =~ /($rac{$a})/i) & ($codon[$i-2] =~ /($rac{$a})/i))
    { $rac{$a2}++;  push(@posit_2x, $i);
    }
   }

   if ($i >= 3)
   {if (($codon[$i-1] =~ /($rac{$a})/i) & ($codon[$i-2] =~ /($rac{$a})/i)
                                       & ($codon[$i-3] =~ /($rac{$a})/i))
    { $rac{$a3}++;  push(@posit_3x, $i);
    }
   }
  }

push(@posit_1x, ("$rac{$a}" . $a1/10));
push(@posit_2x, ("$rac{$a}" . $a2/10));
push(@posit_3x, ("$rac{$a}" . $a3/10));

}

for ($i = 1; $i <= ($seq_length/3); $i++)
{$codon[$i-1] = $codon[$i-1] . " ";}

for ($i = 0; $i <= ($rac{rac_1x}-2); $i++)
        {$rac_1x_positions[$i] = $rac_1x_positions[$i] . ", ";}
for ($i = 0; $i <= ($rac{rac_2x}-2); $i++)
        {$rac_2x_positions[$i] = $rac_2x_positions[$i] . ", ";}
for ($i = 0; $i <= ($rac{rac_3x}-2); $i++)
        {$rac_3x_positions[$i] = $rac_3x_positions[$i] . ", ";}

print "Number of total single rare Arg codons: $rac{rac_1x}\n";
if ($rac{rac_1x} > 0)
{print "occurring at codons: @rac_1x_positions \n";}

print "Number of tandem rare Arg codon double repeats: $rac{rac_2x}\n";
if ($rac{rac_2x} > 0)
{print "occurring at codons:  @rac_2x_positions \n";}

print "Number of tandem rare Arg codon triple repeats: $rac{rac_3x}\n\n";
if ($rac{rac_3x} > 0)
{print "occurring at codons:  @rac_3x_positions \n\n";}

###### preliminary output formatting of rare codons for Leu, Ile, and Pro: ######

print "Too lazy to beautify this new part right now...Results are in order for
Arginine, Leucine, Isoleucine, and Proline, respectively (delimited by numbers
1.1, 2.1, 3.1, 4.1 for singles; 1.2, 2.2, 3.2, 4.2 for doubles; etc.).\n";

print "\nSingle rare codons at positions:\n@posit_1x \n";
print "\nDouble rare codons at positions:\n@posit_2x \n";
print "\nTriple rare codons at positions:\n@posit_3x \n";

###### END OF RaCC SCRIPT ######
```

```
#!/joule2/programs/bin/perl

# This program begins with a sequence alignment, of the sort produced by the GCG
# "bestfit" program. It tabulates the positions that are strongly conserved.
# Then, all strongly conserved residues (i.e. paired by either ':' or '|' in the
# GCG output), ALAs, and GLYs are left alone, while all other residues are
# truncated to ALA as in the "make_poly_ala.pl" program. Extra residues in
# the homology model probe protein that correspond to gaps in the actual protein
# of interest will be deleted.
#
# NOTE: It is assumed that the input PDB file has been edited to contain only 1
# copy of each unique chain that exists in the sequence alignment; and that the
# beginning and ending residues in the PDB file correspond to those at the beginning
# and end of the sequence aligned homology model protein (even though all other
# residues in the PDB file will be unaltered).

# input all of the file names, check that they're accessible

print "\nEnter the name of the GCG .pair sequence alignment file: ";
chomp ($sa_input = <STDIN>);

open (SA_INFILE, $sa_input) || die "Can't open sequence alignment file
\"$sa_input\"\n\n";

print "\nEnter 't' if the homologous protein of known structure is the top
sequence or 'b' if it's the bottom sequence of the sequence alignment: ";
chomp ($top_or_bot = <STDIN>);

print "\nEnter the name of the PDB input file: ";
chomp ($pdb_input = <STDIN>);
if ($pdb_input =~ /\./) {$pdb_output = $pdb_input; $pdb_output =~
s/\./_HOMOLOGY_MODEL\./;}
else {$pdb_output = $pdb_input . "_HOMOLOGY_MODEL\.pdb";}

# process the .pair file to extract conserved residues and those that should be
# deleted

@strings = <SA_INFILE>;

for ($i = 0; $i <= $#strings; $i++)
   {
     if ( $strings[$i] =~ /\s+\d+\s+[A-Ya-y\.]+\s+\d+\s+/ )   { $index = $i-1;
     last; }
   }

for ($j = 0; $j < $index-1; $j++) { shift(@strings); }

$k = $#strings;

if ($k < 5) {$num_row_sets = 1;}
else {$num_row_sets = $k / 4;}

shift(@strings);

#for ($x = 1; $x <= $k; $x++)  {print "\$strings[$x] = $strings[$x]";}
#print "\n\n\n";

if ($num_row_sets == 1) { for ($n=1; $n<=3; $n++) {$prot[$n] = $strings[$n];} }

else
    {
```

```perl
    for ($n = 1; $n <= 3; $n++)
        {
        $prot[$n] = $strings[$n];

        if ($n == 2)
             {
              for ($q = 1; $q < $num_row_sets; $q++)
                 { $z = $n + (4 * $q); substr($strings[$z], 0, 9) = "";
chop($prot[2]);
                 $prot[2] = $prot[2] . $strings[$z]; }
             }

        else {    for ($q = 1; $q < $num_row_sets; $q++)
                 { $z = $n + (4 * $q);
                 $prot[$n] = $prot[$n] . $strings[$z]; }
             }
     }
  }

for ($y=1; $y<=3; $y++)    { $prot[$y] =~ s/(\n|\r)//g; chomp($prot[$y]); }

$prot[1] =~ s/ \d+\s+\d+ //g;
$prot[3] =~ s/ \d+\s+\d+ //g;

if ($top_or_bot eq "b") {$homo_prot = $prot[3]; $your_prot = $prot[1]; }
else {$homo_prot = $prot[1]; $your_prot = $prot[3]; }
$alignment = $prot[2];

# extract beginning and ending residue numbers for new MR search model PDB file:

if ($homo_prot =~ /\s+(\d+)\s+.* (\d+)\s*/) { $res_begin = $1; $res_end = $2; }

else {die "\nSomething is really wrong-- there are no residue numbers in homo_prot\n\n";}

# print out current status:

$homo_prot =~ s/\d/ /g;
$your_prot =~ s/\d/ /g;

substr($homo_prot, 0, 9) = "";
substr($your_prot, 0, 9) = "";
substr($alignment, 0, 9) = "";

# Turn the following lines on to print out the input, re-formatted for the next step:
#print "your_prot =$your_prot\n";
#print "alignment =$alignment\n";
#print "homo_prot =$homo_prot\n";
#print "\n\nBegin residue = $res_begin\n";
#print "\nEnd residue = $res_end\n\n";

@your_prot = split(//, $your_prot);
@alignment = split(//, $alignment);
@homo_prot = split(//, $homo_prot);

# Now make the comparisons and write out the verdict array:

for ($w = 0; $w <= $#homo_prot; $w++)
{
 if ($homo_prot[$w] eq ".")    { $verdict[$w] = "correct_the_index"; next;}

 if (($homo_prot[$w] ne ".") & ($your_prot[$w] eq ".")) {$verdict[$w] = "delete";next;}

 if (($alignment[$w] eq ":") | ($alignment[$w] eq "|")) {$verdict[$w] = "conserve";next;}

 if (($homo_prot[$w] =~ /g/i) | ($homo_prot[$w] =~ /a/i)) {$verdict[$w]="conserve";next;}

  $verdict[$w] = "truncate";
```

```perl
}

for ($j=0; $j <= $#homo_prot; $j++) { if ($homo_prot[$j] =~ /[a-yA-Y\.]/) { $index_end =
$j; }  }

#for ($t=0; $t<=$#homo_prot; $t++)
#{print"\$homo_prot[$t]=$homo_prot[$t] \t $verdict[$t] \n";}
#print"\nEnd value = $index_end\n\n";

# this next section corrects the index values so that the final_homo and final_verdict
array indices
# correspond to the actual homology model protein residue numbers, starting from 1 at the
N'

for ($i = 0; $i <= $index_end; $i++)
{
 if ($verdict[$i] eq "correct_the_index") {$index_corrector++; next;}

 $correct_index = $i + $res_begin - $index_corrector;

 $final_homo_prot[$correct_index] = $homo_prot[$i];
 $final_verdict[$correct_index] = $verdict[$i];
}

# turn the next section on to print out the verdict array, i.e. a list of the verdicts
# (delete, truncate, or conserve) for each homology model protein residue

#for ($t=0; $t<=$#final_homo_prot; $t++)
#{print"\$final_homo_prot[$t]=$final_homo_prot[$t] \t $final_verdict[$t] \n";}

# process the PDB file:

open (PDB_INFILE, $pdb_input) || die "Can't open PDB file \"$pdb_input\"\n\n";
open (PDB_OUTFILE, ">$pdb_output") || die "Can't open PDB file \"$pdb_output\" for
writing\n\n";

print "\n\nOutput file \"$pdb_output\" contains the homology model coordinates.\n\n";

while (<PDB_INFILE>)
{
  if (!/^ATOM +\d+/) {print PDB_OUTFILE "$_"; }

  else

   {

      $_ =~ /\s+(\d+)\s+.*\s+(\d+)\s+/ ; $res_num = $2;

      if ($final_verdict[$res_num] eq "delete") {next;}
#the delete verdict

      if ($final_verdict[$res_num] eq "conserve") {print PDB_OUTFILE "$_"; next;}
#the conserve verdict

      if (  ($final_verdict[$res_num] eq "truncate") & ($_ =~ /^ATOM[ 0-9]{9}(N |CA|CB|C
|O )/)    )
          {s/(VAL|LEU|ILE|MET|PRO|PHE|TRP|SER|THR|ASN|GLN|TYR|CYS|LYS|ARG|HIS|ASP|GLU)/ALA/;
           print PDB_OUTFILE "$_"; next;}

   }
}
```

```
#!/bin/csh -f

# a script to automatically perform several GLRF cross rotation function runs at several
# values of 3 user-input parameters (integration radii, resolution windows, and models)
# IMPORTANT: this works only in conjunction with the program GLRF and the perl script
#            "alter.pl", and assumes you've already synthesized structure factors for the
#            various models (e.g., with sfcal.com) in a file called "models.cal".
#
# command line syntax is:
# " >scripted_glrf.sh model_file rad_low rad_high width_of_resol_win "
#
# The output is a file containing the top 5 peaks from each RF run...

set rad_low rad_high resol_win models_file
set curr_rad curr_res_low curr_res_high curr_model

# set the command line-input variables

set models_file = $argv[1]
set rad_low = $argv[2]
set rad_high = $argv[3]
set resol_win = $argv[4]

# regurgitate the user's command line input. don't worry about checking for
# null inputs and setting default values.

echo ""
echo "VARIATION OF THE 1st PARAMETER (integration radius):"
echo "***************************************************************"
echo "GLRF will use a minimum radius of:"    $rad_low
echo "GLRF will use a maximum radius of:"    $rad_high
echo "GLRF will use these radii incremented by 1.0 Angstroms"
echo ""
echo ""
echo "VARIATION OF THE 2nd PARAMETER (resolution):"
echo "***************************************************************"
echo "GLRF will calculate RFs over a sliding window of resolutions,"
echo "starting at a high resolution limit of 3.5 Anstroms, incrementing"
echo "by 0.5 Angstroms, and with a fixed window width of:"  $resol_win "Angstroms"
echo ""
echo ""
echo "VARIATION OF THE 3rd PARAMETER (MR model):"
echo "***************************************************************"
echo "The filename containing the model names is:"  $models_file
echo "and the models specified in this file are:"
echo ""
cat $models_file
echo ""
echo ""

# now the 3 nested loops that are the core of this script:

set allmodels = `cat $models_file`

bc << eof_bc > temp999
`echo $resol_win` * 10
quit
eof_bc

set resol_win = `awk '{print $1}' temp999 `

set absolut_hi_res = ` expr 35 + $resol_win `

foreach curr_model ( ${allmodels} )
```

```
        set prefix = `(echo $curr_model | awk '{split($1,a,"."); print a[1]}')`
        set curr_rad = $rad_low

        while ($rad_low <= $curr_rad && $curr_rad <= $rad_high)

            set curr_res_low = 35; set curr_res_high = `expr $curr_res_low + $resol_win`

            while (35 <= $curr_res_low && $curr_res_low <= $curr_res_high)

                echo "running alter.pl ${curr_model} radius=${curr_rad}
res.range=${curr_res_low}-${curr_res_high} ..." >> big_run.log
                rm scripted_cross_rf.prt
                rm cross_rf_temp.com
                alter.pl ${curr_model} radius=${curr_rad} res.range=${curr_res_low}-
${curr_res_high}
                chmod 755 cross_rf_temp.com
                cross_rf_temp.com
                egrep "^         +[0-9]+ +[0-9]+ +[0-9]+ +[0-9]+ +[0-9]+\."
scripted_cross_rf.prt | awk '{print $12}' | head -5 >> big_run.log
                set curr_res_low = `expr $curr_res_low + 5`
                set curr_res_high = `expr $curr_res_high + 5`
                if ($curr_res_low >= $absolut_hi_res) break

            end

            set curr_rad = `expr $curr_rad + 1`
            echo ""

        end

        echo "" ; echo "******************************" ; echo ""
end

rm temp999
```

```perl
#!/joule2/programs/bin/perl

open (IN, "scripted_cross_rf.com") || die "Cannot open file \"scripted_cross_rf.com\"\n";
open (OUT, ">>cross_rf_temp.com");

$model = "$ARGV[0]";
$radius = "$ARGV[1]"; $radius =~ /(\d+)/; $radius = $1;
$res_range = "$ARGV[2]"; $res_range =~ /(\d+)-(\d+)/; $res_hi = $1/10; $res_low = $2/10;

print "$model\n$radius\n$res_hi\n$res_low\n" ;

while (<IN>)
        {
        chomp ($_);
        if ($_ =~ /aobsfile/)
         {
          $new = "aobsfile MODELS\/$model";
          print OUT "$new\n";
          next;
         }

        if ($_ =~ /resolution/)
         {
          $new = "resolution $res_low $res_hi";
          print OUT "$new\n";
           next;
         }

        if ($_ =~ /radius/)
         {
          $new = "radius $radius";
          print OUT "$new\n";
           next;
         }

        print OUT "$_\n";
        next;

         }
```

```perl
#!/joule2/programs/bin/perl

# this script processes the output from scripted_glrf.sh into a format good for excel or some
# other spreadsheet.

$infile = $ARGV[0];

open (IN, $infile) || die "Cannot open file \"$infile\"\n";

print "MODEL\t\tInteg_radius\tHI RESOL LIM\tTOP 5 PEAKS in order\tLARGEST diff\n" ;
print "***************************************************************************\n";

while (<IN>)
        {
        chomp ($_);
        if ( $_ =~ /running alter\.pl (\w+\.cal) radius=(\d+) res\.range=(\d+)-\d+/ )
          {
           $model = $1; $radius = $2; $res_high_lim = $3/10;
           $verdict = 1; $peaks = ""; next ;
          }

          else {
                  $peaks = $peaks ."\t". $_ ; $verdict++;
                  if ($verdict == 6)
                          { @top5 = split(/\s+/, $peaks);
# Now calculate the diff btwn peak 1 & peak 2,
# since this is what's really interesting for RF
                            $diff = $top5[1] - $top5[2];
                            print "$model\t$radius\t\t$res_high_lim\t$peaks\t$diff\n"; next ;
                          }
                }

          next ;

        }
```

```
#!/joule2/programs/bin/perl
#
# This is a script that reads in 2 input files: argv[0] is the PDB file
# and argv[1] is the file tabulating conserved residues in the format:
#
# "res#,ss(super strong)|s(trong)|m(edium)|w(eak)" ...
#
# where "res#" is the residue # and "superstrong", "strong", "medium",
# or "weak" specify how strongly that site is conserved.
# The output is a PDB file with all of the B-factors flattened to 20.00,
# except for the "superstrong", "strong", "medium", or "weak" sites, which
# are assigned B-values of 90.00, 70.00, 40.00, or 33.00, respectively.
#
# This is useful for programs like GRASP, which can color a surface by the
# "B-factor" field of the PDB file.
#
# Cameron Mura (July 2001)
# NOTE: make sure occupancies are 1.00 for any atoms with B-factors you want changed

$pdb_in = $ARGV[0];
$site_file = $ARGV[1];
$bfac = "";
@cons = "";
$site_line = "";

open (PDB, $pdb_in) || die "Cannot open file \"$pdb_in\"\n";

while (<PDB>)
        {
        $line = $_;
        chomp ($line);
        $resnum = ""; $bfac = "";
        if ($line =~ /(^(ATOM|HET)\s+\d+\s+[\w\*]+\s+\w\w\w [A-
Z]\s+)(\d+)(.+)(1\.00\s+)(\d+\.\d+)(.*)/)
                {
                $resnum = $3; $bfac = "20.00";
                        $counter = 0;
                        open (SITES, $site_file) || die "Cannot open file \"$site_file\"\n";
                        while (<SITES>)
                        {if ($counter == 0)
                            {
                            $site_line = $_; chomp ($site_line); @cons = "";
                            @cons = split(/,/, $site_line);
                            if ($cons[0] == $resnum and $cons[1] eq "ss" )
                                {$bfac = "95.00"; $resnum++; last;}
                            if ($cons[0] == $resnum and $cons[1] eq "s" )
                                {$bfac = "65.00"; $resnum++; last;}
                            if ($cons[0] == $resnum and $cons[1] eq "m" )
                                {$bfac = "45.00"; $resnum++; last;}
                            if ($cons[0] == $resnum and $cons[1] eq "w" )
                                {$bfac = "30.00"; $resnum++; last;}

                            next;
                            }
                        }
                    close (SITES);

        if ($bfac eq "") { $bfac = "20.00";}
        $crap = $5 . $bfac;
        print "$1$3$4$crap$7\n";
        }
    }
```

```
#!/bin/tcsh -f
# A script to calculate and expand NCS for any type of multimer ...
#
# For example, for the SmAP3 28-mer, subunit B was built most extensively. Now, one can
# use this script to create a 28-mer with each subunit being a copy of 'B'. The script
# figures out the NCS operators and applies them (via "align_v2" and CCP4 "pdbset"
# programs) to get several transformed PDB files ...
#
# command line syntax: >symm.csh inputPDBfile probechain
#
# where "inputPDBfile" is the COMPLETE multimer (e.g. SmAP3 28-mer), and
#       "probechain" is the CHAINID for the monomeric subunit you want to
#                     have copied 28-times over...
# Cameron Mura, May 2002

# get input ready
#
echo "";
set infile = $argv[1]
set probechain = $argv[2]
set prefix = `(echo $infile | awk '{split($1,a,"."); print a[1]}')`
set align_out = $prefix.align_out
echo "INPUT FILE IS:   "$infile; echo ""
echo "CHAINid USED TO EXPAND 28-mer: "$probechain; echo ""
echo "PREFIX FOR SPLIT FILES WILL BE: "$prefix ; echo ""
echo "PAIRWISE ALIGNMENTS WILL BE STORED IN: "$align_out ; echo ""

# split input file into several files separated by chainID
#
al_moleman << EOF_moleman > moleman.log.temp

${infile}
split
${prefix}_
quit
EOF_moleman

# re-name files so chainIDs get renumbered to ASCII equivalents
#
set num_files_deci = `ls ${prefix}_[0-9].pdb | wc -l`
set num_files_char = `ls ${prefix}_[a-z].pdb | wc -l`
set num_files_total = `expr $num_files_deci + $num_files_char`
echo "Number of files_decimal:        "$num_files_deci; echo ""
echo "Number of files_character:  "$num_files_char; echo ""
echo "Number of files_total:  "$num_files_total; echo ""

# must rename the decimal ones before the character ones, otherwise they'll
# definitely get overwritten by the a, b, c, etc...

set x = 1
while ($x <= $num_files_deci)
set rename_value = `expr $x + $num_files_char`
mv ${prefix}_${x}.pdb ${prefix}_${rename_value}.pdb
set x = `expr $x + 1`
end

set index1 = 1
while ($index1 <= $num_files_char)
set temp_ind = `expr $index1 + 96`
set to_rename2 = `perl -e 'print(chr($ARGV[0]),"\n")' $temp_ind `
mv ${prefix}_${to_rename2}.pdb ${prefix}_${index1}.pdb
set index1 = `expr $index1 + 1`
end
```

```
### HERE IS THE ALIGNMENT PART: ###########
###
### include the following line to do both all_atom- and main_chain-mode alignments:
#foreach mode (all_atom main_chain)
foreach mode (main_chain)

if ($mode == "all_atom") set option = 1
if ($mode == "main_chain") set option = 2

echo "=========================================================="; echo;
echo "ALIGNMENT MODE is:" \"$mode\" \(align_v2 option $option\) ; echo;
echo "=========================================================="; echo;

set i1 = 1

while ($i1 <= $num_files_total)

## comment-in the next line to do just an (upper) diagonal matrix of pairs:
#set i2 = $i1
# leave in THIS NEXT LINE to do full symmetric matrix worth of entries:
set i2 = 1

while ($i2 <= $num_files_total)

echo "for aligning" $i2 "to" $i1 "the rmsd is:"

### here's the "align_v2" part, don't mess with it:

align_v2 << eof > temp_${i1}_to_${i2}.log

${option}

${prefix}_${i1}.pdb

PDB
${prefix}_${i2}.pdb

PDB
AUTO
${prefix}_${i2}_to_${i1}.rot.temp
${prefix}_${i2}_to_${i1}.show.temp
stop
eof

cat ${prefix}_${i2}_to_${i1}.show.temp | grep 'rms deviation'
cat ${prefix}_${i2}_to_${i1}.show.temp | grep 'corresponds to a'
echo "------------------------------------------------------------------"; echo;

####### apply the transformation matrix from CCP4:

write_pdbset.pl ${prefix}_${i2}_to_${i1}.show.temp ${prefix}_${i2}.pdb
${prefix}_${i2}_on_${i1}.new.pdb > pdbset_${i2}_to_${i1}.com
chmod 755 pdbset_${i2}_to_${i1}.com
pdbset_${i2}_to_${i1}.com > pdbset_${i2}_to_${i1}.log

echo "------------------------------------------------------------------"; echo;
echo "Transforming subunit ${i2} into position of ${i1} (via PDBSET) ..."; echo;
echo "------------------------------------------------------------------"; echo;
```

```
########### clean-up stuff ###########
rm temp_${i1}_to_${i2}.log
rm pdbset_${i2}_to_${i1}.log
rm ${prefix}_${i2}_to_${i1}.show.temp
rm ${prefix}_${i2}_to_${i1}.rot.temp
rm pdbset_${i2}_to_${i1}.com
####################################

set i2 = `expr $i2 + 1`

end

##############
# include following section to cat together the rotated files and originals into one
# final file:
#cat *to_`echo $i1`_fr.rot gpa_mono_`echo $i1`_fr.pdb > all_7rot_to_`echo $i1`.pdb
#echo "*******************************************************************";echo;
#echo Writing out PDB file of superimposed structures: \"all_7rot_to_`echo $i1`.pdb\";
echo
#echo "*******************************************************************";echo;
##############

set i1 = `expr $i1 + 1`

end

#end
```

```perl
#!/joule2/programs/bin/perl

# A Perl script to convert align_v2-formatted .show file to pdbset.com file
# (just change CELL and SPACEGROUP cards below to generalize this) ...
# Is meant to work in conjunction with "symm.csh" script to expand one subunit
# of the SmAP3 28-mer into 28 copies...was written in generalizable format.
#
# Cameron Mura (May 2002)

$show_in = $ARGV[0];
$pdbin = $ARGV[1];
$pdbout = $ARGV[2];
$on = 0;

open (IN, $show_in) || die "Cannot open file \"$show_in\"\n";
#open (OUT, ">temp_pdbset.com") || die "Cannot open file \"temp_pdbset.com\"\n";

while (<IN>)
        {
         $line = $_;
         chomp ($line);
         if ($line =~ /Rotation matrix applied to latest set\: +([0-9\.\-]+ +[0-9\.\-]+
+[0-9\.\-]+\s*)/)
             {$mat123 = $1; $on++; next;}
          if ($on == 1 & $line =~ / +([0-9\.\-]+ +[0-9\.\-]+ +[0-9\.\-]+\s*)/)
              {$mat456 = $1; $on++; next;          }
          if ($on == 2 & $line =~ / +([0-9\.\-]+ +[0-9\.\-]+ +[0-9\.\-]+\s*)/)
              {$mat789 = $1; $on++; next; }
          if ($on == 3 & $line =~ / +([0-9\.\-]+ +[0-9\.\-]+ +[0-9\.\-]+\s*)/)
              {$mat101112 = $1; $on++; next; }
          if ($on >= 4) { last ; }
         }

# print out the PDBSET lines:
# (note that in the symm.csh script this is re-directed to a PDBSET .com file, which
#  could be done at this stage instead (i.e. in this perl script), if necessary...)

print "#!\/bin\/csh -f\n";
print "pdbset xyzin $pdbin xyzout $pdbout <<eof-1\n";
print "transform\t$mat123 -\n";
print "\t\t$mat456 -\n";
print "\t\t$mat789 -\n";
print "\t\t$mat101112 \n\n";
print "CELL 83.322   172.428    148.108    90.000    89.986    90.000\n";
print "SPACEGROUP P21\n";
print "eof-1\n";

# THE END!
```